\documentclass[twocolumn,pra,showpacs,superscriptaddress,amssymb,amsmath,amsmath]{revtex4-1}
\usepackage{graphicx}
\usepackage{epstopdf}
\usepackage{comment}
\usepackage{hyperref}
\usepackage{bm}

\hypersetup{pdfstartpage=1,pdfmenubar=true,pdfpagemode=None,pdftoolbar=true,
urlcolor=blue,linkcolor=blue,colorlinks = true,citecolor=blue, bookmarksopen=false}

\begin{document}
\title{\textit{Ab initio} electronic structure and prospects for the formation \\ of ultracold calcium--alkali-metal-atom molecular ions}

\author{Wissem Zrafi}
\affiliation{Laboratory of Interfaces and Advanced Materials, Physics Department, Faculty of Science, University of Monastir, Avenue de l'Environnement, 5019 Monastir, Tunisia}
\author{Hela Ladjimi}
\affiliation{Laboratory of Interfaces and Advanced Materials, Physics Department, Faculty of Science, University of Monastir, Avenue de l'Environnement, 5019 Monastir, Tunisia}
\author{Halima Said}
\affiliation{Laboratory of Interfaces and Advanced Materials, Physics Department, Faculty of Science, University of Monastir, Avenue de l'Environnement, 5019 Monastir, Tunisia}
\author{Hamid Berriche}
\email{hamidberriche@yahoo.fr}
\affiliation{Laboratory of Interfaces and Advanced Materials, Physics Department, Faculty of Science, University of Monastir, Avenue de l'Environnement, 5019 Monastir, Tunisia}
\affiliation{Department of Mathematics and Natural Sciences, School of Arts and Sciences, American University of Ras Al Khaimah, Ras Al Khaimah, United Arab Emirates}
\author{Micha\l~Tomza}
\email{michal.tomza@fuw.edu.pl}
\affiliation{Faculty of Physics, University of Warsaw, Pasteura 5, 02-093 Warsaw, Poland}

\date{\today}

\begin{abstract}

Experiments with cold ion-atom mixtures have recently opened the way for the production and application of ultracold molecular ions. Here, in a comparative study, we theoretically investigate ground and several excited electronic states and prospects for the formation of molecular ions composed of a calcium ion and an alkali-metal atom: CaAlk$^{+}$ (Alk=Li, Na, K, Rb, Cs). We use a quantum chemistry approach based on non-empirical pseudopotential, operatorial core-valence correlation, large Gaussian basis sets, and full configuration interaction method for valence electrons. Adiabatic potential energy curves, spectroscopic constants, and transition and permanent electric dipole moments are determined and analyzed for the ground and excited electronic states. We examine the prospects for ion-neutral reactive processes and the production of molecular ions via spontaneous radiative association and laser-induced photoassociation. After that, spontaneous and stimulated blackbody radiation transition rates are calculated and used to obtain radiative lifetimes of vibrational states of the ground and first-excited electronic states. The present results pave the way for the formation and spectroscopy of calcium--alkali-metal-atom molecular ions in modern experiments with cold ion-atom mixtures.

\end{abstract}

\maketitle

\section{Introduction}
Cold mixtures of alkaline-earth-metal ions and alkali-metal atoms have recently emerged as a new field of research at the crossroad of quantum physics and chemistry~\cite{TomzaRMP19,CoteAAMOP16,HarterCP14}. Hybrid systems of trapped ions and ultracold atoms combined in a single experimental setup have opened the way for exciting applications ranging from studying cold collisions~\cite{CotePRA00,ZipkesNature10,RaviNatCommun12,HoltkemeierPRL16,Feldker2019} and controlled chemical reactions~\cite{SikorskyNC18,MakarovPRA03,RatschbacherNatPhys12,FurstPRA18,SikorskyPRL18,CotePRL18,SaitoPRA17} to  quantum simulation of solid-state physics~\cite{GerritsmaPRL12,BissbortPRL13,SchurerPRL17} and quantum computation~\cite{DoerkPRA10}. 
Most of the cold ion-atom experiments use alkaline-earth-metal ions and alkali-metal or alkaline-earth-metal atoms because of their electronic structure favorable for laser cooling. Several cold atomic ion-atom combinations have already been experimentally investigated, including Yb$^+$/Yb~\cite{GrierPRL09}, Ca$^+$/Rb~\cite{HallMP13a}, Ba$^+$/Ca~\cite{SullivanPRL12}, Yb$^+$/Ca~\cite{RellergertPRL11}, Yb$^+$/Rb~\cite{ZipkesNature10}, Ca$^+$/Li~\cite{HazePRA13}, Ca$^+$/Rb~\cite{HallPRL11}, Ca$^+$/Na~\cite{SmithAPB14}, Sr$^+$/Rb~\cite{MeirPRL16}, Yb$^+$/Li~\cite{FurstPRA18}, Ca$^+$/K~\cite{JyothiRSI19}, Ba$^+$/Rb~\cite{SchmidtPRL2020}, and Rb$^+$/Rb~\cite{RaviNatCommun12,HartePRL12}. Ca$^+$ ions are the most common choice because advanced methods of their manipulation and detection have been developed for quantum simulation and computation applications~\cite{SingerRMP10,SchneiderRPP12}.

In cold ion-atom mixtures, diatomic molecular ions can be produced via collision-induced charge-transfer radiative association or light-induced photoassociation~\cite{MakarovPRA03,IdziaszekNJP11,TomzaPRA15a,daSilvaNJP2015,GacesaPRA16,PetrovJCP17}. The radiative association is predicted to dominate charge-transfer processes in most of the mixtures of alkaline-earth-metal ions and alkali-metal atoms. Until now,  CaRb$^+$~\cite{HallPRL11,HallMP13a}, BaRb$^+$~\cite{HallMP13b}, CaYb$^+$~\cite{RellergertPRL11}, CaBa$^+$~\cite{SullivanPRL12} molecular ions were observed as products of cold collisions between respective atomic ions and atoms. On the other hand, Rb$_2^+$~\cite{JyothiPRL16} and Ca$_2^+$~\cite{SullivanPCCP11} molecular ions were formed by the photoionization of ultracold molecules. Molecular ions, which possess a ro-vibrational structure, can likewise be employed to realize cold, controlled ion-atom chemistry~\cite{DeiglmayrPRA12,PuriScience17,KilajNC18,DorflerNC19,PuriNatChem19} and precision measurements~\cite{GermannNatPhys14,Cairncross17}.

A mixture of Ca$^+$ ions and Na atoms was considered in the pioneering proposal, which suggested combining ions trapped in a Paul trap with ultracold atoms~\cite{MakarovPRA03}. Such mixture was later experimentally realized~\cite{SmithAPB14}. Coulomb crystals of Ca$^+$ ions were immersed into ultracold Rb atoms to study radiative charge exchange and molecular ions formation~\cite{HallPRL11,HallMP13a,EberleCPC16}. Ca$^+$ ions were also experimentally studied in mixtures with ultracold Li atoms~\cite{HazePRA13,HazePRA15,HazePRL18,SaitoPRA17}. In fact, Ca$^+$/Li ion-atom combination is one of the most promising systems for reaching the quantum regime of ion-atom collisions~\cite{Feldker2019} due to the favorable mass ration reducing the impact of micromotion-induced heating in hybrid traps~\cite{CetinaPRL12}. Recently, a new apparatus with a mixture of laser-cooled Ca$^+$ ions in a linear Paul trap overlapped with ultracold K atoms in a magneto-optical trap was presented~\cite{JyothiRSI19}. This setup incorporates a high-resolution time-of-flight mass spectrometer designed for radial extraction and detection of reaction products opening the way for detailed studies of the state-selected formation of CaK$^+$ molecular ions. While, the electronic structure of the ground and excited electronic states was already studied for the CaLi$^+$~\cite{Smialkowski2019,Bala2019,SaitoPRA17,HabliMP16,Xie2005,Russon1998,Kimura1983}, CaNa$^+$~\cite{MakarovPRA03,Gacesa2016,JellaliJQSRT18,Smialkowski2019}, and CaRb$^+$~\cite{Tacconi2011,Felix2013,daSilvaNJP2015,Smialkowski2019} molecular ions, to the best of our knowledge, the structure of excited electronic states of the CaK$^+$ and CaCs$^+$ molecular ions has not been presented, yet.

Here, to fill this gap, in a comparative study, we investigate the electronic structure of the group of five diatomic molecular ions composed of a Ca$^+$ ion interacting with an alkali-metal atom: CaAlk$^+$ (Alk=Li, Na, K, Rb, Cs). We calculate ground and several low-lying excited electronic states using a theoretical quantum chemistry approach based on non-empirical pseudopotential, operatorial core-valence correlation, large Gaussian basis sets, and full configuration interaction method for valence electrons. Next, we employ electronic structure data to access prospects for field-free and light-assisted ion-neutral reactive processes and the formation of the considered molecular ions via spontaneous radiative association and laser-induced photoassociation. We discuss similarities and differences between considered systems. Finally, we calculate spontaneous and stimulated blackbody radiation transition rates together with radiative lifetimes of vibrational states of the ground and first excited electronic states.

This paper has the following structure. Section~\ref{sec:met} describes the used computational methods. Section~\ref{sec:res} presents and discusses obtained results, including electronic structure data and spontaneous charge transfer and radiative association rates. The radiative lifetimes of the ground and excited states are also presented there. The experimental implications of the presented calculations are analyzed in detail. To conclude, section~\ref{sec:con} summarizes our work.

\section{Computational details}
\label{sec:met}

In this work, we calculate non-relativistic potential energy curves within the Born-Oppenheimer approximation for the ground and excited electronic states of calcium--alkali-metal-atom molecular ions: CaAlk$^{+}$ (Alk=Li, Na, K, Rb, Cs). To this end, we employ the \textit{ab initio} approach, which was developed and presented previously in several works on alkali hydrides~\cite{Berriche1995,Khelifi2002,Zarfi2006,Khelifi2001}, alkali-metal dimers~\cite{Mabrouk2014,Mabrouk2008,Mabrouk2010,Jendoubi2012}, alkaline-earth-metal hydrides~\cite{Aymar2012, Habli2013,Belayouni2016}, and alkali-metal--alkaline-earth-metal molecular ions~\cite{ElOualhazi2016,Aymar2011}. The investigated CaAlk${}^{+}$ molecular ions, thus, are treated effectively as two-electron systems with efficient non-empirical pseudopotentials in their semi-local form~\cite{Durand1975} used to replace core electrons. Additionally to the pseudopotential treatment, the self-consistent field (SCF) computations are followed by a full valence configuration interaction (FCI) calculations using the CIPCI algorithm (Configuration Interaction by Perturbation of a multiconfiguration wave function Selected Iteratively) of the standard succession of programs developed by the ``Laboratoire de Chimie et Physique de Toulouse''. The core-valence electronic correlations between the polarizable Ca$^{2+}$ core with the valence electrons and polarizable Alk$^+$ cores are included by using core polarization potentials (CPP)~\cite{Müller1984}.

Valence electrons are described with large Gaussian basis sets. For Ca the $(8s,7p,7d)/[8s,7p,5d]$ basis set developed in Ref.~\cite{Habli2013} together with the corresponding $s$, $p$, and $d$ cut-off parameters of the pseudopotential ($\rho_s^{\text{Ca}^+}=1.77405$, $\rho_p^{\text{Ca}^+}=1.81$, $\rho_d^{\text{Ca}^+}=1.691$) optimized in order to reproduce the ionization energy of the Ca${}^{+}$ ion are employed. For Li, Na, K, Rb, and Cs, respectively, the $(9s,8p,6d)/[8s,6p,5d]$, $(7s,6p,5d)/[6s,5p,4d]$, $(8s,6p,3d]/[8s,5p,5d]$, $(7s,4p,5d)/[6s,4p,4d]$, and $(7s,4p,5d)/[6s,4p,4d]$ basis sets are used together with the following cut-off parameters ($\rho_s^\text{Li}=1.434$, $\rho_p^\text{Li}=0.982$, $\rho_d^\text{Li}=0.6$), ($\rho_s^\text{Na}=1.4423$, $\rho_p^\text{Na}=1.625$, $\rho_d^\text{Na}=1.5$), ($\rho_s^\text{K}=2.115$, $\rho_p^\text{K}=2.1125$, $\rho_d^\text{K}=1.983$), ($\rho_s^\text{Rb}=2.5213$, $\rho_p^\text{Rb}=2.2790$, $\rho_d^\text{Rb}=2.5110$), ($\rho_s^\text{Cs}=2.690$, $\rho_p^\text{Cs}=1.850$, $\rho_d^\text{Cs}=2.810$)~\cite{Berriche1995,Khelifi2002,Zarfi2006,Khelifi2001,Mabrouk2014,Mabrouk2008,Mabrouk2010,Jendoubi2012}. The core polarizabilities of the Ca$^+$ ion and alkali atoms are taken to be $\alpha^{\text{Ca}^{2+}}=3.1717\,a^3_0$, $\alpha^{\text{Li}^+}=0.1917\,a^3_0$, $\alpha^{\text{Na}^+}=0.9930\,a^3_0$, $\alpha^{\text{K}^+}=5.457\,a^3_0$, $\alpha^{\text{Rb}^+}=9.245\,a^3_0$ and $\alpha^{\text{Cs}^+}=15.117a^3_0$~\cite{Berriche1995,Khelifi2002,Zarfi2006,Khelifi2001,Mabrouk2014,Mabrouk2008,Mabrouk2010,Jendoubi2012,Habli2013,Mitroy2008}. Employed models reproduce atomic properties with good accuracy when compared with previous theoretical results and experimental measurements~\cite{Berriche1995,Khelifi2002,Zarfi2006,Khelifi2001,Mabrouk2014,Mabrouk2008,Mabrouk2010,Jendoubi2012}.

In the present work, the interaction of the considered alkali-metal and alkaline-earth-metal atoms and ions in the ground and different excited electronic states results in different molecular electronic states of the singlet or triplet $\Sigma^+$, $\Pi$, and $\Delta$ symmetries. The lowest seven atomic thresholds for each of the considered CaAlk$^+$ molecular ions, together with their valence energies and associated molecular electronic states, are collected in Table~\ref{tab:asymptotes}. The calculated energies of Ca$^+$($^2S$)+Alk($^2S$) limits, which describe essential ground-state collisions of alkaline-earth-metal ions with alkali-metal atoms, agree very well (within 5$\,$cm$^{-1}$) with experimental values. Description of the $^1D$ and $^3D$ excited electronic states of the Ca atom is the most challenging with discrepancies of 597$\,$cm$^{-1}$ and 566$\,$cm$^{-1}$ for related atomic limits, respectively. Nevertheless, the overall agreement is good, suggesting good accuracy of molecular calculations.

The spectroscopic constants are extracted from the \textit{ab initio} points interpolated using the cubic spline method. The permanent and transition electric dipole moments are calculated as expectation values of the dipole operator with the calculated electronic wavefunctions. The $z$ axis is chosen along the internuclear axis and is oriented from a Ca atom to an alkali-metal atom. The origin is set in the center of mass. Masses of the most abundant isotopes are assumed within the paper.

The time-independent Schr\"{o}dinger equation for the nuclear motion is solved using the renormalized Numerov algorithm~\cite{JohnsonJCP78} for both bound~\cite{Berriche1995} and continuum states~\cite{TomzaPRA15a}. Rate constants for elastic scattering and inelastic charge-exchange reactive collisions are calculated as implemented and described in Refs.~\cite{TomzaPRA15a,TomzaPRA15b}. The wave functions are propagated to large interatomic distances, and the $K$ and $S$ matrices are extracted by imposing the long-range scattering boundary conditions in terms of the Bessel functions. The elastic rate constants and scattering lengths are obtained from the $S$ matrix for the entrance channel, while inelastic rate constants are computed using the Fermi golden rule type expressions based on the Einstein coefficients between bound and continuum nuclear wave functions of relevant electronic states.

\begin{table*}[tb!]
\caption{Lowest atomic asymptotes, their experimental $E_\text{Exp}$ and calculated $E_\text{Th}$ valence energies, and associated molecular electronic states of the CaAlk$^+$ (Alk=Li, Na, K, Rb, Cs) molecular ions. Experimental energies are averaged on spin-orbit manifold if appropriate. $\Delta E=E_\text{Th}$-$E_\text{exp}$.}\label{tab:asymptotes}
\begin{tabular*}{\textwidth}{@{\extracolsep{\fill}}llrrr}\hline
Asymptote & Molecular states & $E_\text{Exp}\,$(cm$^{-1}$)~\cite{nist2018} &$E_\text{Th}\,$(cm$^{-1})$ & $|\Delta{E}|\,$(cm$^{-1})$ \\ \hline
\multicolumn{5}{c}{\textbf{CaLi$^+$}} \\
Ca($^1S$) + Li$^+$($^1S$) & $^1\Sigma^+$ & -145058 & -144904 & 154 \\ 
Ca$^+$($^2S$) + Li($^2S$) & $^1\Sigma^+$, $^3\Sigma^+$ & -139239 & -139240 & 1 \\ 
Ca($^3P$) + Li$^+$($^1S$) & $^3\Sigma^+$, $^3\Pi$ & -129795 & -129624 & 171 \\ 
Ca$^+$($^2D$) + Li($^2S$) & $^1\Sigma^+$, $^3\Sigma^+$, $^1\Pi$, $^3\Pi$, $^1\Delta$, $^3\Delta$ & -125550 & -125592 & 42 \\ 
Ca($^3D$) + Li$^{+}$($^1S$) & $^3\Sigma^+$, $^3\Pi$, $^3\Delta$ & -124702 & -124136 & 566 \\ 
Ca$^+$($^2S$) + Li($^2P$) & $^1\Sigma^+$, $^3\Sigma^+$, $^1\Pi$, $^3\Pi$ & -124232 & -124337 & 105 \\ 
Ca($^1D$) + Li$^+$($^1S$) & $^1\Sigma^+$, $^1\Pi$, $^1\Delta$ & -123208 & -122611 & 597 \\ 
\hline
\multicolumn{5}{c}{\textbf{CaNa$^+$}} \\ 
Ca($^1S$) + Na$^+$($^1S$) & $^1\Sigma^+$ & -145058 & -144904 & 154 \\ 
Ca$^+$($^2S$) + Na($^2S$) & $^1\Sigma^+$, $^3\Sigma^+$ & -137201 & -137203 & 2 \\ 
Ca($^{3}P$) + Na$^+$($^1S$) & $^3\Sigma^+$, $^3\Pi$ & -129795 & -129624 & 171 \\ 
Ca ($^3D$) + Na$^+$($^1S$) & $^3\Sigma^+$, $^3\Pi$, $^3\Delta$ & -124702 & -124136 & 566 \\ 
Ca$^+$($^2D$) + Na($^2S$) & $^1\Sigma^+$, $^3\Sigma^+$, $^1\Pi$, $^3\Pi$, $^1\Delta$, $^3\Delta$  & -123514 & -123554 & 40 \\ 
Ca($^1D$) + Na$^+$($^1S$) &$^1\Sigma^+$, $^1\Pi$, $^1\Delta$ & -123208 & -122611 & 597 \\ 
Ca($^1P$) + Na$^+$($^1S$) & $^1\Sigma^+$, $^1\Pi$ & -121405 & -121574 & 169 \\ 
\hline
\multicolumn{5}{c}{\textbf{CaK$^+$}} \\ 
Ca($^1S$) + K$^+$($^1S$) & $^1\Sigma^+$ & -145058 & -144904 & 154 \\ 
Ca$^+$($^2S$) + K($^2S$) & $^1\Sigma^+$, $^3\Sigma^+$ & -130760 & -130763 & 3 \\ 
Ca($^3P$) + K$^+$($^1S$) & $^3\Sigma^+$, $^3\Pi$  & -129795 & -129624 & 171 \\ 
Ca($^3D$) + K$^+$($^1S$) & $^3\Sigma^+$, $^3\Pi$, $^3\Delta$ & -124702 & -124136 & 566 \\ 
Ca($^1D$) + K$^+$($^1S$) & $^1\Sigma^+$, $^1\Pi$, $^1\Delta$ & -123208 & -122611 & 597 \\ 
Ca($^1P$) + K$^+$($^1S$) & $^1\Sigma^+$, $^1\Pi$ & -121405 & -121574 & 169 \\ 
Ca$^+$($^1S$) + K($^2P$) & $^1\Sigma^+$, $^3\Sigma^+$, $^1\Pi$, $^3\Pi$ & -117739 & -117757 & 18 \\ 
\hline
\multicolumn{5}{c}{\textbf{CaRb$^+$}} \\ 
Ca($^1S$) + Rb$^+$($^1S$) & $^1\Sigma^+$ & -145058 & -144904 & 154 \\ 
Ca($^3P$) + Rb$^+$($^1S$) & $^3\Sigma^+$, $^3\Pi$  & -129795 & -129624 & 171 \\ 
Ca$^+$($^2S$) + Rb($^2S$) & $^1\Sigma^+$, $^3\Sigma^+$ & -129443 & -129445 & 2 \\ 
Ca($^3D$) + Rb$^+$($^1S$) & $^3\Sigma^+$, $^3\Pi$, $^3\Delta$ & -124702 & -124136 & 566 \\ 
Ca($^1D$) + Rb$^+$($^1S$) &$^1\Sigma^+$, $^1\Pi$, $^1\Delta$ & -123208 & -122611 & 597 \\ 
Ca($^1P$) + Rb$^+$($^1S$) & $^1\Sigma^+$, $^1\Pi$ & -121405 & -121574 & 169 \\ 
Ca$^+$($^2S$) + Rb($^2P$) & $^1\Sigma^+$, $^3\Sigma^+$, $^1\Pi$, $^3\Pi$ & -116705 & -116724 & 19 \\ 
\hline
\multicolumn{5}{c}{\textbf{CaCs$^+$}} \\ 
Ca($^1S$) + Cs$^+$($^1S$) & $^1\Sigma^+$ & -145058 & -144904 & 154 \\ 
Ca($^3P$) + Cs$^+$($^1S$) & $^3\Sigma^+$, $^3\Pi$ & -129795 & -129624 & 171 \\ 
Ca$^+$($^2S$) + Cs($^2S$) & $^1\Sigma^+$, $^3\Sigma^+$ & -127158 & -127163 & 5 \\ 
Ca($^3D$) + Cs$^+$($^1S$) & $^3\Sigma^+$, $^3\Pi$, $^3\Delta$ & -124702 & -124136 & 566 \\
Ca($^1D$) + Cs$^+$($^1S$) & $^1\Sigma^+$, $^1\Pi$, $^1\Delta$ & -123208 & -122611 & 597 \\ 
Ca($^1P$) + Cs$^+$($^1S$) & $^1\Sigma^+$, $^1\Pi$  & -121405 & -121574 & 169 \\ 
Ca$^{+}$($^2S$) + Cs($^2P$) & $^1\Sigma^+$, $^3\Sigma^+$, $^1\Pi$, $^3\Pi$ & -115611 & -115619 & 8 \\ \hline
\end{tabular*}
\end{table*}

\begin{figure}[tb!]
\centering
\includegraphics[width=\columnwidth]{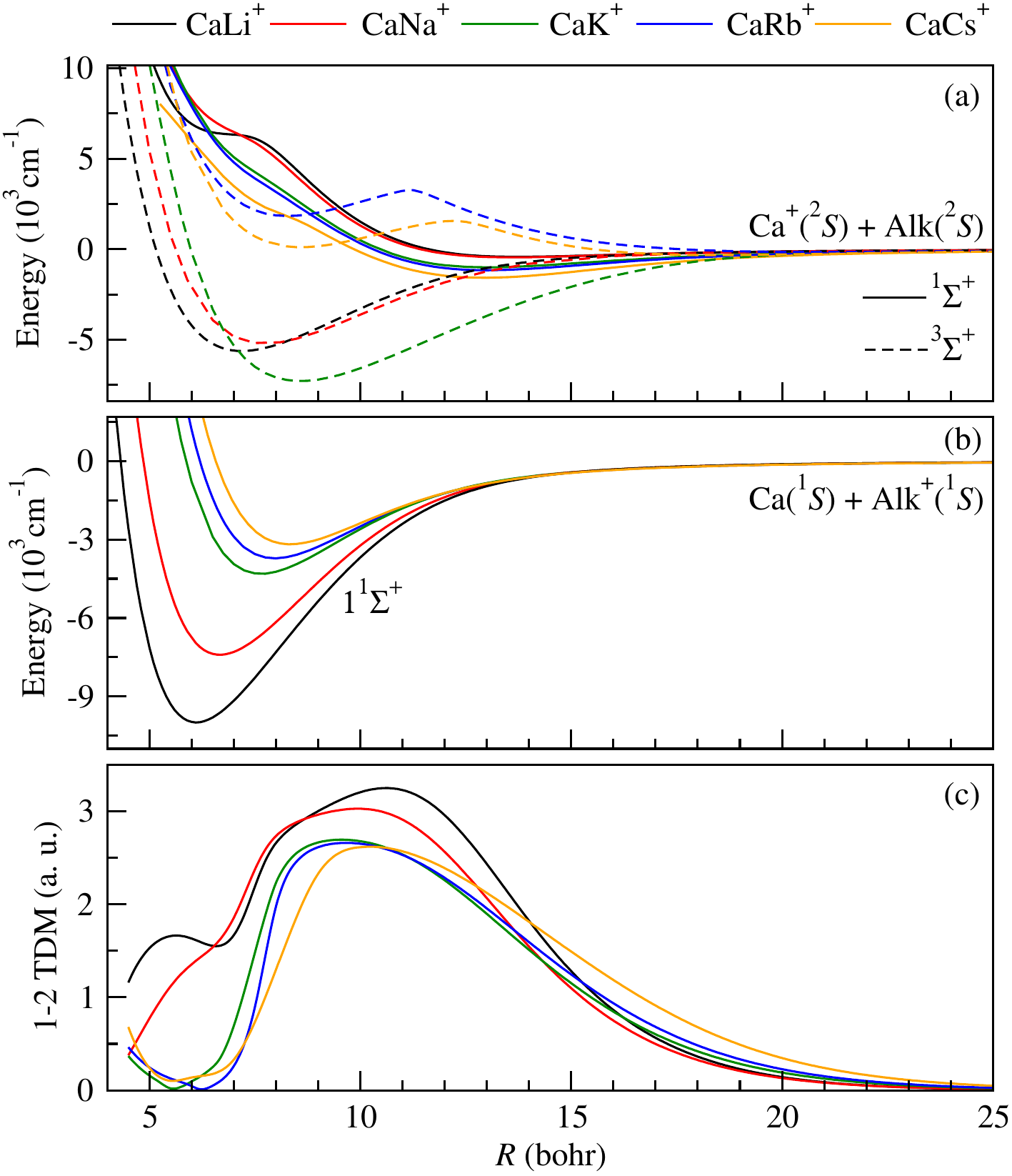}
\caption{(a) Potential energy curves of the $^1\Sigma^+$ and $^3\Sigma^+$ electronic states resulting from the interaction of the Ca$^+$($^2S$) ion and Alk($^2S$) atom (Alk=Li, Na, K, Rb, Cs). (b) Potential energy curves of the ground $^1\Sigma^+$ electronic state resulting from the interaction of the Ca($^1S$) atom and Alk$^+$($^1S$) ion. (c) Transition electric dipole moments between the two lowest $^1\Sigma^+$ states of the CaAlk$^+$ molecular ions.}
\label{fig:CaAlk+}
\end{figure}

\begin{figure}[tb!]
\centering
\includegraphics[width=\columnwidth]{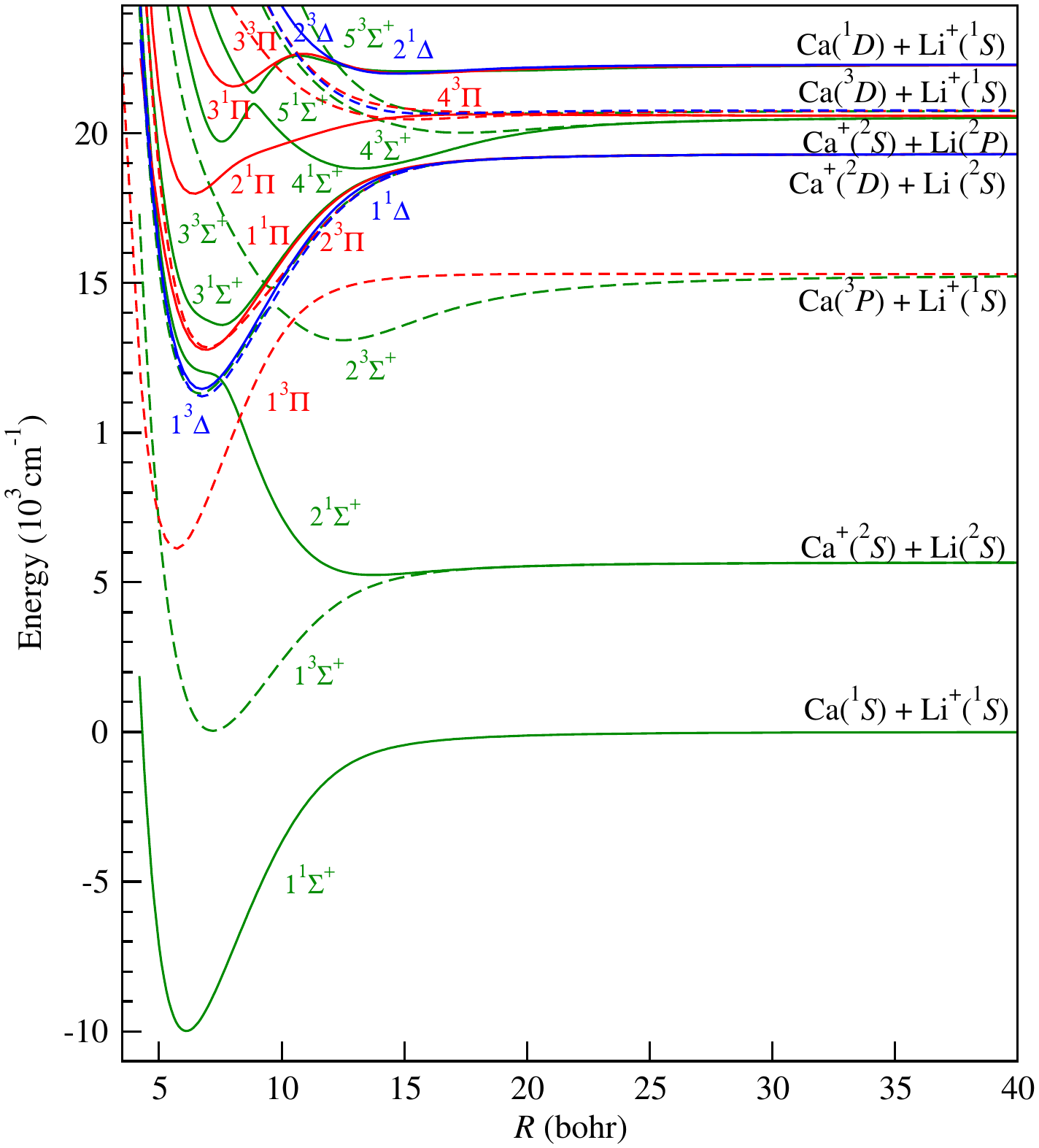}
\caption{Potential energy curves of the CaLi${}^{+}$ molecular ion. Electronic states of the singlet and triplet symmetries are plotted with solid and dashes lines, respectively. Electronic states of $\Sigma^+$, $\Pi$, and $\Delta$ symmetries are plotted with green, red, and blue lines, respectively.}
\label{fig:CaLi+}
\end{figure}

The radiative lifetimes $\tau_{v}$ of vibrational levels $v$, $\tau_{v}={1}/{\Gamma_{v}}$, are calculated from the radiative rates $\Gamma_v=\sum_{v'<v}A_{vv'}+\sum_{v'}B_{vv'}$, which are given by the sums of the Einstein coefficients for the spontaneous emission $A_{vv'}$ and coefficients for the absorption and stimulated emission $B_{vv'}$. The coefficients for the spontaneous emission $A_{vv'}\sim \omega_{vv'}^3 d_{vv'}^2$ are proportional to the third power of the transition frequencies $\omega_{vv'}$ and second power of the transition dipole moments $d_{vv'}$ between the initial $v$ and the final $v'$ vibrational states. The coefficients for the absorption and stimulated emission are proportional to the coefficients for the spontaneous emission and the spectral energy density of the present black body radiation~\cite{Zemke1978,Partridge1981}. The bound-continuum transitions are included either using the Franck-Condon approximation~\cite{Zemke1978} or the sum rule approximation~\cite{Pazyuk1994}, and both methods give the same results.

\section{Results and discussion}
\label{sec:res}

\subsection{Potential energy curves}

\begin{figure}[ptb!]
\centering
\includegraphics[width=\columnwidth]{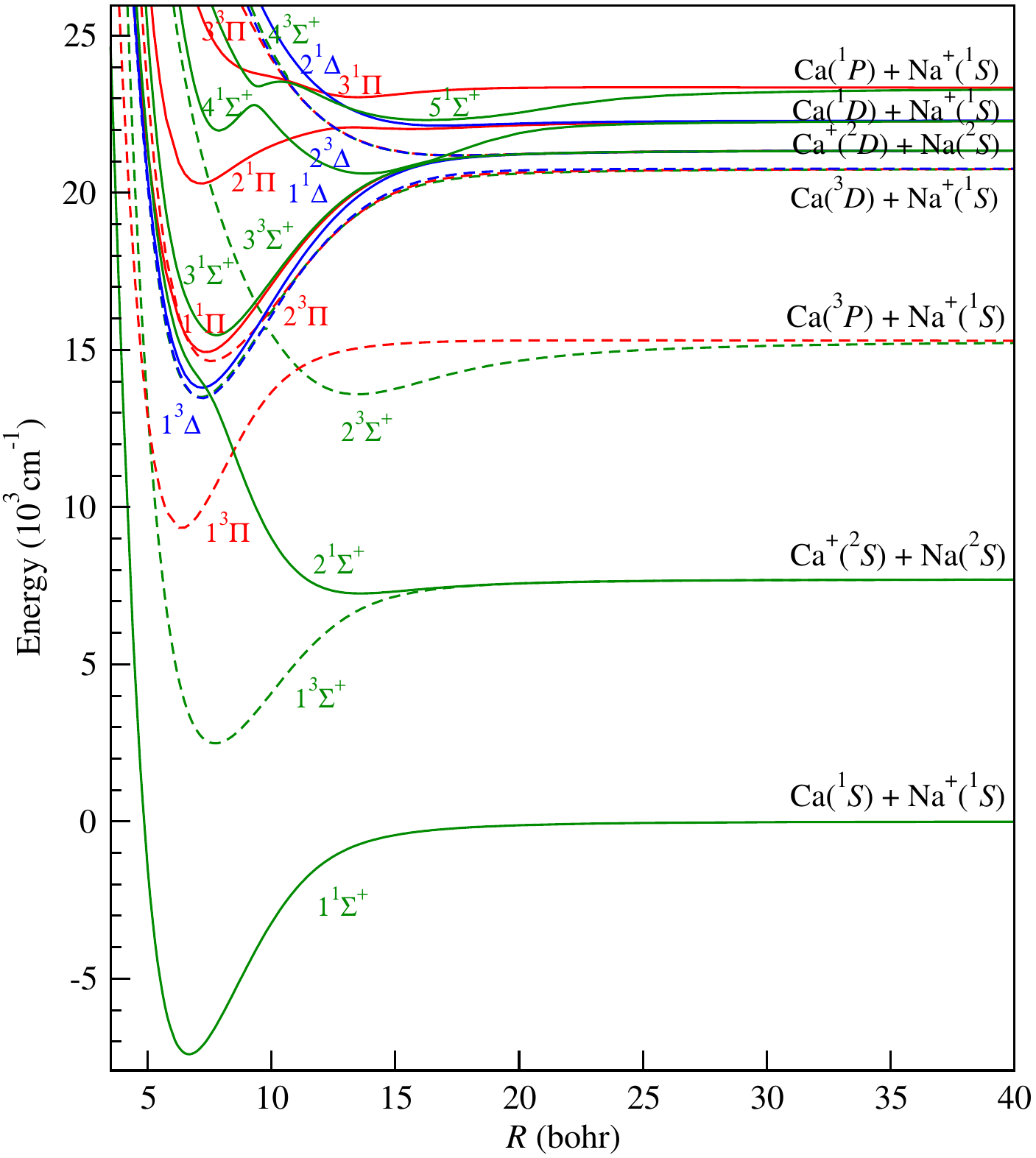}
\caption{Potential energy curves of the CaNa${}^{+}$ molecular ion. Line styles are used as described in Fig.~\ref{fig:CaLi+}.}
\label{fig:CaNa+}
\end{figure}

\begin{figure}[ptb!]
\centering
\includegraphics[width=\columnwidth]{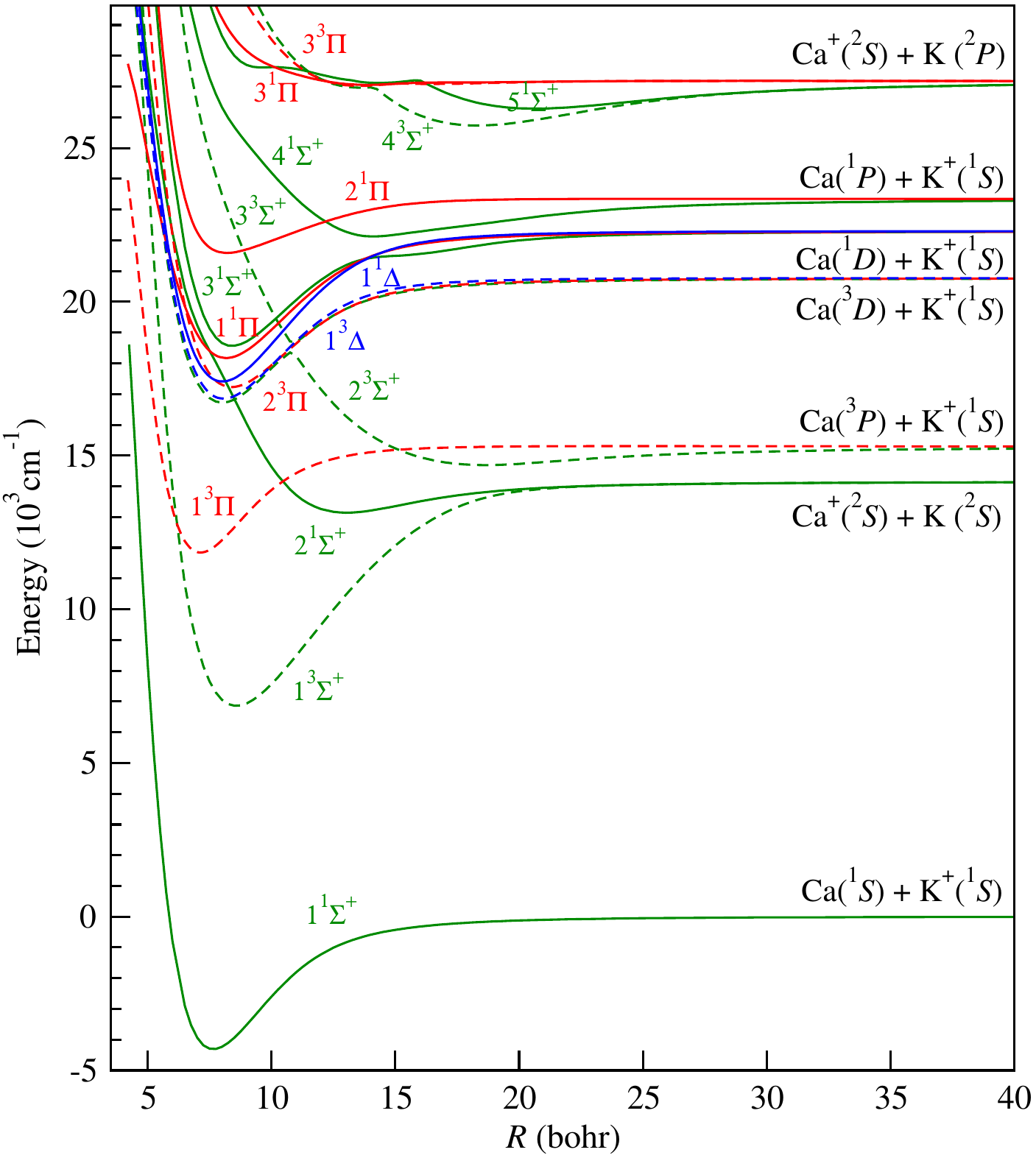}
\caption{Potential energy curves of the CaK${}^{+}$ molecular ion. Line styles are used as described in Fig.~\ref{fig:CaLi+}.}
\label{fig:CaK+}
\end{figure}

\begin{figure}[ptb!]
\centering
\includegraphics[width=\columnwidth]{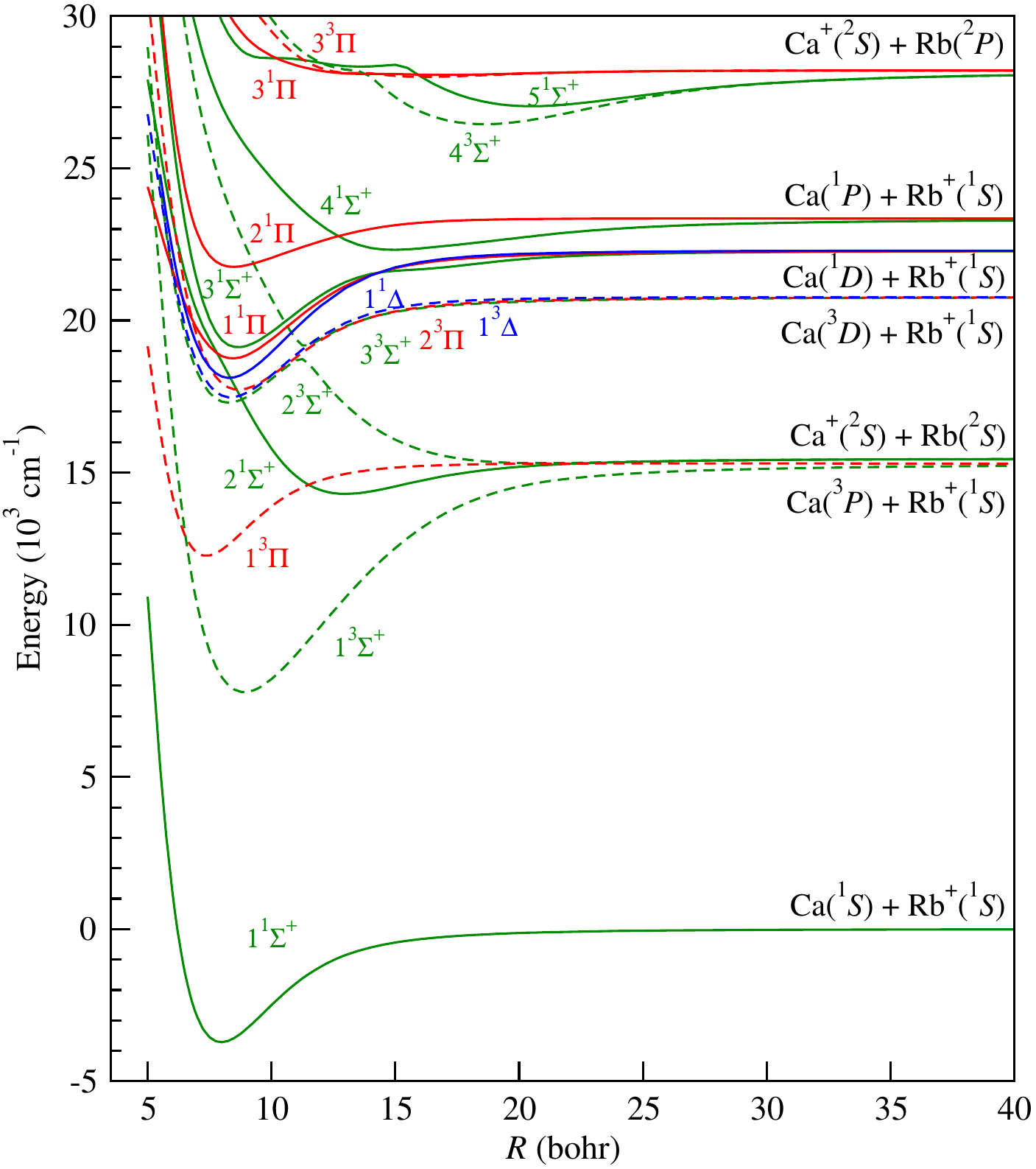}
\caption{Potential energy curves of the CaRb${}^{+}$ molecular ion. Line styles are used as described in Fig.~\ref{fig:CaLi+}.}
\label{fig:CaRb+}
\end{figure}

\begin{figure}[ptb!]
\centering
\includegraphics[width=\columnwidth]{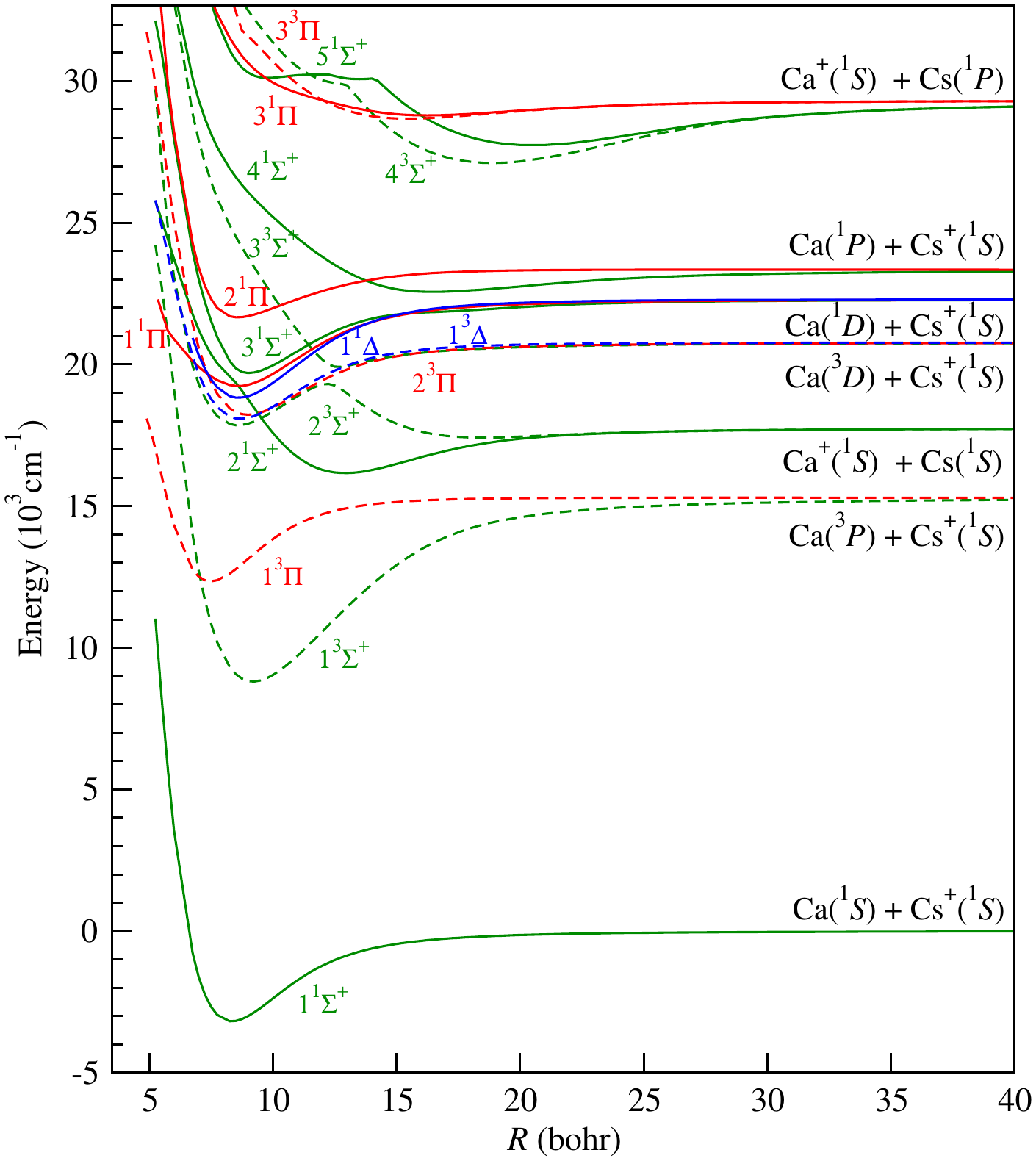}
\caption{Potential energy curves of the CaCs${}^{+}$ molecular ion. Line styles are used as described in Fig.~\ref{fig:CaLi+}.}
\label{fig:CaCs+}
\end{figure}

Potential energy curves (PECs) for the ground and several excited electronic states of the CaLi$^+$, CaNa$^+$, CaK$^+$, CaRb$^+$, and CaCs$^+$ molecular ions are presented in Figs.~\ref{fig:CaAlk+}-\ref{fig:CaCs+}. All electronic states correlated with the seven lowest atomic thresholds of each system are investigated (see Table~\ref{tab:asymptotes}). Thus, several singlet and triplet electronic states of the $\Sigma^+$, $\Pi$, and $\Delta$ spatial symmetries are studied. Spectroscopic characteristics of calculated PECs, i.e.~equilibrium interatomic distances $R_e$, well depths $D_e$, transition energies $T_e$, harmonic constants $\omega_e$, anharmonicity constants $x_e$, and rotational constants $B_e$, are collected in Tables~\ref{tab:CaLi+}-\ref{tab:CaCs+}. Results for excited states of the CaK$^+$ and CaCs$^+$ molecular ions are reported for the first time, while spectroscopic constants for other systems are compared with previous available results. 


Interactions between the ground-state Ca$^+$ ion and ground-state alkali-metal atom are described by the $^1\Sigma^+$ and $^3\Sigma^+$ electronic states and govern ground-state collisions in respective hybrid ion-atom experiments~\cite{TomzaRMP19}. For all considered mixtures, these states are electronically excited, and the radiative charge-transfer and association processes are energetically allowed and may lead to collisional losses~\cite{TomzaPRA15b} leading to the $^1\Sigma^+$ ground electronic state. Therefore in Fig.~\ref{fig:CaAlk+}, we present and compare electronic states correlated with Ca$^+$($^2S$)+Alk($^2S$) and Ca($^1S$)+Alk$^+$($^1S$) atomic thresholds, and transition electric dipole moments between the two lowest $^1\Sigma^+$ electronic states, which drive radiative losses. Corresponding rate constants for reactive collisions are presented in Sec.~\ref{sec:coll}.   

The ground-state proprieties of the CaAlk$^+$ molecular ions are similar to those predicted for the SrAlk$^+$ systems~\cite{Aymar2011}. The $1^1\Sigma^+$ ground electronic state dissociate into Ca($^1S$)+Alk$^+$($^1S$). Therefore, its long-range behavior is very similar for all considered molecular ions and is determined by the induction interaction of the charge of the alkali-metal ion with the polarizability of the Ca atom (see Fig.~\ref{fig:CaAlk+}(b)). The short-range behavior depends more on the involved alkali-metal ion and have covalent bonding nature. The well depth decreases with the mass of the alkali-metal ion from 9986$\,$cm$^{-1}$ for CaLi$^+$ to 3174$\,$cm$^{-1}$ for CaCs$^+$, while the equilibrium distance increases with the mass of the alkali-metal ion from 6.11$\,$bohr for CaLi$^+$ to 8.34$\,$bohr for CaCs$^+$.

The presented ground-state PECs can be compared with recent results calculated with the small-core pseudopotentials and coupled cluster method~\cite{Smialkowski2019}. The well depths obtained with the two methods agree with the mean absolute difference of 68$\,$cm$^{-1}$ (1.7\%), while the equilibrium distances agree with the mean difference of 0.084$\,$bohr (1.1\%). Calculations with large-core pseudopotentials give slightly smaller equilibrium distances and deeper well depths, but the overall good agreement cross validates both approaches and suggests that similar accuracy may be expected for excited electronic states. The agreement with older results collected in Tables~\ref{tab:CaLi+}-\ref{tab:CaCs+} is also satisfactory.

The energy difference between the lowest Ca($^1S$)+Alk$^+$($^1S$) and Ca$^+$($^2S$)+Alk($^2S$) dissociation thresholds increases with the mass of alkali-metal atom from 5819$\,$cm$^{-1}$ for CaLi$^+$ to 17880$\,$cm$^{-1}$ for CaCs$^+$. The well depth of the $2^1\Sigma^+$ state dissociating into Ca$^+$($^2S$)+Alk($^2S$) increases with the mass of alkali-metal atom from 411$\,$cm$^{-1}$ for CaLi$^+$ to 1574$\,$cm$^{-1}$ for CaCs$^+$, while the equilibrium distance decreases slightly from 13.74$\,$bohr for CaLi$^+$ to 12.92$\,$bohr for CaCs$^+$. The $2^1\Sigma^+$ electronic state is relatively shallow because of its non-bonding nature around the equilibrium distance and avoided crossing with the ground $1^1\Sigma^+$ electronic state. 
The $^3\Sigma^+$ electronic state associated with the Ca$^+$($^2S$)+Alk($^2S$) atomic threshold is the lowest triplet state for the CaLi$^+$, CaNa$^+$, and CaK$^+$ molecular ions, while it is the first excited triplet state for the CaRb$^+$ and CaCs$^+$ molecular ions. The change of the order of the Ca$^+$($^2S$)+Alk($^2S$) and Ca($^3P$)+Alk$^+$($^1S$) atomic thresholds in CaRb$^+$ and CaCs$^+$ visibly affects their $^3\Sigma^+$ electronic states dissociating into Ca$^+$($^2S$)+Alk($^2S$), which are much shallower because of avoided crossing with lower lying $^3\Sigma^+$ states (see Fig.~\ref{fig:CaAlk+}(a)). Thus, no clear trend is observed for the lowest  $^3\Sigma^+$ electronic states. For example, the well depth and equilibrium distance of the first $^3\Sigma^+$ state in CaK$^+$ is 7275$\,$cm$^{-1}$ and 8.58$\,$bohr, respectively. 

The density of electronic states increases with the excitation energy, and for all investigated molecular ions, several avoided crossings between excited states of the same electronic symmetry can be found. Strong radial non-adiabatic couplings between involved electronic states can be expected. Some of the excited atomic thresholds are close together that furthermore facilitates interactions between associated electronic states. As a result, several excited states have double-well structures. For example, all $2^3\Sigma^+$ states are significantly repulsive at the short-range distances, partially due to broad avoided crossing with $1^3\Sigma^+$ states, and thus they intersect with attractive $3^3\Sigma^+$ states, forming narrow avoided crossings at the short range. Avoided crossings between $2^1\Sigma^+$, $3^1\Sigma^+$, and $4^1\Sigma^+$ at short- and intermediate-range distances are also pronounced. Additionally, electronic states of different spin and spatial symmetries intersect with each other. These crossings may become avoided crossings if relativistic spin-orbit couplings would be included, which is out of the scope of this paper. Avoided and real crossings may provide mechanism for efficient non-radiative and non-adiabatic charge transfer between ions and atoms in excited electronic states~\cite{RellergertPRL11,Tacconi2011,SayfutyarovaPRA13,LiPRA19}.

\begin{figure}[tb!]
\centering
\includegraphics[width=\columnwidth]{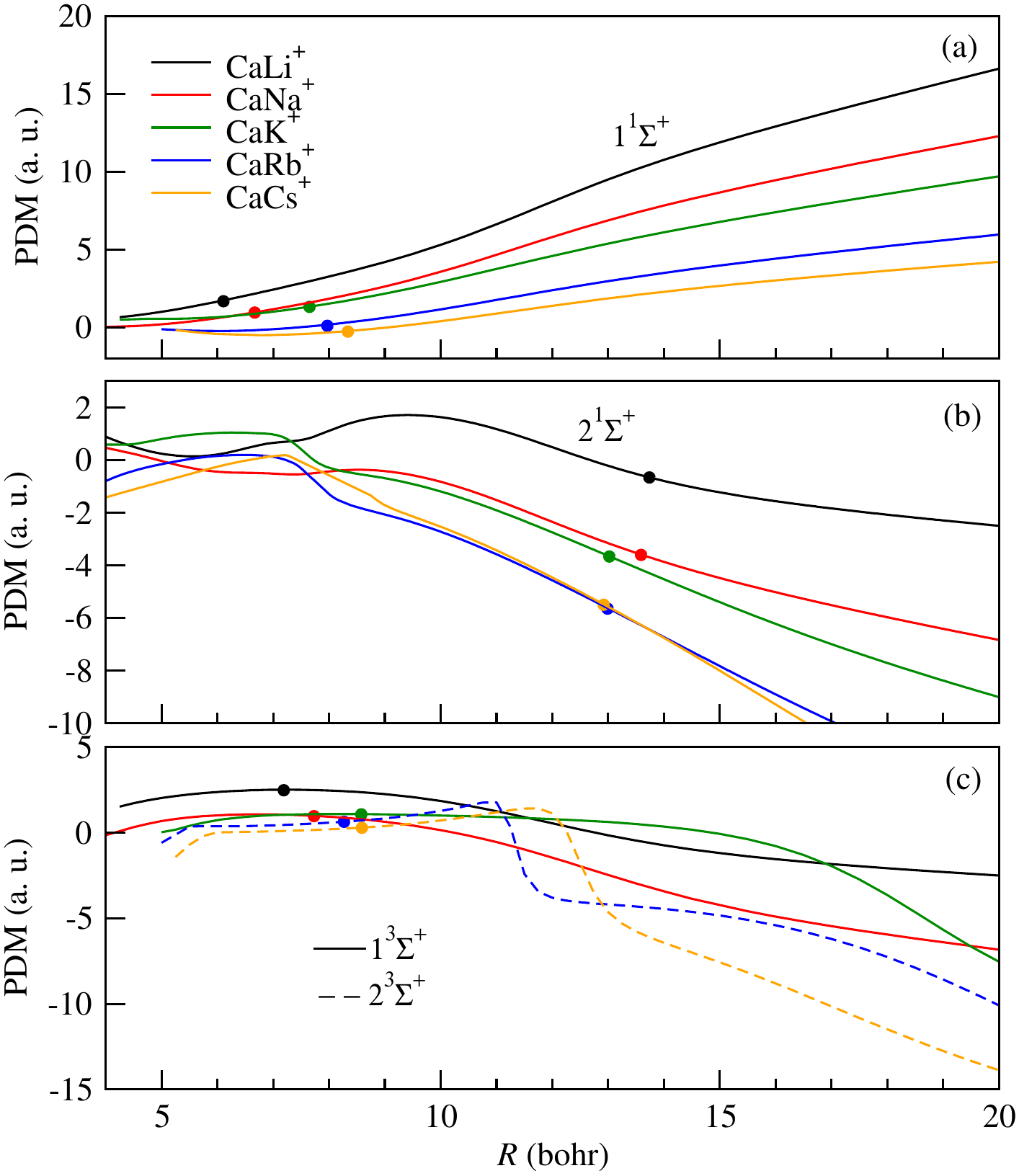}
\caption{Permanent electric dipole moments of the CaAlk$^+$ molecular ions in (a) the ground $^1\Sigma^+$ electronic state, (b) the first excited $^1\Sigma^+$ electronic state, and (c) the $^3\Sigma^+$ electronic state resulting from the interaction of the Ca$^+$($^2S$) ion and Alk($^2S$) atom.}
\label{fig:PDM}
\end{figure}

Spectroscopic constants of calculated excited electronic states for the investigated molecular ions are collected in Tables~\ref{tab:CaLi+}-\ref{tab:CaCs+}. Results for excited electronic states of the CaK$^+$ and CaCs$^+$ molecular ions are reported for the first time, while spectroscopic constants for other systems can be compared with previous available theoretical results~\cite{Smialkowski2019,Bala2019,SaitoPRA17,HabliMP16,Xie2005,Russon1998,
Kimura1983,Smialkowski2019,JellaliJQSRT18,Gacesa2016,MakarovPRA03,daSilvaNJP2015,Felix2013,Tacconi2011}. Similarly as for the ground electronic state, the results for the excited states obtained with different computational methods agree reasonably well. Both well depths and equilibrium distances mostly agree within several percent. In the case of the CaLi$^+$ molecular ion, the present results agree very well with the results of Refs.~\cite{HabliMP16,SaitoPRA17}, while well depths seem to be underestimated in calculations presented in Ref.~\cite{Bala2019}, which employed a single-reference method. For the CaNa$^+$ molecular ion, the present results agree very well with the results of Refs.~\cite{Gacesa2016,JellaliJQSRT18}, while the agreement is worse with calculations presented in Ref.~\cite{MakarovPRA03}, which employed smaller basis sets. In the case of the CaRb$^+$ molecular ion, the present results also agree well with the results of Refs.~\cite{Tacconi2011,Felix2013,daSilvaNJP2015}.

The overall good agreement between the present and previous calculations for the CaLi$^+$, CaNa$^+$, and CaRb$^+$ molecular ions suggests that similar accuracy may be expected for our results for the CaK$^+$ and CaCs$^+$ molecular ions, which have not yet been studied. Electronic structure data for the CaLi$^+$, CaNa$^+$, and CaRb$^+$ molecular ions were successfully employed to guide and interpret experimental measurements~\cite{SmithAPB14,HallPRL11,HallMP13a,EberleCPC16,HazePRA13,HazePRA15,HazePRL18,SaitoPRA17}. Presented potential energy curves for the CaK$^+$ and CaCs$^+$ molecular ions may correspondingly find similar applications, e.g.,~in the context of experimental studies of a mixture of laser-cooled Ca$^+$ ions in a linear Paul trap overlapped with ultracold K atoms in a magneto-optical trap as presented recently in Ref.~\cite{JyothiRSI19}.

\begin{table*}[p!]
\caption{Spectroscopic constants of the ground and excited electronic states of the CaLi$^+$ molecular ion.}\label{tab:CaLi+}
\begin{ruledtabular}
\begin{tabular}{lccccccc}
State & $R_e\,$(bohr) & $D_e\,$(cm$^{-1}$) & $T_{e}\,$(cm${}^{-1}$)& $\omega_e\,$(cm$^{-1}$) & $\omega_e x_e\,$(cm$^{-1}$) & $B_e\,$(cm${}^{-1}$)& Ref.\\
\hline
1$^1\Sigma^+$& 6.11&  9986 & 0 & 244.19 & 1.49&  0.272557 &This work\\
             &6.16 & 9941 & 0 & 246  & - & 0.266 &  \cite{Smialkowski2019}\\
             &6.165 & 10093 & 0  & 245.1 & 1.31 & 0.2650 &  \cite{Bala2019} \\
             & 6.120 & 9973 & 0 & 242 & - & - &   \cite{HabliMP16} \\
             &6.210 & 9641 & 0  &- & - & -&  \cite{Xie2005}\\
             &6.273  &9990 & 0 & 242 & -&  -&  \cite{Russon1998} \\
             &6.200 & 8943 & 0 & 235&  - & - & \cite{Kimura1983}\\
$2^1\Sigma^+$ & 13.74 & 411&  15237 & 44.03 & 1.17 & 0.053897 & This work \\
             &22.41 & 23  &16500 & 6.5 & 0.63 & 0.0201  &  \cite{Bala2019} \\
             &13.68 & 412 & - & 41 & - & - &   \cite{HabliMP16} \\
$3^1\Sigma^+$& 7.56 & 5705 & 23592 & 154.27 & 1.04 & 0.178031 &This work \\
             & 7.42 & 5816 & - & 132 & - & - &   \cite{HabliMP16} \\
$4^1\Sigma^+$&7.55  & 1262 & 29719 & 247.15 & 19.09 & 0.178503& This work \\
             & 7.51 & 1192 & - &  - & - &- &   \cite{HabliMP16} \\
$2^\text{nd}$ min & 13.16  & 1741  & 28820 & 61.82 & 0.71 & 0.058753& This work\\
                   & 13.22 & 1975 & - &  - & - &- &   \cite{HabliMP16} \\
$5^1\Sigma^+$ &8.82 & 1235 & 31379  &370.31  &34.95  &0.130799 &This work\\
$2^\text{nd}$ min &15.00 & 200&  32070 & 22.13 & 0.61 & 0.045223& This work\\
$1^3\Sigma^+$ &7.19 & 5616 & 10033 & 163.19 & 1.19&  0.196826& This work\\
             & 7.18 & 5792 & - &  - & - &- &   \cite{HabliMP16} \\
$2^3\Sigma^+$ & 6.66 & 3911 & 21305&  185.17 & 2.19&  0.229399 & This work\\
             & 6.69 & 3693 & - &  - & - &- &   \cite{HabliMP16} \\
$2^\text{nd}$ min& 12.49 & 2132&  23084 & 72.11 & 0.61 & 0.065255 &This work \\
               & 12.53 & 1807 & - &  - & - &- &   \cite{HabliMP16} \\
$3^3\Sigma^+$ &9.56 & 4456 & 24843&  287.35 & 4.63&  0.111333 & This work \\
             & 9.54 & 4514 & - &  - & - &- &   \cite{HabliMP16} \\
$4^3\Sigma^+$ &17.37&  503 &  30016 & 32.91&  0.54 & 0.033724 & This work \\
              & 16.89 & 325 & - &  - & - &- &   \cite{HabliMP16} \\
$5^3\Sigma^+$ &18.33 & 109 & 30645 & 14 & 0.45 & 0.030284 & This work\\
$1^1\Pi$ & 6.90 & 6541 & 22757 & 172.23 & 1.13 & 0.213718& This work \\
             &6.472  & 1372 &  34548& 214.8 & 1.66 & 0.2397   & \cite{Bala2019} \\
            & 6.69 & 6643 & - &  - & - &- &   \cite{HabliMP16} \\
$2^1\Pi$ & 6.41 & 2599 & 27976 & 184.20  &3.26&  0.247642 &This work \\
          & 6.43 & 2939 & - &  - & - &- &   \cite{HabliMP16} \\
$3^1\Pi$ &  8.04 & 715 & 31560&  149.74 & 7.84  & 0.157408 & This work \\
         & 8.18 & 502 & - &  - & - &- &   \cite{HabliMP16} \\
$2^\text{nd}$ min & 15.18 & 280 & 31994 & 31.87 & 0.91 & 0.044157& This work \\
                  & 16.1 & 162 & - &  - & - &- &   \cite{HabliMP16} \\
$1^3\Pi$  &5.69&  9172&  16116 & 255.88&  1.78 & 0.314279& This work \\
          &5.879&  7621 & 19933 &255.9&  1.57&  0.2904 & \cite{Bala2019}  \\
          & 5.71 & 8982 & - &  - & - &- &   \cite{HabliMP16} \\
$2^3\Pi$  &6.98 & 6460&  22837 & 168.61 & 1.10 & 0.208847 &This work \\
          & 6.87 & 6686 & - &  - & - &- &   \cite{HabliMP16} \\
$3^3\Pi$ & 15.32 & 110& 30465 & 31.94&  2.32 & 0.043353 &This work \\
$1^1\Delta$  &6.73 & 7854 & 21452&  186.25 & 1.10&  0.224652& This work \\
             & 6.71 & 7959 & - &  - & - &- &   \cite{HabliMP16} \\
$2^1\Delta$ & 14.63 & 298&  31988&  37.98 & 1.21  &0.047539& This work \\
$1^3\Delta$  & 6.75 &  8097 & 21207 & 187.75 & 1.09 & 0.223322 &This work \\
             & 6.73 & 8212 & - &  - & - &- &   \cite{HabliMP16} \\
$2^3\Delta$  &16.10&  108 & 30653 & 24.11 & 1.35 & 0.039254 &This work \\
\end{tabular}
\end{ruledtabular}
\end{table*}
                                                                                                                                                                                                                                             \begin{table*}[ptb!]
\caption{Spectroscopic constants of the ground and excited electronic states of the CaNa${}^{+}$ molecular ion.}\label{tab:CaNa+}
\begin{ruledtabular}
\begin{tabular}{lccccccc}
State & $R_e\,$(bohr) & $D_e\,$(cm$^{-1}$) & $T_{e}\,$(cm${}^{-1}$)& $\omega_e\,$(cm$^{-1}$) & $\omega_e x_e\,$(cm$^{-1}$) & $B_e\,$(cm${}^{-1}$)& Ref.\\
\hline
$1^1\Sigma^+$ & 6.67&  7402 & 0 &  135.79&  0.62 & 0.092621& This work \\
       & 6.70 & 7336 & 0 & 137  & - &  0.0918 &  \cite{Smialkowski2019}\\
       & 6.67 & 7441 & 0 & 141.84  & 0.674 &  0.0926 &  \cite{JellaliJQSRT18}\\
       &6.67 & 7488  & 0 &  138.43 & 0.58 & 0.093413& \cite{Gacesa2016} \\
       &6.76 & 6376 & 0 & - & - & - & \cite{MakarovPRA03} \\
$2^1\Sigma^+$ &13.59 & 443 & 14660&  30.02 & 0.51 & 0.022311& This work \\
       & 13.57 & 446 & - & 27.30  & 0.432 &  0.0223 &  \cite{JellaliJQSRT18}\\
       &13.44&  450 & 15069 & 29.86  &0.55  &0.022822& \cite{Gacesa2016} \\
$3^1\Sigma^+$ & 7.77 & 5869 & 22873 & 110.07&  0.52 & 0.068253& This work \\
       & 7.77 & 6007 & - & 102.72  & 0.438 &  0.0582 &  \cite{JellaliJQSRT18}\\
          &7.76&  6311&  23026 & 102.59&  0.43&  0.069019& \cite{Gacesa2016} \\
$4^1\Sigma^+$ & 7.88&  792 & 29387&  132.70 & 14.62 & 0.066361 &This work \\
         &7.95 & 1107&  28339 & 164.03&  1.87 & 0.029906& \cite{Gacesa2016} \\
$2^\text{nd}$ min & 13.73 & 1686 & 28011 &  42.81 & 0.28 & 0.021861 &This work\\
                   &13.23&  1203 & 28243&  48.98 & 1.87 & 0.029906& \cite{Gacesa2016} \\
$5^1\Sigma^+$ & 16.37 & 1018&  29718 & 20.34 & 0.10 & 0.015581 & This work\\
$1^3\Sigma^+$ & 7.73 & 5210 & 9891 & 97.98 & 0.46 & 0.068961& This work \\
              & 7.72 & 5345 & - & 107.7 & 0.532 & 0.0691 &  \cite{JellaliJQSRT18}\\
$2^3\Sigma^+$ &7.15&  1738&  20906 & 109.97&  1.74&  0.080603 & This work \\
              & 7.17 &  1578 & - & 26.67 & 0.112 &  0.0801 &  \cite{JellaliJQSRT18}\\
$2^\text{nd}$ min &13.48 & 1650 & 20994 & 40.71 & 0.25 & 0.022677 &This work \\
                  & 13.53 & 1719 & - & 30.64  & 0.135 & 0.0225 &  \cite{JellaliJQSRT18}\\
$3^3\Sigma^+$ & 9.77 & 4688 & 23471 & 156.42&  1.30 & 0.043169 & This work \\
              & 9.75 & 4426 & - & 120.10  & 0.79 & 0.0433 &  \cite{JellaliJQSRT18}\\
$4^3\Sigma^+$  &17.66 & 153 & 28597 & 12.96 & 0.27 & 0.013212& This work \\
               & 16.55 & 191 & - & 11.80  & 0.173 &  0.0150 &  \cite{JellaliJQSRT18}\\
$1^1\Pi$ &  7.36 & 6415 & 22335 & 101.36  & 0.4 & 0.076069 &This work \\
         & 7.37 & 6608 & - & 105.80  & 0.418 & 0.0758 &  \cite{JellaliJQSRT18}\\
           &7.32 & 6928 & 22352 & 103.04 & 0.38 & 0.077102 & \cite{Gacesa2016} \\
$2^1\Pi$ &  7.14&  1994 & 27697 & 95.76&  1.15 & 0.080829& This work \\
          & 7.15 & 2111 & - & 78.43  & 0.691 &  0.806 &  \cite{JellaliJQSRT18}\\
           &7.10 & 2100 & 27367 & 103.56  &1.00 & 0.082348 & \cite{Gacesa2016} \\
$2^\text{nd}$ min &15.65 & 256 & 29435 & 17.55 & 0.30 & 0.016824& This work \\
           & 16.40 & 248 & -   & 13.58 & 0.254 & 0.0153 &  \cite{JellaliJQSRT18}\\
$3^1\Pi$ & 13.47&  301 & 30446 & 31.73 & 0.84 & 0.022711& This work\\
      & 13.78 & 259  & - & 28.83  & 0.730 & 0.0217 &  \cite{JellaliJQSRT18}\\ 
$1^3\Pi$  &6.36 & 5964 &16733 & 139.08 & 0.81 & 0.101870&  This work \\
          & 6.39 & 5930 & - & 144.50  & 0.861 & 0.1009 &  \cite{JellaliJQSRT18}\\
$2^3\Pi$ & 7.56 & 6115 & 22048 & 94.01 & 0.36 & 0.072097 & This work \\
         & 7.54 & 5895 & - & 101.40  & 0.43 & 0.0724 &  \cite{JellaliJQSRT18}\\
$3^3\Pi$ & 17.48 & 154 & 28597  &14.33 & 0.33 & 0.013486 &This work \\
         & 16.57 & 191 & - & 14.75  & 0.286 & 0.0150 &  \cite{JellaliJQSRT18}\\
$1^1\Delta$ & 7.19 & 7552 & 21200 & 109.13 & 0.39&  0.079708& This work \\
           & 7.19 & 7706 & - & 111.91  & 0.40 & 0.0797 &  \cite{JellaliJQSRT18}\\
$2^1\Delta$ & 16.81&  156 & 29544 & 16.40 & 0.43 & 0.014582& This work \\
            & 17.74 & 104 & - & 13.29  & 0.43 &  0.0131 &  \cite{JellaliJQSRT18}\\
$1^3\Delta$ & 7.20 & 7302 & 20870 & 108.91&  0.41 & 0.079487& This work \\
            & 7.20 & 7063 & - & 111.32  & 0.410 &  0.0794 &  \cite{JellaliJQSRT18}\\
$2^3\Delta$ & 17.66 & 155 & 28597 & 13.63  &0.30 & 0.013212& This work \\  
            & 16.73 & 196 & - & 15.24  & 0.299 & 0.0147 &  \cite{JellaliJQSRT18}\\
\end{tabular}
\end{ruledtabular}
\end{table*}

\begin{table*}[tb!]
\caption{Spectroscopic constants of the ground and excited electronic states of the CaK$^+$ molecular ion.}\label{tab:CaK+}
\begin{ruledtabular}
\begin{tabular}{lccccccc}
State & $R_e\,$(bohr) & $D_e\,$(cm$^{-1}$) & $T_{e}\,$(cm${}^{-1}$)& $\omega_e\,$(cm$^{-1}$) & $\omega_e x_e\,$(cm$^{-1}$) & $B_e\,$(cm${}^{-1}$)& Ref.\\
\hline
$1^1\Sigma^+$ &  7.65&  4306 & 0 &  92.60&  0.49&  0.052002& This work \\
              & 7.71 & 4281 & 0 & 93.2  & - &  0.0513 &  \cite{Smialkowski2019}\\
$2^1\Sigma^+$ &13.02&  992 &  17451&  33.11&  0.27&  0.017953& This work \\
$3^1\Sigma^+$ &8.40&  3718 & 22874&  78.32&  0.41&  0.043131& This work \\
$4^1\Sigma^+$ &14.10&  1177& 26434&  35.18&  0.26&  0.015308& This work \\
$5^1\Sigma^+$ &20.68&  809 &  30590 & 18.18&  0.10&  0.00712 &This work \\
$1^3\Sigma^+$&  8.58&  7275 & 11169 & 78.10 & 0.21&  0.041340& This work \\
$2^3\Sigma^+$&18.66&  559 &  18999&  18.37&  0.15&  0.008740& This work \\
$2^\text{nd}$ min          &8.02&  -1471 & 21029&  76.95&  1.01&  0.047315& This work \\
$3^3\Sigma^+$ &10.89&  2086& 22980&  163.59&  3.21&  0.025662& This work \\
$4^3\Sigma^+$ &18.30&  1366 & 30044&  23.35& 0.10 &  0.009088& This work \\
$1^1\Pi$&  8.17&  4116 & 22482&  72.24&  0.32&  0.045594& This work \\
$2^1\Pi$&  8.20&  1751 & 25900&  56.42&  0.45&  0.045261& This work \\
$3^1\Pi$&  13.86&  111&  31365&  21.27&  1.02&  0.015842& This work \\
$1^3\Pi$&  7.14&  3446 & 16148&  94.98&  0.65&  0.059697& This work \\
$2^3\Pi$&  8.39&  3539&  21529&  71.99& 0.37&  0.043234& This work \\
$3^3\Pi$&  13.57&  152&  31317&  35.31&  2.05&  0.016527& This work \\
$1^1\Delta$ &  8.01&  4887 & 21718&  79.32&  0.32&  0.047433& This work \\
$1^3\Delta$ &8.06&  3918 & 21160&  77.20&  0.38&  0.046847& This work \\ 
\end{tabular}
\end{ruledtabular}
\end{table*}

\begin{table*}[tb!]
\caption{Spectroscopic constants of the ground and excited electronic states of the CaRb$^+$ molecular ion.}\label{tab:CaRb+}
\begin{ruledtabular}
\begin{tabular}{lccccccc}
State & $R_e\,$(bohr) & $D_e\,$(cm$^{-1}$) & $T_{e}\,$(cm${}^{-1}$)& $\omega_e\,$(cm$^{-1}$) & $\omega_e x_e\,$(cm$^{-1}$) & $B_e\,$(cm${}^{-1}$)& Ref.\\ \hline
1${}^{1}$$\Sigma$${}^{+}$&  7.97&  3714&  0 &   73.53&  0.35&  0.034735& This work \\
   & 8.06 & 3666 & 0 & 73.9  & - &  0.0341 &  \cite{Smialkowski2019}\\
   & 7.96&  3851& 0 &  73.02& - & 0.03460& \cite{daSilvaNJP2015} \\
   & 8.26&  3717& 0 &  73.43 &  0.50&  0.032371&\cite{Felix2013} \\
   &  8.0 &  3730&  0 &  -  & - & - & \cite{Tacconi2011}  \\
$2^1\Sigma^+$&  12.99&  1160&  18015&  29.58&  0.15&  0.013076& This work \\
   &12.82&  1284&  17764&  28.20&  0.09&  0.013388& \cite{Felix2013} \\
   &12.86&  1272&  18171&  29.89&  -  &0.01330& \cite{daSilvaNJP2015} \\
   & 13&  1170&  -  &  - & -  & - &   \cite{Tacconi2011} \\
$3^1\Sigma^+$&  8.65&  3175&  22835&  62.04&  0.34&  0.029489& This work \\
$4^1\Sigma^+$&  14.89&  1016&  26035&  22.61&  0.31&  0.009952& This work \\
$5^1\Sigma^+$&  13.61&  2051&  30754&  16.9&  0.31&  0.011912& This work \\
$1^3\Sigma^+$& 8.90&  7494&  11503&  64.21&  0.16&  0.027855& This work \\
      &9.15&  7455&  10806&  77.92&  0.20&  0.026367& \cite{Felix2013} \\
$2^3\Sigma^+$& 8.27&  -1839&  21014&  61.54&  0.49&  0.032261& This work \\
     &8.31&  -1496 & 20128&  -&  -&  0.031976& \cite{Felix2013} \\
$2^\text{nd}$ min &20.82&  137&  19025&  8.37&  0.13&  0.005090& This work \\
       &17.82&  121&  19018&  -& -&  -& \cite{Felix2013} \\
$3^3\Sigma^+$&  11.42&  1481&  22999&  61.54&  0.49&  0.016918& This work \\
$4^3\Sigma^+$&18.57&  1745&  30163&  20.67&  0.04&  0.006398& This work \\
$1^1\Pi$&  8.43&  3542&  22467&  57.48&  0.16&  0.031048& This work \\
$2^1\Pi$&  8.47& 1575&  25476&  46.14&  0.51&  0.030755& This work \\
$3^1\Pi$&  17.0&  131&  31782&  8.70&  0.07&  0.007635& This work \\
$1^3\Pi$&  7.38&  3022&  15982&  75.86&  0.40&  0.040511& This work \\
        &7.63&  3308&  18204&  -&  -&  -& \cite{Felix2013} \\
$2^3\Pi$&  8.67&  3037&  21440&  58.30&  0.21&  0.029353& This work \\
$3^3\Pi$&  16.19&  202&  31718&  12.17&  0.16&  0.008418& This work \\
$1^1\Delta$&  8.29&  4184&  21827&  63.51&  0.18&  0.032105&  This work \\
$1^3\Delta$&  8.35&  3293&  21189&  61.17&  0.24&  0.031646&  This work \\ 
\end{tabular}
\end{ruledtabular}
\end{table*}

\begin{table*}[tb!]
\caption{Spectroscopic constants of the ground and excited electronic states of CaCs$^+$ molecular ion.}\label{tab:CaCs+}
\begin{ruledtabular}
\begin{tabular}{lccccccc}
State & $R_e\,$(bohr) & $D_e\,$(cm$^{-1}$) & $T_{e}\,$(cm${}^{-1}$)& $\omega_e\,$(cm$^{-1}$) & $\omega_e x_e\,$(cm$^{-1}$) & $B_e\,$(cm${}^{-1}$)& Ref.\\
\hline
$1^1\Sigma^+$ &  8.34&  3174&  0 &  64.78&  0.33&  0.028107& This work \\
              & 8.53 & 3017 & 0 & 63.2  & - &  0.0270 &  \cite{Smialkowski2019}\\
$2^1\Sigma^+$ &  12.92&  1574&  19344&  29.78&  0.14&  0.011712& This work \\
$3^1\Sigma^+$ &  9.02&  2584&  22877&  59.85&  0.35&  0.024029& This work \\
$4^1\Sigma^+$ &  16.45&  772&  25739&  15.67&  0.08&  0.007225& This work \\
$5^1\Sigma^+$ &20.50&  1555&  30911&  16.80&  0.05&  0.004652& This work \\
$2^3\Sigma^+$ &  9.20&  6470&  11991&  58.92&  0.13&  0.023098& This work \\
$2^3\Sigma^+$ &  8.59&  -112&  21033&  53.91&  6.48&  0.026495& This work \\
$2^\text{nd}$ min &18.46&  324&  20594&  12.77&  0.13&  0.005737& This work \\
$3^3\Sigma^+$ &  12.65&  849&  23093&  54.35&  0.87&  0.0122217& This work \\
$4^3\Sigma^+$ &  18.88&  2177&  30287&  19.54&  0.04&  0.005485& This work \\
$1^1\Pi$&  8.60 &  3049&  22418&  46.96&  0.18&  0.026433& This work \\
$2^1\Pi$&  8.58 &  1678&  24842&  49.23&  0.36&  0.026557& This work \\
$3^1\Pi$&  16.24 &  505&  31970&  13.42&  0.09&  0.007413& This work \\
$1^3\Pi$&  7.49 &  2941&  15528&  68.65&  0.40&  0.034849& This work \\
$3^3\Pi$&  8.99 &  2534&  21405&  51.33&  0.26&  0.024190& This work \\
$3^3\Pi$&  15.49 &  621&  31854&  15.32&  0.09&  0.008184& This work \\
$1^1\Delta$&  8.62 &  3468&  22006&  55.47&  0.22&  0.026311& This work \\
$1^3\Delta$&  8.69 &  2682&  21264&  53.13&  0.26&  0.025889& This work \\
\end{tabular}
\end{ruledtabular}
\end{table*}

\subsection{Permanent and transition electric dipole moments}

\begin{figure}[tb!]
\centering
\includegraphics[width=\columnwidth]{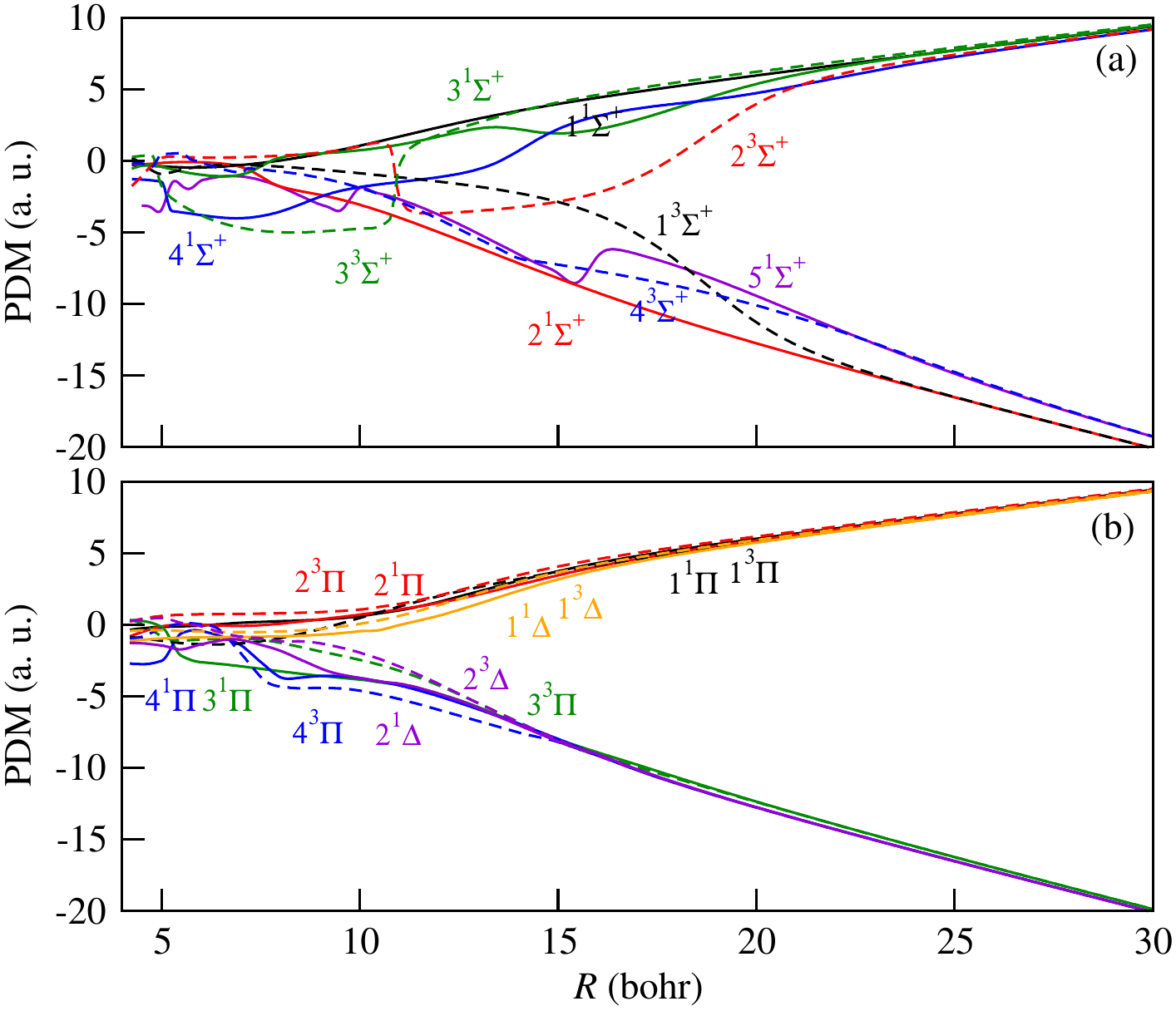}
\caption{Permanent electric dipole moments of (a) $^1\Sigma^+$ and $^3\Sigma^+$, and (b) $^1\Pi$, $^3\Pi$, $^1\Delta$, and $^3\Delta$ electronic states of the CaK$^+$ molecular ion.}
\label{fig:PDM_CaK+}
\end{figure}

\begin{figure}[tb!]
\centering
\includegraphics[width=\columnwidth]{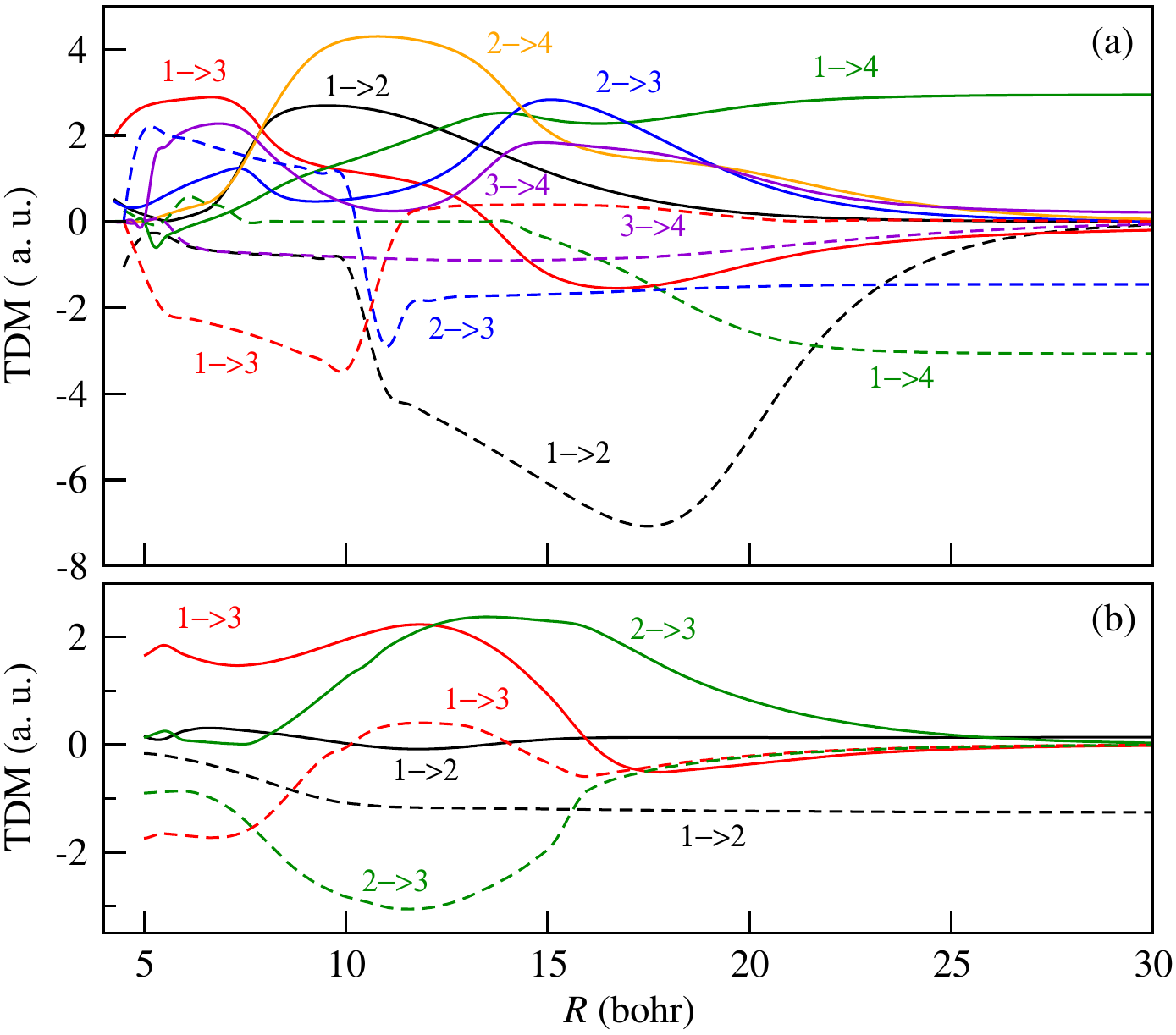}
\caption{Transition electric dipole moments ($n\to m\equiv n^{2S+1}|\Lambda|\to m^{2S+1}|\Lambda|$) between (a) $^1\Sigma^+$ (solid lines) and $^3\Sigma^+$ (dashed lines), and (b) $^1\Pi$ (solid lines) and $^3\Pi$ (dashed lines) electronic states of the CaK$^+$ molecular ion.}
\label{fig:TDM_CaK+}
\end{figure}

The permanent and transition electric dipole moments (PEDMs and TEDMs) determine the interaction of atomic and molecular systems with static and dynamic electric fields, including the laser field. Thus, their knowledge is essential for predicting molecular spectra, lifetimes, and formation schemes. Here, we calculate permanent electric dipole moments for all investigated electronic states of the CaLi$^+$, CaNa$^+$, CaK$^+$, CaRb$^+$, and CaCs$^+$ molecular ions, as well as all transition electric dipole moments between electronic states of the same spin and spatial symmetries.

Transition electric dipole moments between the two lowest $^1\Sigma^+$ states of the CaAlk$^+$ molecular ions are presented in Fig.~\ref{fig:CaAlk+}(c). Their functions have quite similar shapes and values. They govern the radiative charge-transfer and association processes in ground-state collisions between Ca$^+$ ions and alkali-metal atoms which are studied in Sec.~\ref{sec:coll}.

Permanent electric dipole moments of the two lowest $^1\Sigma^+$ electronic states, i.e., $^1\Sigma^+$ states dissociating into the Ca($^1S$)+Alk$^+$($^1S$) and Ca$^+$($^2S$)+Alk($^2S$) atomic thresholds, and of the $^3\Sigma^+$ electronic state associated with the Ca$^+$($^2S$)+Alk($^2S$) atomic threshold are presented in Fig.~\ref{fig:PDM} for all investigated molecular ions. Values of the permanent electric dipole moments for charged molecules depend on the choice of the coordinate-system origin. Here, they are calculated with respect to the center of mass, which is a natural
choice for investigating the rovibrational dynamics. Their absolute values increase with increasing internuclear distance and asymptotically approach the limiting cases where the charge is completely localized at one of the atoms. This behavior is typical for heteronuclear molecular ions and implies that even molecular ions in very weakly bound states have effectively a significant permanent electric dipole moment in contrast to neutral molecules~\cite{TomzaPRA15b}. The difference between the calculated values and the limiting cases is the interaction-induced variation of the permanent electric dipole moment or, in other words, the degree of charge delocalization. Curves for the $^1\Sigma^+$ electronic states are smooth and their asymptotic behaviors reflect the change of the center-of-mass position for different molecular ions. The degree of charge delocalization increases with the mass of the alkali-metal atom according to the increasing difference of the electronegativity of the Ca and alkali-metal atoms. Different asymptotic behaviors for the $1^1\Sigma^+$ and $2^1\Sigma^+$ states reflect the different charge localization for the Ca($^1S$)+Alk$^+$($^1S$) and Ca$^+$($^2S$)+Alk($^2S$) atomic thresholds. Curves for the $^3\Sigma^+$ electronic state of the CaRb$^+$ and CaCs$^+$ molecular ions show irregularities due to avoided crossings with nearby-lying states.

Permanent electric dipole moments of all investigated electronic states of the CaK$^+$ molecular ion are presented in Fig.~\ref{fig:PDM_CaK+}, while transition electric dipole moments between electronic states of this molecular ion are plotted in Fig.~\ref{fig:TDM_CaK+}. PEDMs and TEDMs for other studied molecular ions are collected in Supplemental Material. For PEDMs, two families of curves associated with two possible arrangements of the charge at the Ca$^+$+Alk and Ca+Alk$^+$ atomic thresholds can be identified. The short-range deviations from the asymptotic behavior give information about charge exchange and delocalization due to interatomic interactions. The shapes of calculated PEDM and TEDM curves and their irregularities at short-range distances can be directly associated with avoided crossings between corresponding potential energy curves, that confirms strong interactions between involved electronic states. The knowledge of changing physical character of electronic states may be useful to predict and explain channels of non-radiative charge-transfer processes. TEDMs at large distances drop to zero when two associated atomic thresholds have different charge arrangements or related atomic excitations are dipole-forbidden. They asymptotically tend to the atomic values only in case of atomic thresholds connected by dipole-allowed transitions. 

\begin{figure}[tb!]
\centering
\includegraphics[width=\columnwidth]{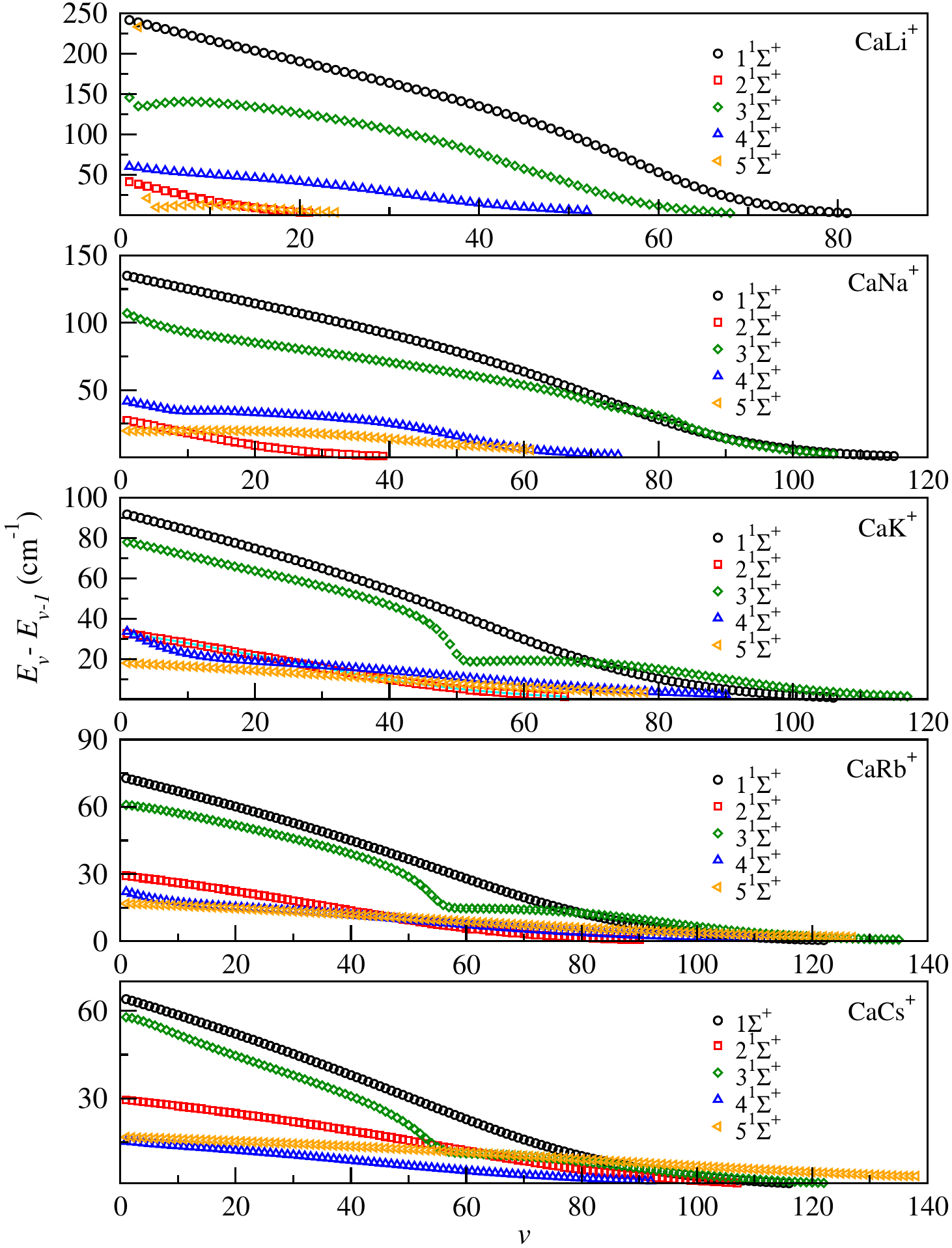}
\caption{Energy spacings between the adjacent vibrational levels ($E_v-E_{v-1}$) of the ground and excited $^1\Sigma^+$ electronic states of the CaAlk$^+$ molecular ions.}
\label{fig:lvls}
\end{figure}

\subsection{Vibrational levels}

We use the present PECs to calculate corresponding vibrational states. In Figure~\ref{fig:lvls}, we present the energy spacings between the adjacent vibrational levels ($E_v-E_{v-1}$) of the ground and excited $^1\Sigma^+$ electronic states of the investigated CaAlk$^+$ molecular ions. For the ground electronic state, there are 86, 103, 114, 125, and 126 vibrational levels for the CaLi$^+$, CaNa$^+$, CaK$^+$, CaRb$^+$, and CaCs$^+$ molecular ions, respectively. The number of vibrational levels increases with the mass of the involved alkali-metal atom, despite the decreasing potential well depth (see Fig.~\ref{fig:CaAlk+}(b)), because the effect of the increasing mass dominates. The spacing between vibrational levels diminishes gradually with a vibrational energy that reflects the strong anharmonicity of the PECs. The overall pattern of energy spacings of different electronic states for different molecular ions is similar. For some states, however, irregularities related to the avoided crossings are visible, e.g., for the $3^1\Sigma^+$ state of CaK$^+$, CaRb$^+$, and CaCs$^{+}$.

\subsection{Spontaneous and light-assisted ion-neutral charge-transfer processes}
\label{sec:coll}

Interactions and collisions of laser-cooled trapped Ca$^+$ ions with ultracold alkali-metal atoms are of the highest importance for experimental realizations of ultracold ion-atom mixtures~\cite{MakarovPRA03,SmithAPB14,HallPRL11,HallMP13a,EberleCPC16,HazePRA13,HazePRA15,HazePRL18,SaitoPRA17,JyothiRSI19}. Even if both the Ca$^+$ ion and alkali-metal atom are in their electronic ground states, the collision- and interaction-induced radiative charge rearrangement is possible in the form of the radiative charge transfer (RCT)
\begin{equation}
\text{Ca}^+ + \text{Alk} \to \text{Ca} + \text{Alk}^+ + \hbar\omega\,,
\end{equation}
where the electron is spontaneously transferred from the alkali-metal atom to the Ca$^+$ ion emitting a photon of energy $\hbar\omega$ and the radiative association (RA)
\begin{equation}
\text{Ca}^+ + \text{Alk} \to \text{Ca}\text{Alk}^+(v,j) + \hbar\omega\,,
\end{equation}
where the CaAlk$^+$ molecular ion in the $(v,j)$ ro-vibrational level of the electronic ground state is spontaneously formed. 

The interaction between the ground-state Ca$^+$ ion and alkali-metal atom both in the $^1\Sigma^+$ and in the $^3\Sigma^+$ states at large distances is dominated by the induction term
\begin{equation}
V(R) \underset{R\to \infty}{=}-\frac{C_4}{R^4}\,,
\end{equation}
where the leading long-range induction coefficient $C_4=\frac{1}{2}e^2\alpha_\text{Alk}$ is given by the static electric dipole polaizability of the alkali-metal atom $\alpha_\text{Alk}$. This long-range interaction determines the characteristic length scale $R_4 = \sqrt{{2\mu C_4}/{\hbar^2}}$ and the related characteristic energy scale $E_4 = {\hbar^2}/{2\mu R_4^2}$~\cite{TomzaRMP19}. These quantities are relevant for ultracold ion-atom collisions because the length scale $R_4$ establishes the order of magnitude of typical ion-atom scattering lengths while the energy scale $E_4$ determines the quantum regime of $s$-wave collisions~\cite{Feldker2019}. Table~\ref{tab:CaAlk+} collects the long-range coefficients, characteristic lengths, and characteristic energies of the ground-state ion-atom interaction for the investigated mixtures. The characteristic lengths are from 1337$\,$bohr for Ca$^+$+Li to 4706$\,$bohr for Ca$^+$+Cs, and they are an order of magnitude larger for ion-atom systems as compared with neutral counterparts. The characteristic energies are from 0.13$\,\mu$K for Ca$^+$+Cs to 8.12$\,\mu$K for Ca$^+$+Li, and they are two orders of magnitude smaller as compared with neutral counterparts. This is one of the reasons, together with inelastic losses and micromotion-induced hearing in the Paul trap~\cite{CetinaPRL12}, why the realization of ion-atom collisions in the quantum regime is very challenging~\cite{Feldker2019}.

\begin{figure}[tb!]
\centering
\includegraphics[width=\columnwidth]{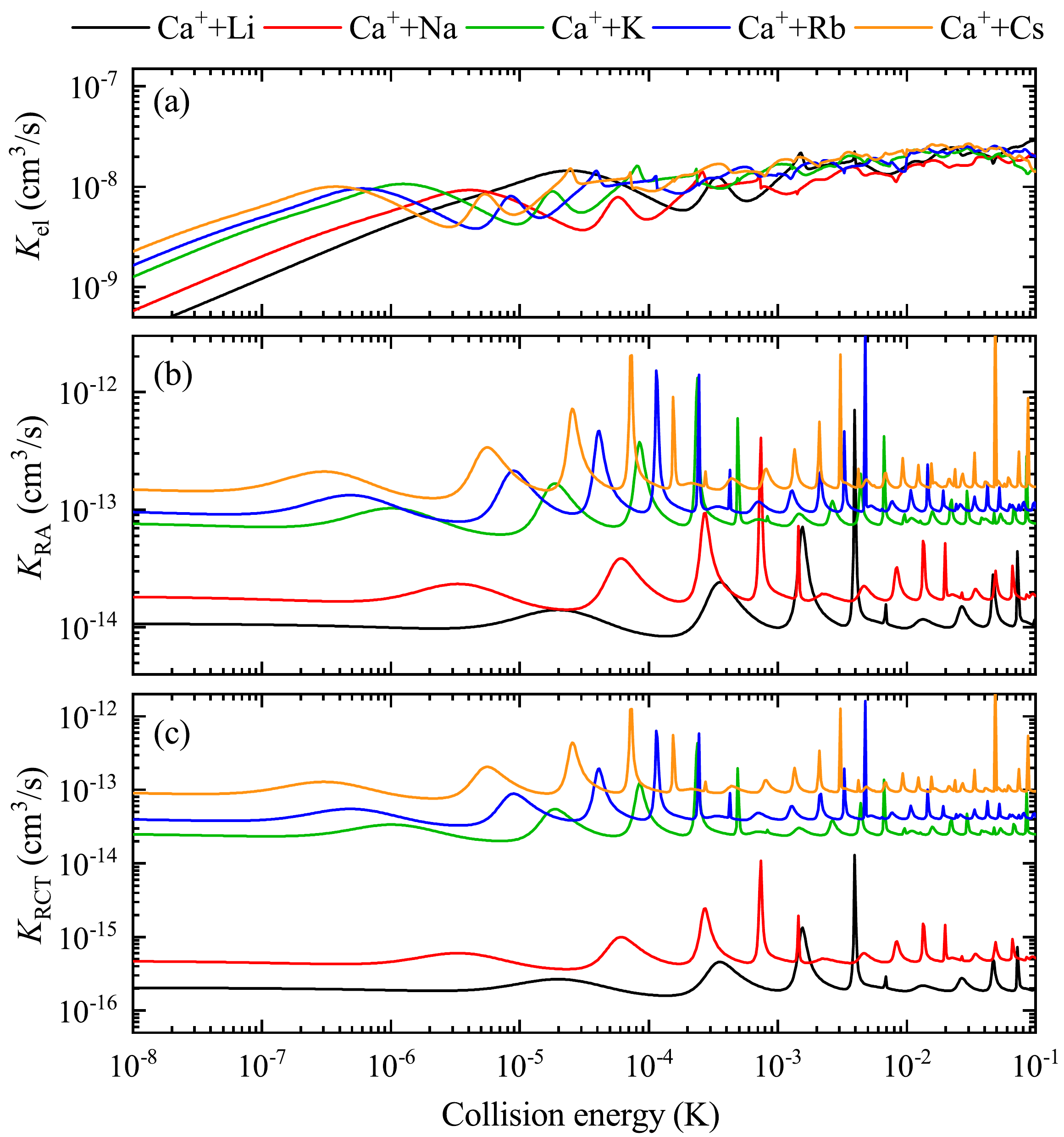}
\caption{Rate constants for collisions between the Ca$^+$ ion and alkali-metal atoms in the $1^1\Sigma^+$ electronic state as a function of the collision energy: (a) elastic scattering, (b) radiative charge transfer, and  (c) radiative association. Scattering lengths are set to $R_4$.}
\label{fig:coll}
\end{figure}

\begin{table}[b!]
\caption{Long-range coefficients~$C_4$, characteristic lengths~$R_4$, and characteristic energies~$E_4$ of the ground-state ion-atom induction interaction for the investigated mixtures.}\label{tab:CaAlk+}
\begin{ruledtabular}
\begin{tabular}{lrrr}
System & $C_4\,$(a.u.) & $R_4\,$(bohr) & $E_4\,$($\mu$K) \\
\hline
$^{40}$Ca$^+$+$^6$Li & 82.2 & 1337 &  8.12  \\
$^{40}$Ca$^+$+$^{23}$Na & 83.2 & 2104 &  1.34 \\
$^{40}$Ca$^+$+$^{39}$K & 145.4 & 3234  & 0.42\\
$^{40}$Ca$^+$+$^{85}$Rb & 159.8 & 3978   & 0.20 \\
$^{40}$Ca$^+$+$^{133}$Cs & 197.8 & 4706  & 0.13 \\
\end{tabular}
\end{ruledtabular}
\end{table}

\begin{table}[b!]
\caption{Rate constants for the Langevin $K_\text{L}$, radiative association $K_\text{RA}$, and radiative charge-transfer $K_\text{RCT}$ collisions in the investigated mixtures.}\label{tab:rates}
\begin{ruledtabular}
\begin{tabular}{lrrr}
System & $K_\text{L}\,$(cm$^3$/s) & $K_\text{RA}\,$(cm$^3$/s) & $K_\text{RCT}\,$(cm$^3$/s) \\
\hline
$^{40}$Ca$^+$+$^6$Li      & $4.73\times 10^{-9}$  & $1.29\times 10^{-14}$  & $2.42\times 10^{-16}$ \\
$^{40}$Ca$^+$+$^{23}$Na   & $3.04\times 10^{-9}$  & $2.09\times 10^{-14}$  & $5.46\times 10^{-16}$ \\
$^{40}$Ca$^+$+$^{39}$K    & $3.46\times 10^{-9}$  & $9.06\times 10^{-14}$  & $2.97\times 10^{-14}$ \\
$^{40}$Ca$^+$+$^{85}$Rb   & $3.09\times 10^{-9}$  & $1.22\times 10^{-13}$  & $5.05\times 10^{-14}$ \\
$^{40}$Ca$^+$+$^{133}$Cs  & $3.23\times 10^{-9}$  & $1.94\times 10^{-13}$  & $1.18\times 10^{-13}$ \\
\end{tabular}
\end{ruledtabular}
\end{table}

\begin{figure}[tb!]
\centering
\includegraphics[width=\columnwidth]{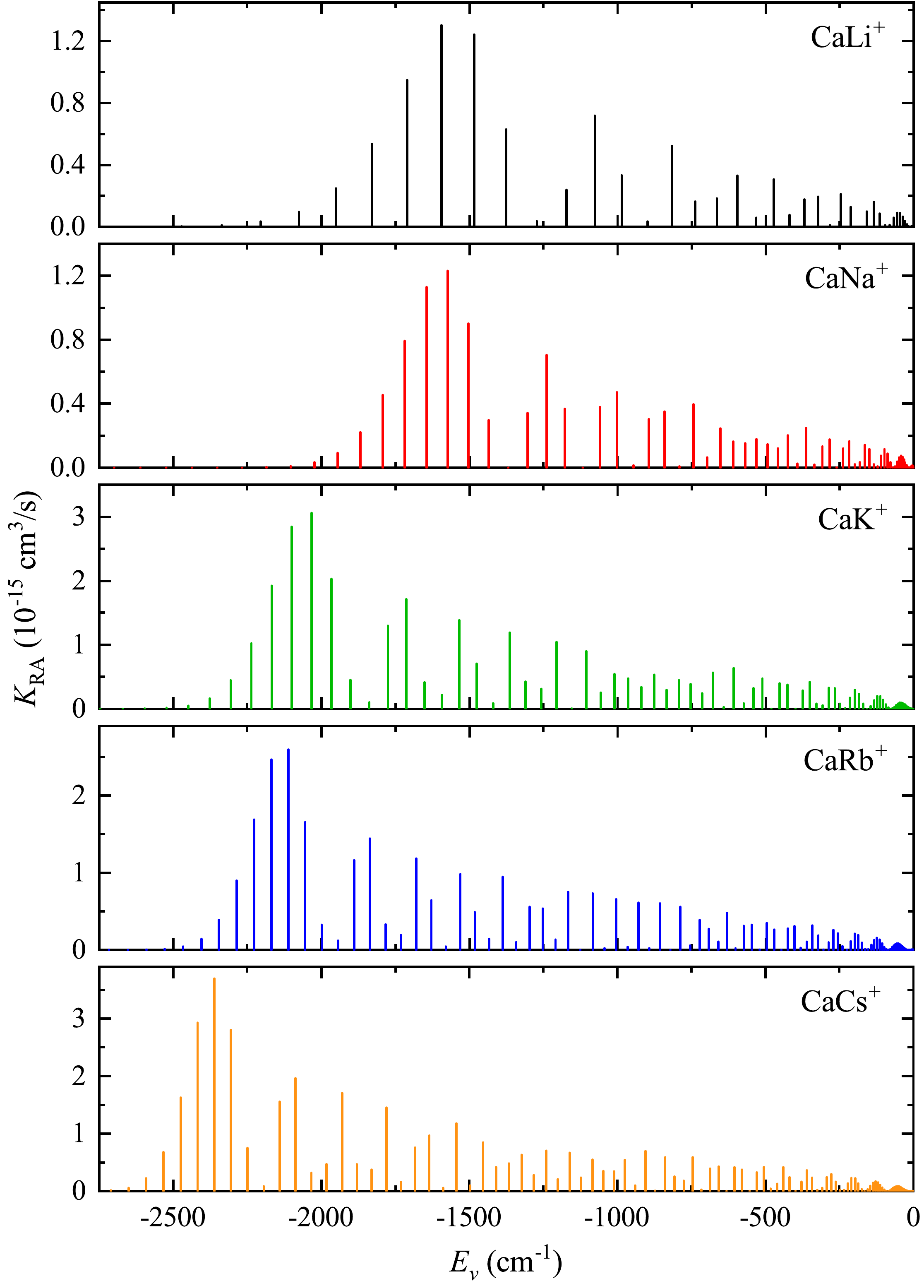}
\caption{Radiative association rate constants as a function of the energy of the final ro-vibrational level in the $1^1\Sigma^+$ ground electronic state of the CaAlk$^+$ molecular ions for the Ca$^+$ ion colliding with the alkali-metal atom in the $2^1\Sigma^+$ state with the scattering length of $R_4$ at the temperature of 1$\,\mu$K.}
\label{fig:KRA}
\end{figure}

To produce and study the considered ion-atom mixtures in the quantum regime, the ion, after initial laser cooling, should be subsequently cooled sympathetically via elastic collisions with surrounding ultracold neutral gas~\cite{HazePRL18,Feldker2019,SchmidtPRL2020}. Such a scheme is feasible only if rates for elastic scattering are significantly larger than rates for inelastic collisions. Therefore, in Fig.~\ref{fig:coll}, we present rate constants for elastic and radiative inelastic collisions between the Ca$^+$ ion and alkali-metal atoms in the $2^1\Sigma^+$ electronic state as a function of the collision energy. We assume typical scattering length of $a_s=R_4$ in the entrance channel, while the results do not depend on the scattering length in the exit channel. Rate constants for small collision energies and pattern of shape resonances depend strongly on the scattering length, but the overall magnitude of rate constants does not depend on it. In the range of investigated collision energies, the rate constants for the elastic scattering $K_\text{el}$ for all systems have similar values of around $10^{-8}\,$cm$^3$/s. Because the same scattering length (in unites of the characteristic length) is assumed for all Ca$^+$+Alk mixtures, the pattern of shape resonances is very similar for all systems, however positions of shape resonances are scaled according to the characteristic energies. Similarly to other ion-atom systems~\cite{Krych2011,IdziaszekNJP11,SayfutyarovaPRA13,TomzaPRA15a,daSilvaNJP2015}, shape resonances are more pronounced for inelastic rate constants, however if the thermal distribution of collision energies is assumed, the thermal averaging removes energy dependence for temperatures larger than 1$\,$mK in agreement with predictions of the classical Langevin capture theory~\cite{Levine09}. The magnitude of the rate constants for the radiative association $K_\text{RA}$ and charge transfer $K_\text{RCT}$ depends on the system. In Table~\ref{tab:rates}, we collect thermally averaged rate constants for the radiative association and radiative charge-transfer collisions in the investigated mixtures compared with the Langevin rate constants $K_\text{L}=2\pi\sqrt{2C_4/\mu}$. Similarly as for other alkaline-earth-metal--alkali-metal ion-atom systems~\cite{TomzaPRA15a,daSilvaNJP2015,SaitoPRA17}, the rate constants for the radiative losses are at least $10^4$ times smaller than Langevin and elastic rate constants. The radiative rates constants increase with the mass of the alkali-metal atom according to the increasing energy of an emitted photon. At the same time, the radiative association is 53, 38, 3.1, 2.4, and 1.6 times more probable than the radiative charge transfer for Ca$^+$+Li, Ca$^+$+Na, Ca$^+$+K, Ca$^+$+Rb, and Ca$^+$+Cs collisions, respectively.

Radiative association rate constants as a function of the energy of the final ro-vibrational level in the $1^1\Sigma^+$ electronic ground state for the Ca$^+$ ion colliding with the alkali-metal atoms in the $2^1\Sigma^+$ state are presented in Fig.~\ref{fig:KRA}. For all systems, the formation of molecular ions in vibrational levels from the middle of the spectrum is the most probable. For example, the formation of molecular ions in vibrational levels with the vibrational quantum number around $v=46$, $v=28$, and $v=13$ and the binding energy around 1595$\,$cm$^{-1}$, 2033$\,$cm$^{-1}$, and 2361$\,$cm$^{-1}$ are the most probable for CaLi$^+$, CaK$^+$, and CaCs$^+$, respectively. The molecular formation probability decreases gradually for decreasing binding energies and is strongly suppressed for binding energies larger than 2500$\,$cm$^{-1}$ because of the interplay between Franck-Condon factors between vibrational levels of the $2^1\Sigma^+$ and $2^1\Sigma^+$ electronic states and transition electric dipole moment between them (see Fig.~\ref{fig:CaAlk+}). Interestingly, more deeply bound molecular ions can be formed for heavier alkali-metal atom despite they have smaller potential well depths.    

In a field-free case, where all spin orientations are present, the described above reactive collisions governed by the $2^1\Sigma^+$ electronic state constitute 25\% of scattering. Remaining collisions are governed by the $^3\Sigma^+$ state, which for Ca$^+$+Li, Ca$^+$+Na, and Ca$^+$+K is free from radiative losses. For Ca$^+$+Rb and Ca$^+$+Cs the radiative losses are also possible from the $^3\Sigma^+$ state but they should be much less probable than radiative losses from the $^1\Sigma^+$ state because of much smaller energies of realized photons. The radiative association and charge transfer are expected to be a dominant loss mechanism for the ground-state Ca$^+$+Li and Ca$^+$+Na collisions, because the entrance atomic threshold is well separated from lower and higher lying electronic states in these systems. In the case of Ca$^+$+K collisions, the radiative processes should also be most important however the coupling with the $1^3\Pi$ state may affect them. In the case of Ca$^+$($^2S$)+Rb($^2S$) collisions, nonradiative charge-transfer losses were observed~\cite{Tacconi2011,HallMP13a} as a dominant mechanism because of strong nonadiabatic couplings with below nearby-lying electronic states associated with the Ca($^3P$)+Rb$^+$($^1S$) atomic threshold. In the case of  Ca$^+$+Cs collisions, more balanced interplay between radiative and nonradiative processes can be expected. Detailed studies of the nonradiative collisional dynamics are out of the scope of this paper.

If the Ca$^+$ ions or alkali-metal atoms are excited by a laser field, the light-induced charge-transfer and association processes are possible
\begin{equation}
\begin{split}
\text{Ca}^+ + \text{Alk} + \hbar\omega' &\to \text{Ca} + \text{Alk}^+  \,,\\
\text{Ca}^+ + \text{Alk} + \hbar\omega' &\to \text{Ca}\text{Alk}^+(v,j) \,,
\end{split}
\end{equation}
where a laser field can be employed to directly stimulate the transition to the ground electronic state~\cite{TomzaPRA15a,PetrovJCP17} or to excite the ion-atom system to higher excited states~\cite{GacesaPRA16}. In the latter case, both radiative and nonradiative deexcition processes can happen depending on the structure of excited electronic states. Nonradiative deexcition can be driven by nonadiabatic couplings between electronic states of the same symmetry or spin-orbit couplings between electronic states of different symmetry. If the ion-atom system is excited to higher-lying atomic thresholds, then the sequence of radiative and nonradiative deexcitations trough intermediate excited electronic states can also be envisioned~\cite{GacesaPRA16}.

The calculated potential energy curves and transition electric dipole moments can be employed to predict and interpret experimental measurements. Various rate constants for charge-transfer collisions between excited-state Ca$^+$ ions and Li~\cite{SaitoPRA17}, Na~\cite{SmithAPB14}, and Rb~\cite{HallMP13a} atoms were measured and rationalized based on the structure of real and avoided crossings between involved molecular electronic states at short distances. For all investigated systems, exciting both Ca$^+$ ion to the $^2D$ or $^2P$ states and alkali-metal atom to the $^2P$ state significantly enhances the charge-transfer rate constants from negligible to significant fraction of the Langevin rate constant. For example, in Fig.~\ref{fig:CaLi+} and Fig.~\ref{fig:CaNa+}, the Ca$^+$($^2D$)+Li($^2S$) and Ca$^+$($^2D$)+Na($^2S$) atomic thresholds and associated molecular electronic states are closely surrounded by several electronic states. Similar light-assisted enhancements of charge transfer can also be expected for Ca$^+$+K and Ca$^+$+Cs collisions.

\begin{figure}[tb!]
\centering
\includegraphics[width=\columnwidth]{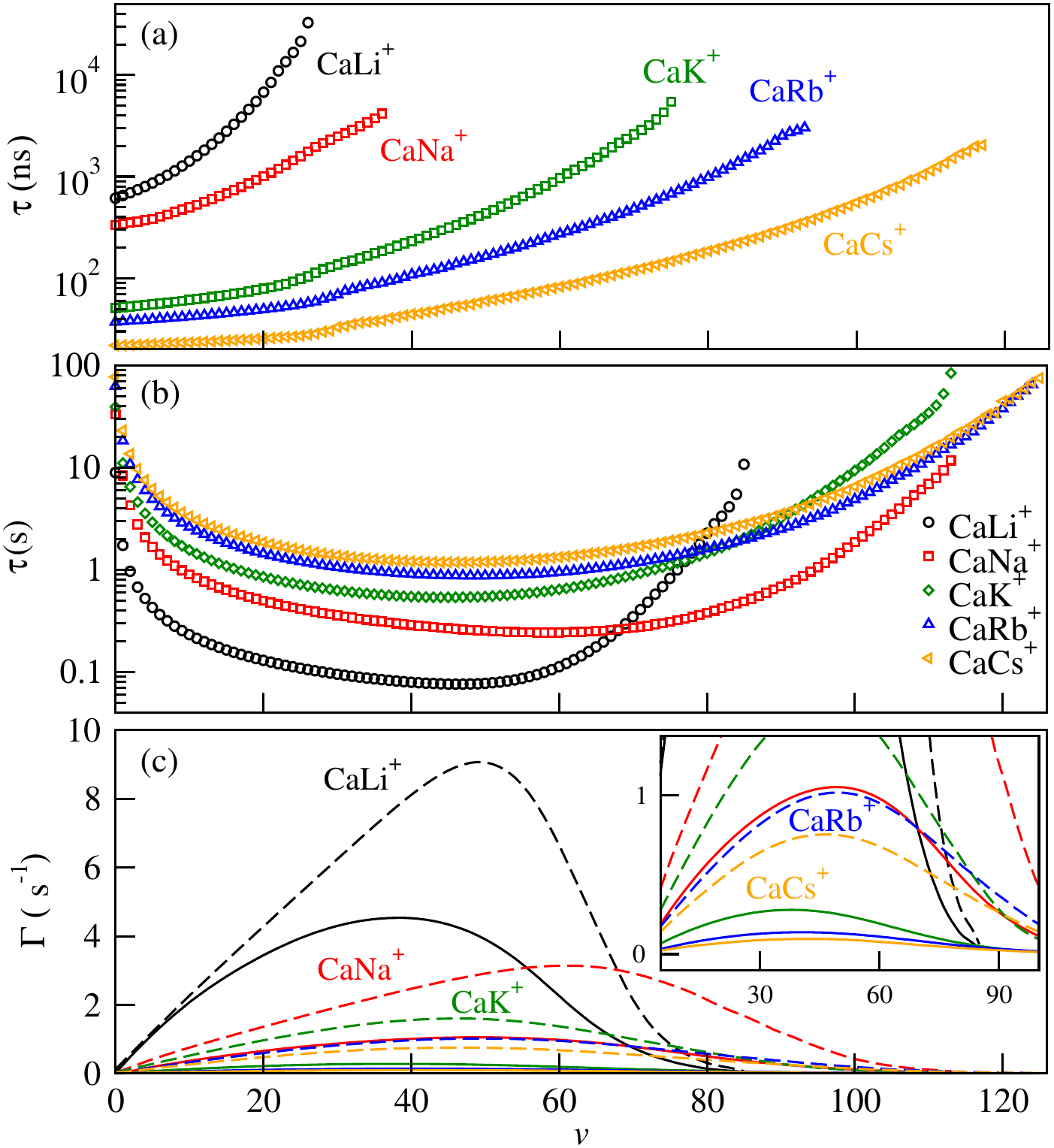}
\caption{Radiative lifetimes of the vibrational levels of the (a) first-excited and (b) ground $^1\Sigma^+$ electronic states of the CaAlk$^+$ molecular ions. (c) The spontaneous (full lines) and black-body-radiation stimulated (dashed lines) transition rates for the vibrational levels of the ground $^1\Sigma^+$ electronic state of the CaAlk$^+$ molecular ions.
}
\label{fig:times}
\end{figure}

The interplay between photoassociation into excited weakly bound molecular ions and subsequent deexcition to the ground-state molecular ions or competitive dissociative charge transfer can be expected for collisions in the laser field~\cite{TomzaPRA15a,GacesaPRA16,PetrovJCP17}. The Ca($^2S$)+Alk($^2P$) atomic thresholds are relatively well separated from other thresholds in the considered systems. This opens the way for photoassociation spectroscopy and molecular ion formation similar as in alkali-metal gases~\cite{JonesRMP06}. The existence of several charge-transferred atomic thresholds may also allow for short-range photoassociation schemes. For example, the Ca($^3P$)+Li$^+$($^1S$) and Ca($^3P$)+Na$^+$($^1S$) atomic thresholds are well separated from other asymptotes, and the relevant $1^3\Sigma^+$ and $2^3\Sigma^+$ electronic states have a similar shape and are connected by the large transition dipole moment at the short range. Finally, magnetic Feshbach resonances in the ground electronic state can be employed to enhance molecular ion formation rates. Detailed studies of the photoassociation and magnetoassociation schemes, however, are out of the scope of this paper.

\subsection{Radiative lifetimes}

We use the present PECs, PEDMs, and TEDMs to calculate the lifetimes of vibrational states of the ground and first excited $^1\Sigma^+$ electronic states of the considered CaAlk$^+$ molecular ions. These lifetimes may be useful to assess prospect for the formation and spectroscopy of calcium--alkali-metal-atom molecular ions in modern experiments with cold ion-atom mixtures. 

The lifetimes of vibrational levels of the first excited $^1\Sigma^+$ electronic state are presented in Fig.~\ref{fig:times}(a) and are governed by the transition electric dipole moment to the ground electronic state associated with emitting an optical photon. This transition moment is significant at short internuclear distances and decreases exponentially with the increasing internuclear distance (see Fig.~\ref{fig:CaAlk+}(c)). Therefore, the lowest vibrational levels have lifetimes in the range of tens to hundreds of nanosecond (between 22$\,$ns for CaCs$^+$ and 335$\,$ns for CaLi$^+$ for the lowest vibrational level), while the most weakly bound levels have lifetimes exceeding microseconds. The lifetimes increase with the vibrational number and decrease with the increasing mass of the alkali-metal atom.

The lifetimes of vibrational levels of the ground $^1\Sigma^+$ electronic state are presented in Fig.~\ref{fig:times}(b) and are governed by its permanent electric dipole moment responsible for weak transitions between different vibrational levels associated with emitting microwave photons. In the case of these transitions, both relatively weak spontaneous emission and stimulated by the black body radiation absorption and emission have to be included. We assume the black body radiation spectrum with the temperature of 300$\,$K. The lifetimes of the lowest and the most-weakly-bound vibrational levels exceed ten seconds (between 9$\,$s for CaLi$^+$ and 77$\,s$ CaCs$^+$ for the lowest vibrational level), while other levels have lifetimes of the order of one second. The interplay between the spontaneous and stimulated transitions can be seen in Fig.~\ref{fig:times}(c), where we compare the spontaneous and stimulated transition rates for the vibrational levels of the ground $^1\Sigma^+$ electronic state of the CaAlk$^+$ molecular ions. The present lifetimes have similar characteristics as comparable results for other neutral and ionic dimers~\cite{Zemke2004,Fedorov2017}.

\section{Conclusion} 
\label{sec:con}

Motivated by recent experimental studies on ultracold mixtures of Ca$^+$ ions immerse in alkali-metal atoms, in a comparative study, we have investigated the electronic structure and the prospects for the formation of the molecular ions composed of a calcium ion and an alkali-metal atom: CaAlk$^+$ (Alk=Li, Na, K, Rb, Cs).  We have used the theoretical quantum chemistry approach based on non-empirical pseudopotential, operatorial core-valence correlation, large Gaussian basis sets, and full configuration interaction method for valence electrons. We have calculated adiabatic potential energy curves, spectroscopic constants, and transition as well as permanent electric dipole moments for the ground and several excited singlet and triplet electronic states of the $\Sigma^+$, $\Pi$, and $\Delta$ spatial symmetries. Next, the electronic structure data have been employed to examine the prospects for the ion-neutral reactive processes and production of molecular ions via spontaneous radiative association and laser-induced photoassociation. Finally, we have calculated the radiative lifetimes of vibrational states of the ground and first excited electronic states.

Our results are in good agreement with the previous theoretical studies of the electronic structure of the ground and excited electronic states of the CaLi$^+$~\cite{Smialkowski2019,Bala2019,SaitoPRA17,HabliMP16,Xie2005,Russon1998,Kimura1983}, CaNa$^+$~\cite{MakarovPRA03,Gacesa2016,JellaliJQSRT18,Smialkowski2019}, and CaRb$^+$~\cite{Tacconi2011,Felix2013,daSilvaNJP2015,Smialkowski2019} molecular ions, which confirms the accuracy of the employed computational approach. The structure of the excited electronic states of the CaK$^+$ and CaCs$^+$ molecular ions is reported here for the first time. The rate constants for the radiative charge transfer and association in the ground-state collisions of the Ca$^+$ ion and alkali-metal atom are predicted to be much smaller than the rate constants for elastic scattering for all the considered systems. They are also predicted to increase with the mass of the alkali-metal atom. For the ground-state Ca$^+$+K collisions, radiative losses should be the main source of losses, negligible for buffer gas cooling or other applications. For the ground-state Ca$^+$+Cs collisions, the interplay between radiative and nonradiative charger-transfer processes is expected. The radiative association leads to the formation of ground-state molecular ions with binding vibrational energies in the range of 1500-2500$\,$cm$^{-1}$ and is predicted to be more probable than radiative charge transfer. For all the systems, the excited-state inelastic collisions are expected to be much faster than the ground-state ones. Based on the electronic structure, photoassociation schemes based on both short-range and long-range excitations can be envisioned. The radiative lifetimes of vibrational states of the ground and first excited electronic states are found in the range of 0.1-100$\,$s and 10$\,$ns-10$\,\mu$s, respectively. The present results may be useful and pave the way for the formation and spectroscopy of calcium--alkali-metal-atom molecular ions in modern experiments with cold ion-atom mixtures. In the future, the presented computational scheme will be employed to study excited electronic states in triatomic molecular ions. 

The full potential energy curves, permanent and transition electric dipole moments as a function of interatomic distance in the numerical form are available for all investigated systems from the authors upon request.

\begin{acknowledgments}
M.T.~was supported by the National Science Centre Poland (Sonata Grant nr. 2015/19/D/ST4/02173), the Foundation for Polish Science within the First Team programme cofinanced by the European Union under the European Regional Development Fund, and PL-Grid Infrastructure.
\end{acknowledgments}

\bibliography{bibliography}

\begin{thebibliography}{90}%
\makeatletter
\providecommand \@ifxundefined [1]{%
 \@ifx{#1\undefined}
}%
\providecommand \@ifnum [1]{%
 \ifnum #1\expandafter \@firstoftwo
 \else \expandafter \@secondoftwo
 \fi
}%
\providecommand \@ifx [1]{%
 \ifx #1\expandafter \@firstoftwo
 \else \expandafter \@secondoftwo
 \fi
}%
\providecommand \natexlab [1]{#1}%
\providecommand \enquote  [1]{``#1''}%
\providecommand \bibnamefont  [1]{#1}%
\providecommand \bibfnamefont [1]{#1}%
\providecommand \citenamefont [1]{#1}%
\providecommand \href@noop [0]{\@secondoftwo}%
\providecommand \href [0]{\begingroup \@sanitize@url \@href}%
\providecommand \@href[1]{\@@startlink{#1}\@@href}%
\providecommand \@@href[1]{\endgroup#1\@@endlink}%
\providecommand \@sanitize@url [0]{\catcode `\\12\catcode `\$12\catcode
  `\&12\catcode `\#12\catcode `\^12\catcode `\_12\catcode `\%12\relax}%
\providecommand \@@startlink[1]{}%
\providecommand \@@endlink[0]{}%
\providecommand \url  [0]{\begingroup\@sanitize@url \@url }%
\providecommand \@url [1]{\endgroup\@href {#1}{\urlprefix }}%
\providecommand \urlprefix  [0]{URL }%
\providecommand \Eprint [0]{\href }%
\providecommand \doibase [0]{http://dx.doi.org/}%
\providecommand \selectlanguage [0]{\@gobble}%
\providecommand \bibinfo  [0]{\@secondoftwo}%
\providecommand \bibfield  [0]{\@secondoftwo}%
\providecommand \translation [1]{[#1]}%
\providecommand \BibitemOpen [0]{}%
\providecommand \bibitemStop [0]{}%
\providecommand \bibitemNoStop [0]{.\EOS\space}%
\providecommand \EOS [0]{\spacefactor3000\relax}%
\providecommand \BibitemShut  [1]{\csname bibitem#1\endcsname}%
\let\auto@bib@innerbib\@empty
\bibitem [{\citenamefont {Tomza}\ \emph {et~al.}(2019)\citenamefont {Tomza},
  \citenamefont {Jachymski}, \citenamefont {Gerritsma}, \citenamefont
  {Negretti}, \citenamefont {Calarco}, \citenamefont {Idziaszek},\ and\
  \citenamefont {Julienne}}]{TomzaRMP19}%
  \BibitemOpen
  \bibfield  {author} {\bibinfo {author} {\bibfnamefont {M.}~\bibnamefont
  {Tomza}}, \bibinfo {author} {\bibfnamefont {K.}~\bibnamefont {Jachymski}},
  \bibinfo {author} {\bibfnamefont {R.}~\bibnamefont {Gerritsma}}, \bibinfo
  {author} {\bibfnamefont {A.}~\bibnamefont {Negretti}}, \bibinfo {author}
  {\bibfnamefont {T.}~\bibnamefont {Calarco}}, \bibinfo {author} {\bibfnamefont
  {Z.}~\bibnamefont {Idziaszek}}, \ and\ \bibinfo {author} {\bibfnamefont
  {P.~S.}\ \bibnamefont {Julienne}},\ }\href {\doibase
  10.1103/RevModPhys.91.035001} {\bibfield  {journal} {\bibinfo  {journal}
  {Rev. Mod. Phys.}\ }\textbf {\bibinfo {volume} {91}},\ \bibinfo {pages}
  {035001} (\bibinfo {year} {2019})}\BibitemShut {NoStop}%
\bibitem [{\citenamefont {C{\^o}t{\'e}}(2016)}]{CoteAAMOP16}%
  \BibitemOpen
  \bibfield  {author} {\bibinfo {author} {\bibfnamefont {R.}~\bibnamefont
  {C{\^o}t{\'e}}},\ }\href {\doibase 10.1016/bs.aamop.2016.04.004} {\bibfield
  {journal} {\bibinfo  {journal} {Adv. At. Mol. Opt. Phys.}\ }\textbf {\bibinfo
  {volume} {65}},\ \bibinfo {pages} {67} (\bibinfo {year} {2016})}\BibitemShut
  {NoStop}%
\bibitem [{\citenamefont {H\"arter}\ and\ \citenamefont
  {Denschlag}(2014)}]{HarterCP14}%
  \BibitemOpen
  \bibfield  {author} {\bibinfo {author} {\bibfnamefont {A.}~\bibnamefont
  {H\"arter}}\ and\ \bibinfo {author} {\bibfnamefont {J.~H.}\ \bibnamefont
  {Denschlag}},\ }\href {\doibase 10.1080/00107514.2013.854618} {\bibfield
  {journal} {\bibinfo  {journal} {Contemp. Phys.}\ }\textbf {\bibinfo {volume}
  {55}},\ \bibinfo {pages} {33} (\bibinfo {year} {2014})}\BibitemShut {NoStop}%
\bibitem [{\citenamefont {C\^ot\'e}\ and\ \citenamefont
  {Dalgarno}(2000)}]{CotePRA00}%
  \BibitemOpen
  \bibfield  {author} {\bibinfo {author} {\bibfnamefont {R.}~\bibnamefont
  {C\^ot\'e}}\ and\ \bibinfo {author} {\bibfnamefont {A.}~\bibnamefont
  {Dalgarno}},\ }\href {\doibase 10.1103/PhysRevA.62.012709} {\bibfield
  {journal} {\bibinfo  {journal} {Phys. Rev. A}\ }\textbf {\bibinfo {volume}
  {62}},\ \bibinfo {pages} {012709} (\bibinfo {year} {2000})}\BibitemShut
  {NoStop}%
\bibitem [{\citenamefont {Zipkes}\ \emph {et~al.}(2010)\citenamefont {Zipkes},
  \citenamefont {Palzer}, \citenamefont {Sias},\ and\ \citenamefont
  {K{\"o}hl}}]{ZipkesNature10}%
  \BibitemOpen
  \bibfield  {author} {\bibinfo {author} {\bibfnamefont {C.}~\bibnamefont
  {Zipkes}}, \bibinfo {author} {\bibfnamefont {S.}~\bibnamefont {Palzer}},
  \bibinfo {author} {\bibfnamefont {C.}~\bibnamefont {Sias}}, \ and\ \bibinfo
  {author} {\bibfnamefont {M.}~\bibnamefont {K{\"o}hl}},\ }\href {\doibase
  10.1038/nature08865} {\bibfield  {journal} {\bibinfo  {journal} {Nature}\
  }\textbf {\bibinfo {volume} {464}},\ \bibinfo {pages} {388} (\bibinfo {year}
  {2010})}\BibitemShut {NoStop}%
\bibitem [{\citenamefont {Ravi}\ \emph {et~al.}(2012)\citenamefont {Ravi},
  \citenamefont {Lee}, \citenamefont {Sharma}, \citenamefont {Werth},\ and\
  \citenamefont {Rangwala}}]{RaviNatCommun12}%
  \BibitemOpen
  \bibfield  {author} {\bibinfo {author} {\bibfnamefont {K.}~\bibnamefont
  {Ravi}}, \bibinfo {author} {\bibfnamefont {S.}~\bibnamefont {Lee}}, \bibinfo
  {author} {\bibfnamefont {A.}~\bibnamefont {Sharma}}, \bibinfo {author}
  {\bibfnamefont {G.}~\bibnamefont {Werth}}, \ and\ \bibinfo {author}
  {\bibfnamefont {S.}~\bibnamefont {Rangwala}},\ }\href {\doibase
  10.1038/ncomms2131} {\bibfield  {journal} {\bibinfo  {journal} {Nat.
  Commun.}\ }\textbf {\bibinfo {volume} {3}},\ \bibinfo {pages} {1126}
  (\bibinfo {year} {2012})}\BibitemShut {NoStop}%
\bibitem [{\citenamefont {H\"oltkemeier}\ \emph {et~al.}(2016)\citenamefont
  {H\"oltkemeier}, \citenamefont {Weckesser}, \citenamefont {L\'opez-Carrera},\
  and\ \citenamefont {Weidem\"uller}}]{HoltkemeierPRL16}%
  \BibitemOpen
  \bibfield  {author} {\bibinfo {author} {\bibfnamefont {B.}~\bibnamefont
  {H\"oltkemeier}}, \bibinfo {author} {\bibfnamefont {P.}~\bibnamefont
  {Weckesser}}, \bibinfo {author} {\bibfnamefont {H.}~\bibnamefont
  {L\'opez-Carrera}}, \ and\ \bibinfo {author} {\bibfnamefont {M.}~\bibnamefont
  {Weidem\"uller}},\ }\href {\doibase 10.1103/PhysRevLett.116.233003}
  {\bibfield  {journal} {\bibinfo  {journal} {Phys. Rev. Lett.}\ }\textbf
  {\bibinfo {volume} {116}},\ \bibinfo {pages} {233003} (\bibinfo {year}
  {2016})}\BibitemShut {NoStop}%
\bibitem [{\citenamefont {Feldker}\ \emph {et~al.}(2020)\citenamefont
  {Feldker}, \citenamefont {F{\"u}rst}, \citenamefont {Hirzler}, \citenamefont
  {Ewald}, \citenamefont {Mazzanti}, \citenamefont {Wiater}, \citenamefont
  {Tomza},\ and\ \citenamefont {Gerritsma}}]{Feldker2019}%
  \BibitemOpen
  \bibfield  {author} {\bibinfo {author} {\bibfnamefont {T.}~\bibnamefont
  {Feldker}}, \bibinfo {author} {\bibfnamefont {H.}~\bibnamefont {F{\"u}rst}},
  \bibinfo {author} {\bibfnamefont {H.}~\bibnamefont {Hirzler}}, \bibinfo
  {author} {\bibfnamefont {N.}~\bibnamefont {Ewald}}, \bibinfo {author}
  {\bibfnamefont {M.}~\bibnamefont {Mazzanti}}, \bibinfo {author}
  {\bibfnamefont {D.}~\bibnamefont {Wiater}}, \bibinfo {author} {\bibfnamefont
  {M.}~\bibnamefont {Tomza}}, \ and\ \bibinfo {author} {\bibfnamefont
  {R.}~\bibnamefont {Gerritsma}},\ }\href {\doibase 10.1038/s41567-019-0772-5}
  {\bibfield  {journal} {\bibinfo  {journal} {Nat. Phys.}\ } (\bibinfo {year}
  {2020}),\ 10.1038/s41567-019-0772-5}\BibitemShut {NoStop}%
\bibitem [{\citenamefont {Sikorsky}\ \emph
  {et~al.}(2018{\natexlab{a}})\citenamefont {Sikorsky}, \citenamefont {Meir},
  \citenamefont {Ben-shlomi}, \citenamefont {Akerman},\ and\ \citenamefont
  {Ozeri}}]{SikorskyNC18}%
  \BibitemOpen
  \bibfield  {author} {\bibinfo {author} {\bibfnamefont {T.}~\bibnamefont
  {Sikorsky}}, \bibinfo {author} {\bibfnamefont {Z.}~\bibnamefont {Meir}},
  \bibinfo {author} {\bibfnamefont {R.}~\bibnamefont {Ben-shlomi}}, \bibinfo
  {author} {\bibfnamefont {N.}~\bibnamefont {Akerman}}, \ and\ \bibinfo
  {author} {\bibfnamefont {R.}~\bibnamefont {Ozeri}},\ }\href {\doibase
  10.1038/s41467-018-03373-y} {\bibfield  {journal} {\bibinfo  {journal} {Nat.
  Comm.}\ }\textbf {\bibinfo {volume} {9}},\ \bibinfo {pages} {920} (\bibinfo
  {year} {2018}{\natexlab{a}})}\BibitemShut {NoStop}%
\bibitem [{\citenamefont {Makarov}\ \emph {et~al.}(2003)\citenamefont
  {Makarov}, \citenamefont {C\^ot\'e}, \citenamefont {Michels},\ and\
  \citenamefont {Smith}}]{MakarovPRA03}%
  \BibitemOpen
  \bibfield  {author} {\bibinfo {author} {\bibfnamefont {O.~P.}\ \bibnamefont
  {Makarov}}, \bibinfo {author} {\bibfnamefont {R.}~\bibnamefont {C\^ot\'e}},
  \bibinfo {author} {\bibfnamefont {H.}~\bibnamefont {Michels}}, \ and\
  \bibinfo {author} {\bibfnamefont {W.~W.}\ \bibnamefont {Smith}},\ }\href
  {\doibase 10.1103/PhysRevA.67.042705} {\bibfield  {journal} {\bibinfo
  {journal} {Phys. Rev. A}\ }\textbf {\bibinfo {volume} {67}},\ \bibinfo
  {pages} {042705} (\bibinfo {year} {2003})}\BibitemShut {NoStop}%
\bibitem [{\citenamefont {Ratschbacher}\ \emph {et~al.}(2012)\citenamefont
  {Ratschbacher}, \citenamefont {Zipkes}, \citenamefont {Sias},\ and\
  \citenamefont {Kohl}}]{RatschbacherNatPhys12}%
  \BibitemOpen
  \bibfield  {author} {\bibinfo {author} {\bibfnamefont {L.}~\bibnamefont
  {Ratschbacher}}, \bibinfo {author} {\bibfnamefont {C.}~\bibnamefont
  {Zipkes}}, \bibinfo {author} {\bibfnamefont {C.}~\bibnamefont {Sias}}, \ and\
  \bibinfo {author} {\bibfnamefont {M.}~\bibnamefont {Kohl}},\ }\href {\doibase
  10.1038/nphys2373} {\bibfield  {journal} {\bibinfo  {journal} {Nat. Phys.}\
  }\textbf {\bibinfo {volume} {8}},\ \bibinfo {pages} {649} (\bibinfo {year}
  {2012})}\BibitemShut {NoStop}%
\bibitem [{\citenamefont {F\"urst}\ \emph {et~al.}(2018)\citenamefont
  {F\"urst}, \citenamefont {Feldker}, \citenamefont {Ewald}, \citenamefont
  {Joger}, \citenamefont {Tomza},\ and\ \citenamefont
  {Gerritsma}}]{FurstPRA18}%
  \BibitemOpen
  \bibfield  {author} {\bibinfo {author} {\bibfnamefont {H.}~\bibnamefont
  {F\"urst}}, \bibinfo {author} {\bibfnamefont {T.}~\bibnamefont {Feldker}},
  \bibinfo {author} {\bibfnamefont {N.~V.}\ \bibnamefont {Ewald}}, \bibinfo
  {author} {\bibfnamefont {J.}~\bibnamefont {Joger}}, \bibinfo {author}
  {\bibfnamefont {M.}~\bibnamefont {Tomza}}, \ and\ \bibinfo {author}
  {\bibfnamefont {R.}~\bibnamefont {Gerritsma}},\ }\href {\doibase
  10.1103/PhysRevA.98.012713} {\bibfield  {journal} {\bibinfo  {journal} {Phys.
  Rev. A}\ }\textbf {\bibinfo {volume} {98}},\ \bibinfo {pages} {012713}
  (\bibinfo {year} {2018})}\BibitemShut {NoStop}%
\bibitem [{\citenamefont {Sikorsky}\ \emph
  {et~al.}(2018{\natexlab{b}})\citenamefont {Sikorsky}, \citenamefont {Morita},
  \citenamefont {Meir}, \citenamefont {Buchachenko}, \citenamefont
  {Ben-shlomi}, \citenamefont {Akerman}, \citenamefont {Narevicius},
  \citenamefont {Tscherbul},\ and\ \citenamefont {Ozeri}}]{SikorskyPRL18}%
  \BibitemOpen
  \bibfield  {author} {\bibinfo {author} {\bibfnamefont {T.}~\bibnamefont
  {Sikorsky}}, \bibinfo {author} {\bibfnamefont {M.}~\bibnamefont {Morita}},
  \bibinfo {author} {\bibfnamefont {Z.}~\bibnamefont {Meir}}, \bibinfo {author}
  {\bibfnamefont {A.~A.}\ \bibnamefont {Buchachenko}}, \bibinfo {author}
  {\bibfnamefont {R.}~\bibnamefont {Ben-shlomi}}, \bibinfo {author}
  {\bibfnamefont {N.}~\bibnamefont {Akerman}}, \bibinfo {author} {\bibfnamefont
  {E.}~\bibnamefont {Narevicius}}, \bibinfo {author} {\bibfnamefont {T.~V.}\
  \bibnamefont {Tscherbul}}, \ and\ \bibinfo {author} {\bibfnamefont
  {R.}~\bibnamefont {Ozeri}},\ }\href {\doibase 10.1103/PhysRevLett.121.173402}
  {\bibfield  {journal} {\bibinfo  {journal} {Phys. Rev. Lett.}\ }\textbf
  {\bibinfo {volume} {121}},\ \bibinfo {pages} {173402} (\bibinfo {year}
  {2018}{\natexlab{b}})}\BibitemShut {NoStop}%
\bibitem [{\citenamefont {C\^ot\'e}\ and\ \citenamefont
  {Simbotin}(2018)}]{CotePRL18}%
  \BibitemOpen
  \bibfield  {author} {\bibinfo {author} {\bibfnamefont {R.}~\bibnamefont
  {C\^ot\'e}}\ and\ \bibinfo {author} {\bibfnamefont {I.}~\bibnamefont
  {Simbotin}},\ }\href {\doibase 10.1103/PhysRevLett.121.173401} {\bibfield
  {journal} {\bibinfo  {journal} {Phys. Rev. Lett.}\ }\textbf {\bibinfo
  {volume} {121}},\ \bibinfo {pages} {173401} (\bibinfo {year}
  {2018})}\BibitemShut {NoStop}%
\bibitem [{\citenamefont {Saito}\ \emph {et~al.}(2017)\citenamefont {Saito},
  \citenamefont {Haze}, \citenamefont {Sasakawa}, \citenamefont {Nakai},
  \citenamefont {Raoult}, \citenamefont {Da~Silva}, \citenamefont {Dulieu},\
  and\ \citenamefont {Mukaiyama}}]{SaitoPRA17}%
  \BibitemOpen
  \bibfield  {author} {\bibinfo {author} {\bibfnamefont {R.}~\bibnamefont
  {Saito}}, \bibinfo {author} {\bibfnamefont {S.}~\bibnamefont {Haze}},
  \bibinfo {author} {\bibfnamefont {M.}~\bibnamefont {Sasakawa}}, \bibinfo
  {author} {\bibfnamefont {R.}~\bibnamefont {Nakai}}, \bibinfo {author}
  {\bibfnamefont {M.}~\bibnamefont {Raoult}}, \bibinfo {author} {\bibfnamefont
  {H.}~\bibnamefont {Da~Silva}}, \bibinfo {author} {\bibfnamefont
  {O.}~\bibnamefont {Dulieu}}, \ and\ \bibinfo {author} {\bibfnamefont
  {T.}~\bibnamefont {Mukaiyama}},\ }\href {\doibase 10.1103/PhysRevA.95.032709}
  {\bibfield  {journal} {\bibinfo  {journal} {Phys. Rev. A}\ }\textbf {\bibinfo
  {volume} {95}},\ \bibinfo {pages} {032709} (\bibinfo {year}
  {2017})}\BibitemShut {NoStop}%
\bibitem [{\citenamefont {Gerritsma}\ \emph {et~al.}(2012)\citenamefont
  {Gerritsma}, \citenamefont {Negretti}, \citenamefont {Doerk}, \citenamefont
  {Idziaszek}, \citenamefont {Calarco},\ and\ \citenamefont
  {Schmidt-Kaler}}]{GerritsmaPRL12}%
  \BibitemOpen
  \bibfield  {author} {\bibinfo {author} {\bibfnamefont {R.}~\bibnamefont
  {Gerritsma}}, \bibinfo {author} {\bibfnamefont {A.}~\bibnamefont {Negretti}},
  \bibinfo {author} {\bibfnamefont {H.}~\bibnamefont {Doerk}}, \bibinfo
  {author} {\bibfnamefont {Z.}~\bibnamefont {Idziaszek}}, \bibinfo {author}
  {\bibfnamefont {T.}~\bibnamefont {Calarco}}, \ and\ \bibinfo {author}
  {\bibfnamefont {F.}~\bibnamefont {Schmidt-Kaler}},\ }\href {\doibase
  10.1103/PhysRevLett.109.080402} {\bibfield  {journal} {\bibinfo  {journal}
  {Phys. Rev. Lett.}\ }\textbf {\bibinfo {volume} {109}},\ \bibinfo {pages}
  {080402} (\bibinfo {year} {2012})}\BibitemShut {NoStop}%
\bibitem [{\citenamefont {Bissbort}\ \emph {et~al.}(2013)\citenamefont
  {Bissbort}, \citenamefont {Cocks}, \citenamefont {Negretti}, \citenamefont
  {Idziaszek}, \citenamefont {Calarco}, \citenamefont {Schmidt-Kaler},
  \citenamefont {Hofstetter},\ and\ \citenamefont {Gerritsma}}]{BissbortPRL13}%
  \BibitemOpen
  \bibfield  {author} {\bibinfo {author} {\bibfnamefont {U.}~\bibnamefont
  {Bissbort}}, \bibinfo {author} {\bibfnamefont {D.}~\bibnamefont {Cocks}},
  \bibinfo {author} {\bibfnamefont {A.}~\bibnamefont {Negretti}}, \bibinfo
  {author} {\bibfnamefont {Z.}~\bibnamefont {Idziaszek}}, \bibinfo {author}
  {\bibfnamefont {T.}~\bibnamefont {Calarco}}, \bibinfo {author} {\bibfnamefont
  {F.}~\bibnamefont {Schmidt-Kaler}}, \bibinfo {author} {\bibfnamefont
  {W.}~\bibnamefont {Hofstetter}}, \ and\ \bibinfo {author} {\bibfnamefont
  {R.}~\bibnamefont {Gerritsma}},\ }\href {\doibase
  10.1103/PhysRevLett.111.080501} {\bibfield  {journal} {\bibinfo  {journal}
  {Phys. Rev. Lett.}\ }\textbf {\bibinfo {volume} {111}},\ \bibinfo {pages}
  {080501} (\bibinfo {year} {2013})}\BibitemShut {NoStop}%
\bibitem [{\citenamefont {Schurer}\ \emph {et~al.}(2017)\citenamefont
  {Schurer}, \citenamefont {Negretti},\ and\ \citenamefont
  {Schmelcher}}]{SchurerPRL17}%
  \BibitemOpen
  \bibfield  {author} {\bibinfo {author} {\bibfnamefont {J.~M.}\ \bibnamefont
  {Schurer}}, \bibinfo {author} {\bibfnamefont {A.}~\bibnamefont {Negretti}}, \
  and\ \bibinfo {author} {\bibfnamefont {P.}~\bibnamefont {Schmelcher}},\
  }\href {\doibase 10.1103/PhysRevLett.119.063001} {\bibfield  {journal}
  {\bibinfo  {journal} {Phys. Rev. Lett.}\ }\textbf {\bibinfo {volume} {119}},\
  \bibinfo {pages} {063001} (\bibinfo {year} {2017})}\BibitemShut {NoStop}%
\bibitem [{\citenamefont {Doerk}\ \emph {et~al.}(2010)\citenamefont {Doerk},
  \citenamefont {Idziaszek},\ and\ \citenamefont {Calarco}}]{DoerkPRA10}%
  \BibitemOpen
  \bibfield  {author} {\bibinfo {author} {\bibfnamefont {H.}~\bibnamefont
  {Doerk}}, \bibinfo {author} {\bibfnamefont {Z.}~\bibnamefont {Idziaszek}}, \
  and\ \bibinfo {author} {\bibfnamefont {T.}~\bibnamefont {Calarco}},\ }\href
  {\doibase 10.1103/PhysRevA.81.012708} {\bibfield  {journal} {\bibinfo
  {journal} {Phys. Rev. A}\ }\textbf {\bibinfo {volume} {81}},\ \bibinfo
  {pages} {012708} (\bibinfo {year} {2010})}\BibitemShut {NoStop}%
\bibitem [{\citenamefont {Grier}\ \emph {et~al.}(2009)\citenamefont {Grier},
  \citenamefont {Cetina}, \citenamefont {Oru\ifmmode \check{c}\else
  \v{c}\fi{}evi\ifmmode~\acute{c}\else \'{c}\fi{}},\ and\ \citenamefont
  {Vuleti\ifmmode~\acute{c}\else \'{c}\fi{}}}]{GrierPRL09}%
  \BibitemOpen
  \bibfield  {author} {\bibinfo {author} {\bibfnamefont {A.~T.}\ \bibnamefont
  {Grier}}, \bibinfo {author} {\bibfnamefont {M.}~\bibnamefont {Cetina}},
  \bibinfo {author} {\bibfnamefont {F.}~\bibnamefont {Oru\ifmmode
  \check{c}\else \v{c}\fi{}evi\ifmmode~\acute{c}\else \'{c}\fi{}}}, \ and\
  \bibinfo {author} {\bibfnamefont {V.}~\bibnamefont
  {Vuleti\ifmmode~\acute{c}\else \'{c}\fi{}}},\ }\href {\doibase
  10.1103/PhysRevLett.102.223201} {\bibfield  {journal} {\bibinfo  {journal}
  {Phys. Rev. Lett.}\ }\textbf {\bibinfo {volume} {102}},\ \bibinfo {pages}
  {223201} (\bibinfo {year} {2009})}\BibitemShut {NoStop}%
\bibitem [{\citenamefont {Hall}\ \emph
  {et~al.}(2013{\natexlab{a}})\citenamefont {Hall}, \citenamefont {Eberle},
  \citenamefont {Hegi}, \citenamefont {Raoult}, \citenamefont {Aymar},
  \citenamefont {Dulieu},\ and\ \citenamefont {Willitsch}}]{HallMP13a}%
  \BibitemOpen
  \bibfield  {author} {\bibinfo {author} {\bibfnamefont {F.~H.}\ \bibnamefont
  {Hall}}, \bibinfo {author} {\bibfnamefont {P.}~\bibnamefont {Eberle}},
  \bibinfo {author} {\bibfnamefont {G.}~\bibnamefont {Hegi}}, \bibinfo {author}
  {\bibfnamefont {M.}~\bibnamefont {Raoult}}, \bibinfo {author} {\bibfnamefont
  {M.}~\bibnamefont {Aymar}}, \bibinfo {author} {\bibfnamefont
  {O.}~\bibnamefont {Dulieu}}, \ and\ \bibinfo {author} {\bibfnamefont
  {S.}~\bibnamefont {Willitsch}},\ }\href {\doibase
  10.1080/00268976.2013.780107} {\bibfield  {journal} {\bibinfo  {journal}
  {Mol. Phys.}\ }\textbf {\bibinfo {volume} {111}},\ \bibinfo {pages} {2020}
  (\bibinfo {year} {2013}{\natexlab{a}})}\BibitemShut {NoStop}%
\bibitem [{\citenamefont {Sullivan}\ \emph {et~al.}(2012)\citenamefont
  {Sullivan}, \citenamefont {Rellergert}, \citenamefont {Kotochigova},\ and\
  \citenamefont {Hudson}}]{SullivanPRL12}%
  \BibitemOpen
  \bibfield  {author} {\bibinfo {author} {\bibfnamefont {S.~T.}\ \bibnamefont
  {Sullivan}}, \bibinfo {author} {\bibfnamefont {W.~G.}\ \bibnamefont
  {Rellergert}}, \bibinfo {author} {\bibfnamefont {S.}~\bibnamefont
  {Kotochigova}}, \ and\ \bibinfo {author} {\bibfnamefont {E.~R.}\ \bibnamefont
  {Hudson}},\ }\href {\doibase 10.1103/PhysRevLett.109.223002} {\bibfield
  {journal} {\bibinfo  {journal} {Phys. Rev. Lett.}\ }\textbf {\bibinfo
  {volume} {109}},\ \bibinfo {pages} {223002} (\bibinfo {year}
  {2012})}\BibitemShut {NoStop}%
\bibitem [{\citenamefont {Rellergert}\ \emph {et~al.}(2011)\citenamefont
  {Rellergert}, \citenamefont {Sullivan}, \citenamefont {Kotochigova},
  \citenamefont {Petrov}, \citenamefont {Chen}, \citenamefont {Schowalter},\
  and\ \citenamefont {Hudson}}]{RellergertPRL11}%
  \BibitemOpen
  \bibfield  {author} {\bibinfo {author} {\bibfnamefont {W.~G.}\ \bibnamefont
  {Rellergert}}, \bibinfo {author} {\bibfnamefont {S.~T.}\ \bibnamefont
  {Sullivan}}, \bibinfo {author} {\bibfnamefont {S.}~\bibnamefont
  {Kotochigova}}, \bibinfo {author} {\bibfnamefont {A.}~\bibnamefont {Petrov}},
  \bibinfo {author} {\bibfnamefont {K.}~\bibnamefont {Chen}}, \bibinfo {author}
  {\bibfnamefont {S.~J.}\ \bibnamefont {Schowalter}}, \ and\ \bibinfo {author}
  {\bibfnamefont {E.~R.}\ \bibnamefont {Hudson}},\ }\href {\doibase
  10.1103/PhysRevLett.107.243201} {\bibfield  {journal} {\bibinfo  {journal}
  {Phys. Rev. Lett.}\ }\textbf {\bibinfo {volume} {107}},\ \bibinfo {pages}
  {243201} (\bibinfo {year} {2011})}\BibitemShut {NoStop}%
\bibitem [{\citenamefont {Haze}\ \emph {et~al.}(2013)\citenamefont {Haze},
  \citenamefont {Hata}, \citenamefont {Fujinaga},\ and\ \citenamefont
  {Mukaiyama}}]{HazePRA13}%
  \BibitemOpen
  \bibfield  {author} {\bibinfo {author} {\bibfnamefont {S.}~\bibnamefont
  {Haze}}, \bibinfo {author} {\bibfnamefont {S.}~\bibnamefont {Hata}}, \bibinfo
  {author} {\bibfnamefont {M.}~\bibnamefont {Fujinaga}}, \ and\ \bibinfo
  {author} {\bibfnamefont {T.}~\bibnamefont {Mukaiyama}},\ }\href {\doibase
  10.1103/PhysRevA.87.052715} {\bibfield  {journal} {\bibinfo  {journal} {Phys.
  Rev. A}\ }\textbf {\bibinfo {volume} {87}},\ \bibinfo {pages} {052715}
  (\bibinfo {year} {2013})}\BibitemShut {NoStop}%
\bibitem [{\citenamefont {Hall}\ \emph {et~al.}(2011)\citenamefont {Hall},
  \citenamefont {Aymar}, \citenamefont {Bouloufa-Maafa}, \citenamefont
  {Dulieu},\ and\ \citenamefont {Willitsch}}]{HallPRL11}%
  \BibitemOpen
  \bibfield  {author} {\bibinfo {author} {\bibfnamefont {F.~H.~J.}\
  \bibnamefont {Hall}}, \bibinfo {author} {\bibfnamefont {M.}~\bibnamefont
  {Aymar}}, \bibinfo {author} {\bibfnamefont {N.}~\bibnamefont
  {Bouloufa-Maafa}}, \bibinfo {author} {\bibfnamefont {O.}~\bibnamefont
  {Dulieu}}, \ and\ \bibinfo {author} {\bibfnamefont {S.}~\bibnamefont
  {Willitsch}},\ }\href {\doibase 10.1103/PhysRevLett.107.243202} {\bibfield
  {journal} {\bibinfo  {journal} {Phys. Rev. Lett.}\ }\textbf {\bibinfo
  {volume} {107}},\ \bibinfo {pages} {243202} (\bibinfo {year}
  {2011})}\BibitemShut {NoStop}%
\bibitem [{\citenamefont {Smith}\ \emph {et~al.}(2014)\citenamefont {Smith},
  \citenamefont {Goodman}, \citenamefont {Sivarajah}, \citenamefont {Wells},
  \citenamefont {Banerjee}, \citenamefont {C\^ot\'e}, \citenamefont {Michels},
  \citenamefont {Mongtomery},\ and\ \citenamefont {Narducci}}]{SmithAPB14}%
  \BibitemOpen
  \bibfield  {author} {\bibinfo {author} {\bibfnamefont {W.}~\bibnamefont
  {Smith}}, \bibinfo {author} {\bibfnamefont {D.}~\bibnamefont {Goodman}},
  \bibinfo {author} {\bibfnamefont {I.}~\bibnamefont {Sivarajah}}, \bibinfo
  {author} {\bibfnamefont {J.}~\bibnamefont {Wells}}, \bibinfo {author}
  {\bibfnamefont {S.}~\bibnamefont {Banerjee}}, \bibinfo {author}
  {\bibfnamefont {R.}~\bibnamefont {C\^ot\'e}}, \bibinfo {author}
  {\bibfnamefont {H.}~\bibnamefont {Michels}}, \bibinfo {author} {\bibfnamefont
  {J.~A.}\ \bibnamefont {Mongtomery}}, \ and\ \bibinfo {author} {\bibfnamefont
  {F.}~\bibnamefont {Narducci}},\ }\href {\doibase 10.1007/s00340-013-5672-2}
  {\bibfield  {journal} {\bibinfo  {journal} {Appl. Phys. B}\ }\textbf
  {\bibinfo {volume} {114}},\ \bibinfo {pages} {75} (\bibinfo {year}
  {2014})}\BibitemShut {NoStop}%
\bibitem [{\citenamefont {Meir}\ \emph {et~al.}(2016)\citenamefont {Meir},
  \citenamefont {Sikorsky}, \citenamefont {Ben-shlomi}, \citenamefont
  {Akerman}, \citenamefont {Dallal},\ and\ \citenamefont {Ozeri}}]{MeirPRL16}%
  \BibitemOpen
  \bibfield  {author} {\bibinfo {author} {\bibfnamefont {Z.}~\bibnamefont
  {Meir}}, \bibinfo {author} {\bibfnamefont {T.}~\bibnamefont {Sikorsky}},
  \bibinfo {author} {\bibfnamefont {R.}~\bibnamefont {Ben-shlomi}}, \bibinfo
  {author} {\bibfnamefont {N.}~\bibnamefont {Akerman}}, \bibinfo {author}
  {\bibfnamefont {Y.}~\bibnamefont {Dallal}}, \ and\ \bibinfo {author}
  {\bibfnamefont {R.}~\bibnamefont {Ozeri}},\ }\href {\doibase
  10.1103/PhysRevLett.117.243401} {\bibfield  {journal} {\bibinfo  {journal}
  {Phys. Rev. Lett.}\ }\textbf {\bibinfo {volume} {117}},\ \bibinfo {pages}
  {243401} (\bibinfo {year} {2016})}\BibitemShut {NoStop}%
\bibitem [{\citenamefont {Jyothi}\ \emph {et~al.}(2019)\citenamefont {Jyothi},
  \citenamefont {Egodapitiya}, \citenamefont {Bondurant}, \citenamefont {Jia},
  \citenamefont {Pretzsch}, \citenamefont {Chiappina}, \citenamefont {Shu},\
  and\ \citenamefont {Brown}}]{JyothiRSI19}%
  \BibitemOpen
  \bibfield  {author} {\bibinfo {author} {\bibfnamefont {S.}~\bibnamefont
  {Jyothi}}, \bibinfo {author} {\bibfnamefont {K.~N.}\ \bibnamefont
  {Egodapitiya}}, \bibinfo {author} {\bibfnamefont {B.}~\bibnamefont
  {Bondurant}}, \bibinfo {author} {\bibfnamefont {Z.}~\bibnamefont {Jia}},
  \bibinfo {author} {\bibfnamefont {E.}~\bibnamefont {Pretzsch}}, \bibinfo
  {author} {\bibfnamefont {P.}~\bibnamefont {Chiappina}}, \bibinfo {author}
  {\bibfnamefont {G.}~\bibnamefont {Shu}}, \ and\ \bibinfo {author}
  {\bibfnamefont {K.~R.}\ \bibnamefont {Brown}},\ }\href {\doibase
  10.1063/1.5121431} {\bibfield  {journal} {\bibinfo  {journal} {Rev. Sci.
  Instrum.}\ }\textbf {\bibinfo {volume} {90}},\ \bibinfo {pages} {103201}
  (\bibinfo {year} {2019})}\BibitemShut {NoStop}%
\bibitem [{\citenamefont {Schmidt}\ \emph {et~al.}(2020)\citenamefont
  {Schmidt}, \citenamefont {Weckesser}, \citenamefont {Thielemann},
  \citenamefont {Schaetz},\ and\ \citenamefont {Karpa}}]{SchmidtPRL2020}%
  \BibitemOpen
  \bibfield  {author} {\bibinfo {author} {\bibfnamefont {J.}~\bibnamefont
  {Schmidt}}, \bibinfo {author} {\bibfnamefont {P.}~\bibnamefont {Weckesser}},
  \bibinfo {author} {\bibfnamefont {F.}~\bibnamefont {Thielemann}}, \bibinfo
  {author} {\bibfnamefont {T.}~\bibnamefont {Schaetz}}, \ and\ \bibinfo
  {author} {\bibfnamefont {L.}~\bibnamefont {Karpa}},\ }\href {\doibase
  10.1103/PhysRevLett.124.053402} {\bibfield  {journal} {\bibinfo  {journal}
  {Phys. Rev. Lett.}\ }\textbf {\bibinfo {volume} {124}},\ \bibinfo {pages}
  {053402} (\bibinfo {year} {2020})}\BibitemShut {NoStop}%
\bibitem [{\citenamefont {H\"arter}\ \emph {et~al.}(2012)\citenamefont
  {H\"arter}, \citenamefont {Kr\"ukow}, \citenamefont {Brunner}, \citenamefont
  {Schnitzler}, \citenamefont {Schmid},\ and\ \citenamefont
  {Hecker~Denschlag}}]{HartePRL12}%
  \BibitemOpen
  \bibfield  {author} {\bibinfo {author} {\bibfnamefont {A.}~\bibnamefont
  {H\"arter}}, \bibinfo {author} {\bibfnamefont {A.}~\bibnamefont {Kr\"ukow}},
  \bibinfo {author} {\bibfnamefont {A.}~\bibnamefont {Brunner}}, \bibinfo
  {author} {\bibfnamefont {W.}~\bibnamefont {Schnitzler}}, \bibinfo {author}
  {\bibfnamefont {S.}~\bibnamefont {Schmid}}, \ and\ \bibinfo {author}
  {\bibfnamefont {J.}~\bibnamefont {Hecker~Denschlag}},\ }\href {\doibase
  10.1103/PhysRevLett.109.123201} {\bibfield  {journal} {\bibinfo  {journal}
  {Phys. Rev. Lett.}\ }\textbf {\bibinfo {volume} {109}},\ \bibinfo {pages}
  {123201} (\bibinfo {year} {2012})}\BibitemShut {NoStop}%
\bibitem [{\citenamefont {Singer}\ \emph {et~al.}(2010)\citenamefont {Singer},
  \citenamefont {Poschinger}, \citenamefont {Murphy}, \citenamefont {Ivanov},
  \citenamefont {Ziesel}, \citenamefont {Calarco},\ and\ \citenamefont
  {Schmidt-Kaler}}]{SingerRMP10}%
  \BibitemOpen
  \bibfield  {author} {\bibinfo {author} {\bibfnamefont {K.}~\bibnamefont
  {Singer}}, \bibinfo {author} {\bibfnamefont {U.}~\bibnamefont {Poschinger}},
  \bibinfo {author} {\bibfnamefont {M.}~\bibnamefont {Murphy}}, \bibinfo
  {author} {\bibfnamefont {P.}~\bibnamefont {Ivanov}}, \bibinfo {author}
  {\bibfnamefont {F.}~\bibnamefont {Ziesel}}, \bibinfo {author} {\bibfnamefont
  {T.}~\bibnamefont {Calarco}}, \ and\ \bibinfo {author} {\bibfnamefont
  {F.}~\bibnamefont {Schmidt-Kaler}},\ }\href {\doibase
  10.1103/RevModPhys.82.2609} {\bibfield  {journal} {\bibinfo  {journal} {Rev.
  Mod. Phys.}\ }\textbf {\bibinfo {volume} {82}},\ \bibinfo {pages} {2609}
  (\bibinfo {year} {2010})}\BibitemShut {NoStop}%
\bibitem [{\citenamefont {Schneider}\ \emph {et~al.}(2012)\citenamefont
  {Schneider}, \citenamefont {Porras},\ and\ \citenamefont
  {Schaetz}}]{SchneiderRPP12}%
  \BibitemOpen
  \bibfield  {author} {\bibinfo {author} {\bibfnamefont {C.}~\bibnamefont
  {Schneider}}, \bibinfo {author} {\bibfnamefont {D.}~\bibnamefont {Porras}}, \
  and\ \bibinfo {author} {\bibfnamefont {T.}~\bibnamefont {Schaetz}},\ }\href
  {\doibase 10.1088/0034-4885/75/2/024401} {\bibfield  {journal} {\bibinfo
  {journal} {Rep. Prog. Phys.}\ }\textbf {\bibinfo {volume} {75}},\ \bibinfo
  {pages} {024401} (\bibinfo {year} {2012})}\BibitemShut {NoStop}%
\bibitem [{\citenamefont {Idziaszek}\ \emph {et~al.}(2011)\citenamefont
  {Idziaszek}, \citenamefont {Simoni}, \citenamefont {Calarco},\ and\
  \citenamefont {Julienne}}]{IdziaszekNJP11}%
  \BibitemOpen
  \bibfield  {author} {\bibinfo {author} {\bibfnamefont {Z.}~\bibnamefont
  {Idziaszek}}, \bibinfo {author} {\bibfnamefont {A.}~\bibnamefont {Simoni}},
  \bibinfo {author} {\bibfnamefont {T.}~\bibnamefont {Calarco}}, \ and\
  \bibinfo {author} {\bibfnamefont {P.~S.}\ \bibnamefont {Julienne}},\ }\href
  {\doibase 10.1088/1367-2630/13/8/083005} {\bibfield  {journal} {\bibinfo
  {journal} {New J. Phys.}\ }\textbf {\bibinfo {volume} {13}},\ \bibinfo
  {pages} {083005} (\bibinfo {year} {2011})}\BibitemShut {NoStop}%
\bibitem [{\citenamefont {Tomza}\ \emph {et~al.}(2015)\citenamefont {Tomza},
  \citenamefont {Koch},\ and\ \citenamefont {Moszynski}}]{TomzaPRA15a}%
  \BibitemOpen
  \bibfield  {author} {\bibinfo {author} {\bibfnamefont {M.}~\bibnamefont
  {Tomza}}, \bibinfo {author} {\bibfnamefont {C.~P.}\ \bibnamefont {Koch}}, \
  and\ \bibinfo {author} {\bibfnamefont {R.}~\bibnamefont {Moszynski}},\ }\href
  {\doibase 10.1103/PhysRevA.91.042706} {\bibfield  {journal} {\bibinfo
  {journal} {Phys. Rev. A}\ }\textbf {\bibinfo {volume} {91}},\ \bibinfo
  {pages} {042706} (\bibinfo {year} {2015})}\BibitemShut {NoStop}%
\bibitem [{\citenamefont {da~Silva~Jr}\ \emph {et~al.}(2015)\citenamefont
  {da~Silva~Jr}, \citenamefont {Raoult}, \citenamefont {Aymar},\ and\
  \citenamefont {Dulieu}}]{daSilvaNJP2015}%
  \BibitemOpen
  \bibfield  {author} {\bibinfo {author} {\bibfnamefont {H.}~\bibnamefont
  {da~Silva~Jr}}, \bibinfo {author} {\bibfnamefont {M.}~\bibnamefont {Raoult}},
  \bibinfo {author} {\bibfnamefont {M.}~\bibnamefont {Aymar}}, \ and\ \bibinfo
  {author} {\bibfnamefont {O.}~\bibnamefont {Dulieu}},\ }\href {\doibase
  10.1088/1367-2630/17/4/045015} {\bibfield  {journal} {\bibinfo  {journal}
  {New J. Phys.}\ }\textbf {\bibinfo {volume} {17}},\ \bibinfo {pages} {045015}
  (\bibinfo {year} {2015})}\BibitemShut {NoStop}%
\bibitem [{\citenamefont {Gacesa}\ \emph
  {et~al.}(2016{\natexlab{a}})\citenamefont {Gacesa}, \citenamefont
  {Montgomery}, \citenamefont {Michels},\ and\ \citenamefont
  {C\^ot\'e}}]{GacesaPRA16}%
  \BibitemOpen
  \bibfield  {author} {\bibinfo {author} {\bibfnamefont {M.}~\bibnamefont
  {Gacesa}}, \bibinfo {author} {\bibfnamefont {J.~A.}\ \bibnamefont
  {Montgomery}}, \bibinfo {author} {\bibfnamefont {H.~H.}\ \bibnamefont
  {Michels}}, \ and\ \bibinfo {author} {\bibfnamefont {R.}~\bibnamefont
  {C\^ot\'e}},\ }\href {\doibase 10.1103/PhysRevA.94.013407} {\bibfield
  {journal} {\bibinfo  {journal} {Phys. Rev. A}\ }\textbf {\bibinfo {volume}
  {94}},\ \bibinfo {pages} {013407} (\bibinfo {year}
  {2016}{\natexlab{a}})}\BibitemShut {NoStop}%
\bibitem [{\citenamefont {Petrov}\ \emph {et~al.}(2017)\citenamefont {Petrov},
  \citenamefont {Makrides},\ and\ \citenamefont {Kotochigova}}]{PetrovJCP17}%
  \BibitemOpen
  \bibfield  {author} {\bibinfo {author} {\bibfnamefont {A.}~\bibnamefont
  {Petrov}}, \bibinfo {author} {\bibfnamefont {C.}~\bibnamefont {Makrides}}, \
  and\ \bibinfo {author} {\bibfnamefont {S.}~\bibnamefont {Kotochigova}},\
  }\href {\doibase 10.1063/1.4976972} {\bibfield  {journal} {\bibinfo
  {journal} {J. Chem. Phys.}\ }\textbf {\bibinfo {volume} {146}},\ \bibinfo
  {pages} {084304} (\bibinfo {year} {2017})}\BibitemShut {NoStop}%
\bibitem [{\citenamefont {Hall}\ \emph
  {et~al.}(2013{\natexlab{b}})\citenamefont {Hall}, \citenamefont {Aymar},
  \citenamefont {Raoult}, \citenamefont {Dulieu},\ and\ \citenamefont
  {Willitsch}}]{HallMP13b}%
  \BibitemOpen
  \bibfield  {author} {\bibinfo {author} {\bibfnamefont {F.~H.}\ \bibnamefont
  {Hall}}, \bibinfo {author} {\bibfnamefont {M.}~\bibnamefont {Aymar}},
  \bibinfo {author} {\bibfnamefont {M.}~\bibnamefont {Raoult}}, \bibinfo
  {author} {\bibfnamefont {O.}~\bibnamefont {Dulieu}}, \ and\ \bibinfo {author}
  {\bibfnamefont {S.}~\bibnamefont {Willitsch}},\ }\href {\doibase
  10.1080/00268976.2013.770930} {\bibfield  {journal} {\bibinfo  {journal}
  {Mol. Phys.}\ }\textbf {\bibinfo {volume} {111}},\ \bibinfo {pages} {1683}
  (\bibinfo {year} {2013}{\natexlab{b}})}\BibitemShut {NoStop}%
\bibitem [{\citenamefont {Jyothi}\ \emph {et~al.}(2016)\citenamefont {Jyothi},
  \citenamefont {Ray}, \citenamefont {Dutta}, \citenamefont {Allouche},
  \citenamefont {Vexiau}, \citenamefont {Dulieu},\ and\ \citenamefont
  {Rangwala}}]{JyothiPRL16}%
  \BibitemOpen
  \bibfield  {author} {\bibinfo {author} {\bibfnamefont {S.}~\bibnamefont
  {Jyothi}}, \bibinfo {author} {\bibfnamefont {T.}~\bibnamefont {Ray}},
  \bibinfo {author} {\bibfnamefont {S.}~\bibnamefont {Dutta}}, \bibinfo
  {author} {\bibfnamefont {A.~R.}\ \bibnamefont {Allouche}}, \bibinfo {author}
  {\bibfnamefont {R.}~\bibnamefont {Vexiau}}, \bibinfo {author} {\bibfnamefont
  {O.}~\bibnamefont {Dulieu}}, \ and\ \bibinfo {author} {\bibfnamefont {S.~A.}\
  \bibnamefont {Rangwala}},\ }\href {\doibase 10.1103/PhysRevLett.117.213002}
  {\bibfield  {journal} {\bibinfo  {journal} {Phys. Rev. Lett.}\ }\textbf
  {\bibinfo {volume} {117}},\ \bibinfo {pages} {213002} (\bibinfo {year}
  {2016})}\BibitemShut {NoStop}%
\bibitem [{\citenamefont {Sullivan}\ \emph {et~al.}(2011)\citenamefont
  {Sullivan}, \citenamefont {Rellergert}, \citenamefont {Kotochigova},
  \citenamefont {Chen}, \citenamefont {Schowalter},\ and\ \citenamefont
  {Hudson}}]{SullivanPCCP11}%
  \BibitemOpen
  \bibfield  {author} {\bibinfo {author} {\bibfnamefont {S.~T.}\ \bibnamefont
  {Sullivan}}, \bibinfo {author} {\bibfnamefont {W.~G.}\ \bibnamefont
  {Rellergert}}, \bibinfo {author} {\bibfnamefont {S.}~\bibnamefont
  {Kotochigova}}, \bibinfo {author} {\bibfnamefont {K.}~\bibnamefont {Chen}},
  \bibinfo {author} {\bibfnamefont {S.~J.}\ \bibnamefont {Schowalter}}, \ and\
  \bibinfo {author} {\bibfnamefont {E.~R.}\ \bibnamefont {Hudson}},\ }\href
  {\doibase 10.1039/C1CP21205B} {\bibfield  {journal} {\bibinfo  {journal}
  {Phys. Chem. Chem. Phys.}\ }\textbf {\bibinfo {volume} {13}},\ \bibinfo
  {pages} {18859} (\bibinfo {year} {2011})}\BibitemShut {NoStop}%
\bibitem [{\citenamefont {Deiglmayr}\ \emph {et~al.}(2012)\citenamefont
  {Deiglmayr}, \citenamefont {G\"oritz}, \citenamefont {Best}, \citenamefont
  {Weidem\"uller},\ and\ \citenamefont {Wester}}]{DeiglmayrPRA12}%
  \BibitemOpen
  \bibfield  {author} {\bibinfo {author} {\bibfnamefont {J.}~\bibnamefont
  {Deiglmayr}}, \bibinfo {author} {\bibfnamefont {A.}~\bibnamefont {G\"oritz}},
  \bibinfo {author} {\bibfnamefont {T.}~\bibnamefont {Best}}, \bibinfo {author}
  {\bibfnamefont {M.}~\bibnamefont {Weidem\"uller}}, \ and\ \bibinfo {author}
  {\bibfnamefont {R.}~\bibnamefont {Wester}},\ }\href {\doibase
  10.1103/PhysRevA.86.043438} {\bibfield  {journal} {\bibinfo  {journal} {Phys.
  Rev. A}\ }\textbf {\bibinfo {volume} {86}},\ \bibinfo {pages} {043438}
  (\bibinfo {year} {2012})}\BibitemShut {NoStop}%
\bibitem [{\citenamefont {Puri}\ \emph {et~al.}(2017)\citenamefont {Puri},
  \citenamefont {Mills}, \citenamefont {Schneider}, \citenamefont {Simbotin},
  \citenamefont {Montgomery}, \citenamefont {C{\^o}t{\'e}}, \citenamefont
  {Suits},\ and\ \citenamefont {Hudson}}]{PuriScience17}%
  \BibitemOpen
  \bibfield  {author} {\bibinfo {author} {\bibfnamefont {P.}~\bibnamefont
  {Puri}}, \bibinfo {author} {\bibfnamefont {M.}~\bibnamefont {Mills}},
  \bibinfo {author} {\bibfnamefont {C.}~\bibnamefont {Schneider}}, \bibinfo
  {author} {\bibfnamefont {I.}~\bibnamefont {Simbotin}}, \bibinfo {author}
  {\bibfnamefont {J.~A.}\ \bibnamefont {Montgomery}}, \bibinfo {author}
  {\bibfnamefont {R.}~\bibnamefont {C{\^o}t{\'e}}}, \bibinfo {author}
  {\bibfnamefont {A.~G.}\ \bibnamefont {Suits}}, \ and\ \bibinfo {author}
  {\bibfnamefont {E.~R.}\ \bibnamefont {Hudson}},\ }\href {\doibase
  10.1126/science.aan4701} {\bibfield  {journal} {\bibinfo  {journal}
  {Science}\ }\textbf {\bibinfo {volume} {357}},\ \bibinfo {pages} {1370}
  (\bibinfo {year} {2017})}\BibitemShut {NoStop}%
\bibitem [{\citenamefont {Kilaj}\ \emph {et~al.}(2018)\citenamefont {Kilaj},
  \citenamefont {Gao}, \citenamefont {R{\"o}sch}, \citenamefont {Rivero},
  \citenamefont {K{\"u}pper},\ and\ \citenamefont {Willitsch}}]{KilajNC18}%
  \BibitemOpen
  \bibfield  {author} {\bibinfo {author} {\bibfnamefont {A.}~\bibnamefont
  {Kilaj}}, \bibinfo {author} {\bibfnamefont {H.}~\bibnamefont {Gao}}, \bibinfo
  {author} {\bibfnamefont {D.}~\bibnamefont {R{\"o}sch}}, \bibinfo {author}
  {\bibfnamefont {U.}~\bibnamefont {Rivero}}, \bibinfo {author} {\bibfnamefont
  {J.}~\bibnamefont {K{\"u}pper}}, \ and\ \bibinfo {author} {\bibfnamefont
  {S.}~\bibnamefont {Willitsch}},\ }\href {\doibase 10.1038/s41467-018-04483-3}
  {\bibfield  {journal} {\bibinfo  {journal} {Nat. Comm.}\ }\textbf {\bibinfo
  {volume} {9}},\ \bibinfo {pages} {2096} (\bibinfo {year} {2018})}\BibitemShut
  {NoStop}%
\bibitem [{\citenamefont {D{\"o}rfler}\ \emph {et~al.}(2019)\citenamefont
  {D{\"o}rfler}, \citenamefont {Eberle}, \citenamefont {Koner}, \citenamefont
  {Tomza}, \citenamefont {Meuwly},\ and\ \citenamefont
  {Willitsch}}]{DorflerNC19}%
  \BibitemOpen
  \bibfield  {author} {\bibinfo {author} {\bibfnamefont {A.~D.}\ \bibnamefont
  {D{\"o}rfler}}, \bibinfo {author} {\bibfnamefont {P.}~\bibnamefont {Eberle}},
  \bibinfo {author} {\bibfnamefont {D.}~\bibnamefont {Koner}}, \bibinfo
  {author} {\bibfnamefont {M.}~\bibnamefont {Tomza}}, \bibinfo {author}
  {\bibfnamefont {M.}~\bibnamefont {Meuwly}}, \ and\ \bibinfo {author}
  {\bibfnamefont {S.}~\bibnamefont {Willitsch}},\ }\href {\doibase
  10.1038/s41467-019-13218-x} {\bibfield  {journal} {\bibinfo  {journal} {Nat.
  Commun.}\ }\textbf {\bibinfo {volume} {10}},\ \bibinfo {pages} {5429}
  (\bibinfo {year} {2019})}\BibitemShut {NoStop}%
\bibitem [{\citenamefont {Puri}\ \emph {et~al.}(2019)\citenamefont {Puri},
  \citenamefont {Mills}, \citenamefont {Simbotin}, \citenamefont {Montgomery},
  \citenamefont {C{\^o}t{\'e}}, \citenamefont {Schneider}, \citenamefont
  {Suits},\ and\ \citenamefont {Hudson}}]{PuriNatChem19}%
  \BibitemOpen
  \bibfield  {author} {\bibinfo {author} {\bibfnamefont {P.}~\bibnamefont
  {Puri}}, \bibinfo {author} {\bibfnamefont {M.}~\bibnamefont {Mills}},
  \bibinfo {author} {\bibfnamefont {I.}~\bibnamefont {Simbotin}}, \bibinfo
  {author} {\bibfnamefont {J.~A.}\ \bibnamefont {Montgomery}}, \bibinfo
  {author} {\bibfnamefont {R.}~\bibnamefont {C{\^o}t{\'e}}}, \bibinfo {author}
  {\bibfnamefont {C.}~\bibnamefont {Schneider}}, \bibinfo {author}
  {\bibfnamefont {A.~G.}\ \bibnamefont {Suits}}, \ and\ \bibinfo {author}
  {\bibfnamefont {E.~R.}\ \bibnamefont {Hudson}},\ }\href {\doibase
  10.1038/s41557-019-0264-3} {\bibfield  {journal} {\bibinfo  {journal} {Nat.
  Chem.}\ ,\ \bibinfo {pages} {615}} (\bibinfo {year} {2019})}\BibitemShut
  {NoStop}%
\bibitem [{\citenamefont {Germann}\ \emph {et~al.}(2014)\citenamefont
  {Germann}, \citenamefont {Tong},\ and\ \citenamefont
  {Willitsch}}]{GermannNatPhys14}%
  \BibitemOpen
  \bibfield  {author} {\bibinfo {author} {\bibfnamefont {M.}~\bibnamefont
  {Germann}}, \bibinfo {author} {\bibfnamefont {X.}~\bibnamefont {Tong}}, \
  and\ \bibinfo {author} {\bibfnamefont {S.}~\bibnamefont {Willitsch}},\ }\href
  {\doibase 10.1038/nphys3085} {\bibfield  {journal} {\bibinfo  {journal} {Nat.
  Phys.}\ }\textbf {\bibinfo {volume} {10}},\ \bibinfo {pages} {820} (\bibinfo
  {year} {2014})}\BibitemShut {NoStop}%
\bibitem [{\citenamefont {Cairncross}\ \emph {et~al.}(2017)\citenamefont
  {Cairncross}, \citenamefont {Gresh}, \citenamefont {Grau}, \citenamefont
  {Cossel}, \citenamefont {Roussy}, \citenamefont {Ni}, \citenamefont {Zhou},
  \citenamefont {Ye},\ and\ \citenamefont {Cornell}}]{Cairncross17}%
  \BibitemOpen
  \bibfield  {author} {\bibinfo {author} {\bibfnamefont {W.~B.}\ \bibnamefont
  {Cairncross}}, \bibinfo {author} {\bibfnamefont {D.~N.}\ \bibnamefont
  {Gresh}}, \bibinfo {author} {\bibfnamefont {M.}~\bibnamefont {Grau}},
  \bibinfo {author} {\bibfnamefont {K.~C.}\ \bibnamefont {Cossel}}, \bibinfo
  {author} {\bibfnamefont {T.~S.}\ \bibnamefont {Roussy}}, \bibinfo {author}
  {\bibfnamefont {Y.}~\bibnamefont {Ni}}, \bibinfo {author} {\bibfnamefont
  {Y.}~\bibnamefont {Zhou}}, \bibinfo {author} {\bibfnamefont {J.}~\bibnamefont
  {Ye}}, \ and\ \bibinfo {author} {\bibfnamefont {E.~A.}\ \bibnamefont
  {Cornell}},\ }\href {\doibase 10.1103/PhysRevLett.119.153001} {\bibfield
  {journal} {\bibinfo  {journal} {Phys. Rev. Lett.}\ }\textbf {\bibinfo
  {volume} {119}},\ \bibinfo {pages} {153001} (\bibinfo {year}
  {2017})}\BibitemShut {NoStop}%
\bibitem [{\citenamefont {Eberle}\ \emph {et~al.}(2016)\citenamefont {Eberle},
  \citenamefont {D\"orfler}, \citenamefont {Von~Planta}, \citenamefont {Ravi},\
  and\ \citenamefont {Willitsch}}]{EberleCPC16}%
  \BibitemOpen
  \bibfield  {author} {\bibinfo {author} {\bibfnamefont {P.}~\bibnamefont
  {Eberle}}, \bibinfo {author} {\bibfnamefont {A.~D.}\ \bibnamefont
  {D\"orfler}}, \bibinfo {author} {\bibfnamefont {C.}~\bibnamefont
  {Von~Planta}}, \bibinfo {author} {\bibfnamefont {K.}~\bibnamefont {Ravi}}, \
  and\ \bibinfo {author} {\bibfnamefont {S.}~\bibnamefont {Willitsch}},\ }\href
  {\doibase 10.1002/cphc.201600643} {\bibfield  {journal} {\bibinfo  {journal}
  {ChemPhysChem}\ }\textbf {\bibinfo {volume} {17}},\ \bibinfo {pages} {3769}
  (\bibinfo {year} {2016})}\BibitemShut {NoStop}%
\bibitem [{\citenamefont {Haze}\ \emph {et~al.}(2015)\citenamefont {Haze},
  \citenamefont {Saito}, \citenamefont {Fujinaga},\ and\ \citenamefont
  {Mukaiyama}}]{HazePRA15}%
  \BibitemOpen
  \bibfield  {author} {\bibinfo {author} {\bibfnamefont {S.}~\bibnamefont
  {Haze}}, \bibinfo {author} {\bibfnamefont {R.}~\bibnamefont {Saito}},
  \bibinfo {author} {\bibfnamefont {M.}~\bibnamefont {Fujinaga}}, \ and\
  \bibinfo {author} {\bibfnamefont {T.}~\bibnamefont {Mukaiyama}},\ }\href
  {\doibase 10.1103/PhysRevA.91.032709} {\bibfield  {journal} {\bibinfo
  {journal} {Phys. Rev. A}\ }\textbf {\bibinfo {volume} {91}},\ \bibinfo
  {pages} {032709} (\bibinfo {year} {2015})}\BibitemShut {NoStop}%
\bibitem [{\citenamefont {Haze}\ \emph {et~al.}(2018)\citenamefont {Haze},
  \citenamefont {Sasakawa}, \citenamefont {Saito}, \citenamefont {Nakai},\ and\
  \citenamefont {Mukaiyama}}]{HazePRL18}%
  \BibitemOpen
  \bibfield  {author} {\bibinfo {author} {\bibfnamefont {S.}~\bibnamefont
  {Haze}}, \bibinfo {author} {\bibfnamefont {M.}~\bibnamefont {Sasakawa}},
  \bibinfo {author} {\bibfnamefont {R.}~\bibnamefont {Saito}}, \bibinfo
  {author} {\bibfnamefont {R.}~\bibnamefont {Nakai}}, \ and\ \bibinfo {author}
  {\bibfnamefont {T.}~\bibnamefont {Mukaiyama}},\ }\href {\doibase
  10.1103/PhysRevLett.120.043401} {\bibfield  {journal} {\bibinfo  {journal}
  {Phys. Rev. Lett.}\ }\textbf {\bibinfo {volume} {120}},\ \bibinfo {pages}
  {043401} (\bibinfo {year} {2018})}\BibitemShut {NoStop}%
\bibitem [{\citenamefont {Cetina}\ \emph {et~al.}(2012)\citenamefont {Cetina},
  \citenamefont {Grier},\ and\ \citenamefont {Vuletic}}]{CetinaPRL12}%
  \BibitemOpen
  \bibfield  {author} {\bibinfo {author} {\bibfnamefont {M.}~\bibnamefont
  {Cetina}}, \bibinfo {author} {\bibfnamefont {A.~T.}\ \bibnamefont {Grier}}, \
  and\ \bibinfo {author} {\bibfnamefont {V.}~\bibnamefont {Vuletic}},\ }\href
  {\doibase 10.1103/PhysRevLett.109.253201} {\bibfield  {journal} {\bibinfo
  {journal} {Phys. Rev. Lett.}\ }\textbf {\bibinfo {volume} {109}},\ \bibinfo
  {pages} {253201} (\bibinfo {year} {2012})}\BibitemShut {NoStop}%
\bibitem [{\citenamefont {\ifmmode~\acute{S}\else \'{S}\fi{}mia\l{}kowski}\
  and\ \citenamefont {Tomza}(2020)}]{Smialkowski2019}%
  \BibitemOpen
  \bibfield  {author} {\bibinfo {author} {\bibfnamefont {M.}~\bibnamefont
  {\ifmmode~\acute{S}\else \'{S}\fi{}mia\l{}kowski}}\ and\ \bibinfo {author}
  {\bibfnamefont {M.}~\bibnamefont {Tomza}},\ }\href {\doibase
  10.1103/PhysRevA.101.012501} {\bibfield  {journal} {\bibinfo  {journal}
  {Phys. Rev. A}\ }\textbf {\bibinfo {volume} {101}},\ \bibinfo {pages}
  {012501} (\bibinfo {year} {2020})}\BibitemShut {NoStop}%
\bibitem [{\citenamefont {Bala}\ \emph {et~al.}(2019)\citenamefont {Bala},
  \citenamefont {Nataraj}, \citenamefont {Abe},\ and\ \citenamefont
  {Kajita}}]{Bala2019}%
  \BibitemOpen
  \bibfield  {author} {\bibinfo {author} {\bibfnamefont {R.}~\bibnamefont
  {Bala}}, \bibinfo {author} {\bibfnamefont {H.}~\bibnamefont {Nataraj}},
  \bibinfo {author} {\bibfnamefont {M.}~\bibnamefont {Abe}}, \ and\ \bibinfo
  {author} {\bibfnamefont {M.}~\bibnamefont {Kajita}},\ }\href {\doibase
  10.1080/00268976.2018.1539258} {\bibfield  {journal} {\bibinfo  {journal}
  {Mol. Phys.}\ }\textbf {\bibinfo {volume} {117}},\ \bibinfo {pages} {712}
  (\bibinfo {year} {2019})}\BibitemShut {NoStop}%
\bibitem [{\citenamefont {Habli}\ \emph {et~al.}(2016)\citenamefont {Habli},
  \citenamefont {Mejrissi}, \citenamefont {Ghalla}, \citenamefont {Yaghmour},
  \citenamefont {Oujia},\ and\ \citenamefont {Gad\'ea}}]{HabliMP16}%
  \BibitemOpen
  \bibfield  {author} {\bibinfo {author} {\bibfnamefont {H.}~\bibnamefont
  {Habli}}, \bibinfo {author} {\bibfnamefont {L.}~\bibnamefont {Mejrissi}},
  \bibinfo {author} {\bibfnamefont {H.}~\bibnamefont {Ghalla}}, \bibinfo
  {author} {\bibfnamefont {S.~J.}\ \bibnamefont {Yaghmour}}, \bibinfo {author}
  {\bibfnamefont {B.}~\bibnamefont {Oujia}}, \ and\ \bibinfo {author}
  {\bibfnamefont {F.~X.}\ \bibnamefont {Gad\'ea}},\ }\href {\doibase
  10.1080/00268976.2016.1140843} {\bibfield  {journal} {\bibinfo  {journal}
  {Mol. Phys.}\ }\textbf {\bibinfo {volume} {114}},\ \bibinfo {pages} {1568}
  (\bibinfo {year} {2016})}\BibitemShut {NoStop}%
\bibitem [{\citenamefont {Xie}\ and\ \citenamefont {Gong}(2005)}]{Xie2005}%
  \BibitemOpen
  \bibfield  {author} {\bibinfo {author} {\bibfnamefont {R.-H.}\ \bibnamefont
  {Xie}}\ and\ \bibinfo {author} {\bibfnamefont {J.}~\bibnamefont {Gong}},\
  }\href {\doibase 10.1103/PhysRevLett.95.263202} {\bibfield  {journal}
  {\bibinfo  {journal} {Phys. Rev. Lett.}\ }\textbf {\bibinfo {volume} {95}},\
  \bibinfo {pages} {263202} (\bibinfo {year} {2005})}\BibitemShut {NoStop}%
\bibitem [{\citenamefont {Russon}\ \emph {et~al.}(1998)\citenamefont {Russon},
  \citenamefont {Rothschopf}, \citenamefont {Morse}, \citenamefont {Boldyrev},\
  and\ \citenamefont {Simons}}]{Russon1998}%
  \BibitemOpen
  \bibfield  {author} {\bibinfo {author} {\bibfnamefont {L.}~\bibnamefont
  {Russon}}, \bibinfo {author} {\bibfnamefont {G.}~\bibnamefont {Rothschopf}},
  \bibinfo {author} {\bibfnamefont {M.}~\bibnamefont {Morse}}, \bibinfo
  {author} {\bibfnamefont {A.}~\bibnamefont {Boldyrev}}, \ and\ \bibinfo
  {author} {\bibfnamefont {J.}~\bibnamefont {Simons}},\ }\href {\doibase
  10.1063/1.477317} {\bibfield  {journal} {\bibinfo  {journal} {J. Chem.
  Phys.}\ }\textbf {\bibinfo {volume} {109}},\ \bibinfo {pages} {6655}
  (\bibinfo {year} {1998})}\BibitemShut {NoStop}%
\bibitem [{\citenamefont {Kimura}\ \emph {et~al.}(1983)\citenamefont {Kimura},
  \citenamefont {Sato},\ and\ \citenamefont {Olson}}]{Kimura1983}%
  \BibitemOpen
  \bibfield  {author} {\bibinfo {author} {\bibfnamefont {M.}~\bibnamefont
  {Kimura}}, \bibinfo {author} {\bibfnamefont {H.}~\bibnamefont {Sato}}, \ and\
  \bibinfo {author} {\bibfnamefont {R.}~\bibnamefont {Olson}},\ }\href
  {\doibase 10.1103/PhysRevA.28.2085} {\bibfield  {journal} {\bibinfo
  {journal} {Phys. Rev. A.}\ }\textbf {\bibinfo {volume} {28}},\ \bibinfo
  {pages} {2085} (\bibinfo {year} {1983})}\BibitemShut {NoStop}%
\bibitem [{\citenamefont {Gacesa}\ \emph
  {et~al.}(2016{\natexlab{b}})\citenamefont {Gacesa}, \citenamefont
  {Montgomery~Jr}, \citenamefont {Michels},\ and\ \citenamefont
  {C{\^o}t{\'e}}}]{Gacesa2016}%
  \BibitemOpen
  \bibfield  {author} {\bibinfo {author} {\bibfnamefont {M.}~\bibnamefont
  {Gacesa}}, \bibinfo {author} {\bibfnamefont {J.~A.}\ \bibnamefont
  {Montgomery~Jr}}, \bibinfo {author} {\bibfnamefont {H.~H.}\ \bibnamefont
  {Michels}}, \ and\ \bibinfo {author} {\bibfnamefont {R.}~\bibnamefont
  {C{\^o}t{\'e}}},\ }\href {\doibase 10.1103/PhysRevA.94.013407} {\bibfield
  {journal} {\bibinfo  {journal} {Phys. Rev. A.}\ }\textbf {\bibinfo {volume}
  {94}},\ \bibinfo {pages} {013407} (\bibinfo {year}
  {2016}{\natexlab{b}})}\BibitemShut {NoStop}%
\bibitem [{\citenamefont {Jellali}\ \emph {et~al.}(2018)\citenamefont
  {Jellali}, \citenamefont {Habli}, \citenamefont {Mejrissi}, \citenamefont
  {Hamdi}, \citenamefont {Oujia},\ and\ \citenamefont
  {Gad\'ea}}]{JellaliJQSRT18}%
  \BibitemOpen
  \bibfield  {author} {\bibinfo {author} {\bibfnamefont {S.}~\bibnamefont
  {Jellali}}, \bibinfo {author} {\bibfnamefont {H.}~\bibnamefont {Habli}},
  \bibinfo {author} {\bibfnamefont {L.}~\bibnamefont {Mejrissi}}, \bibinfo
  {author} {\bibfnamefont {R.}~\bibnamefont {Hamdi}}, \bibinfo {author}
  {\bibfnamefont {B.}~\bibnamefont {Oujia}}, \ and\ \bibinfo {author}
  {\bibfnamefont {F.~X.}\ \bibnamefont {Gad\'ea}},\ }\href {\doibase
  10.1016/j.jqsrt.2018.01.025} {\bibfield  {journal} {\bibinfo  {journal} {J.
  Quant. Spectrosc. Radiat. Transf.}\ }\textbf {\bibinfo {volume} {209}},\
  \bibinfo {pages} {45} (\bibinfo {year} {2018})}\BibitemShut {NoStop}%
\bibitem [{\citenamefont {Tacconi}\ \emph {et~al.}(2011)\citenamefont
  {Tacconi}, \citenamefont {Gianturco},\ and\ \citenamefont
  {Belyaev}}]{Tacconi2011}%
  \BibitemOpen
  \bibfield  {author} {\bibinfo {author} {\bibfnamefont {M.}~\bibnamefont
  {Tacconi}}, \bibinfo {author} {\bibfnamefont {F.}~\bibnamefont {Gianturco}},
  \ and\ \bibinfo {author} {\bibfnamefont {A.}~\bibnamefont {Belyaev}},\ }\href
  {\doibase 10.1039/c1cp20916g} {\bibfield  {journal} {\bibinfo  {journal}
  {Phys. Chem. Chem. Phys.}\ }\textbf {\bibinfo {volume} {13}},\ \bibinfo
  {pages} {19156–19164} (\bibinfo {year} {2011})}\BibitemShut {NoStop}%
\bibitem [{\citenamefont {Hall}\ \emph
  {et~al.}(2013{\natexlab{c}})\citenamefont {Hall}, \citenamefont {Eberle},
  \citenamefont {Hegi}, \citenamefont {Raoult}, \citenamefont {Aymar},
  \citenamefont {Dulieu},\ and\ \citenamefont {Willitsch}}]{Felix2013}%
  \BibitemOpen
  \bibfield  {author} {\bibinfo {author} {\bibfnamefont {F.~H.}\ \bibnamefont
  {Hall}}, \bibinfo {author} {\bibfnamefont {P.}~\bibnamefont {Eberle}},
  \bibinfo {author} {\bibfnamefont {G.}~\bibnamefont {Hegi}}, \bibinfo {author}
  {\bibfnamefont {M.}~\bibnamefont {Raoult}}, \bibinfo {author} {\bibfnamefont
  {M.}~\bibnamefont {Aymar}}, \bibinfo {author} {\bibfnamefont
  {O.}~\bibnamefont {Dulieu}}, \ and\ \bibinfo {author} {\bibfnamefont
  {S.}~\bibnamefont {Willitsch}},\ }\href {\doibase
  10.1080/00268976.2013.780107} {\bibfield  {journal} {\bibinfo  {journal}
  {Mol. Phys.}\ }\textbf {\bibinfo {volume} {111}},\ \bibinfo {pages} {2020}
  (\bibinfo {year} {2013}{\natexlab{c}})}\BibitemShut {NoStop}%
\bibitem [{\citenamefont {Berriche}(shed)}]{Berriche1995}%
  \BibitemOpen
  \bibfield  {author} {\bibinfo {author} {\bibfnamefont {H.}~\bibnamefont
  {Berriche}},\ }\emph {\bibinfo {title} {Etude spectroscopique ab initio de
  LiH et LiH+ au-delà de l'approximation de Born Oppenheimer}},\ \href@noop {}
  {Ph.D. thesis},\ \bibinfo  {school} {Paul Sabatier University} (\bibinfo
  {year} {1995 (\textit{unpublished})})\BibitemShut {NoStop}%
\bibitem [{\citenamefont {Khelifi}\ \emph {et~al.}(2002)\citenamefont
  {Khelifi}, \citenamefont {Zrafi}, \citenamefont {Oujia},\ and\ \citenamefont
  {Gadea}}]{Khelifi2002}%
  \BibitemOpen
  \bibfield  {author} {\bibinfo {author} {\bibfnamefont {N.}~\bibnamefont
  {Khelifi}}, \bibinfo {author} {\bibfnamefont {W.}~\bibnamefont {Zrafi}},
  \bibinfo {author} {\bibfnamefont {B.}~\bibnamefont {Oujia}}, \ and\ \bibinfo
  {author} {\bibfnamefont {F.~X.}\ \bibnamefont {Gadea}},\ }\href {\doibase
  10.1103/PhysRevA.65.042513} {\bibfield  {journal} {\bibinfo  {journal} {Phys.
  Rev. A.}\ }\textbf {\bibinfo {volume} {65}},\ \bibinfo {pages} {425131}
  (\bibinfo {year} {2002})}\BibitemShut {NoStop}%
\bibitem [{\citenamefont {Zrafi}\ \emph {et~al.}(2006)\citenamefont {Zrafi},
  \citenamefont {Oujia},\ and\ \citenamefont {Gadea}}]{Zarfi2006}%
  \BibitemOpen
  \bibfield  {author} {\bibinfo {author} {\bibfnamefont {W.}~\bibnamefont
  {Zrafi}}, \bibinfo {author} {\bibfnamefont {B.}~\bibnamefont {Oujia}}, \ and\
  \bibinfo {author} {\bibfnamefont {F.~X.}\ \bibnamefont {Gadea}},\ }\href
  {\doibase 10.1088/0953-4075/39/18/011} {\bibfield  {journal} {\bibinfo
  {journal} {J. Phys. B: At. Mol. Opt. Phys.}\ }\textbf {\bibinfo {volume}
  {39}},\ \bibinfo {pages} {3815} (\bibinfo {year} {2006})}\BibitemShut
  {NoStop}%
\bibitem [{\citenamefont {Khelifi}\ \emph {et~al.}(2001)\citenamefont
  {Khelifi}, \citenamefont {Oujia},\ and\ \citenamefont {Gadea}}]{Khelifi2001}%
  \BibitemOpen
  \bibfield  {author} {\bibinfo {author} {\bibfnamefont {N.}~\bibnamefont
  {Khelifi}}, \bibinfo {author} {\bibfnamefont {B.}~\bibnamefont {Oujia}}, \
  and\ \bibinfo {author} {\bibfnamefont {F.~X.}\ \bibnamefont {Gadea}},\ }\href
  {\doibase 10.1063/1.1436467} {\bibfield  {journal} {\bibinfo  {journal} {J.
  Chem. Phys.}\ }\textbf {\bibinfo {volume} {116}},\ \bibinfo {pages} {2879}
  (\bibinfo {year} {2001})}\BibitemShut {NoStop}%
\bibitem [{\citenamefont {Mabrouk}\ and\ \citenamefont
  {Berriche}(2014)}]{Mabrouk2014}%
  \BibitemOpen
  \bibfield  {author} {\bibinfo {author} {\bibfnamefont {N.}~\bibnamefont
  {Mabrouk}}\ and\ \bibinfo {author} {\bibfnamefont {H.}~\bibnamefont
  {Berriche}},\ }\href {\doibase 10.1021/jp5043427} {\bibfield  {journal}
  {\bibinfo  {journal} {J. Phys. Chem. A}\ }\textbf {\bibinfo {volume} {118}},\
  \bibinfo {pages} {8828−8841} (\bibinfo {year} {2014})}\BibitemShut
  {NoStop}%
\bibitem [{\citenamefont {Mabrouk}\ and\ \citenamefont
  {Berriche}(2008)}]{Mabrouk2008}%
  \BibitemOpen
  \bibfield  {author} {\bibinfo {author} {\bibfnamefont {N.}~\bibnamefont
  {Mabrouk}}\ and\ \bibinfo {author} {\bibfnamefont {H.}~\bibnamefont
  {Berriche}},\ }\href {\doibase 10.1088/0953-4075/41/15/155101} {\bibfield
  {journal} {\bibinfo  {journal} {J. Phys. B: At. Mol. Opt. Phys.}\ }\textbf
  {\bibinfo {volume} {41}},\ \bibinfo {pages} {155101} (\bibinfo {year}
  {2008})}\BibitemShut {NoStop}%
\bibitem [{\citenamefont {Mabrouk}\ \emph {et~al.}(2010)\citenamefont
  {Mabrouk}, \citenamefont {Berriche}, \citenamefont {Ouada},\ and\
  \citenamefont {Gadea}}]{Mabrouk2010}%
  \BibitemOpen
  \bibfield  {author} {\bibinfo {author} {\bibfnamefont {N.}~\bibnamefont
  {Mabrouk}}, \bibinfo {author} {\bibfnamefont {H.}~\bibnamefont {Berriche}},
  \bibinfo {author} {\bibfnamefont {H.~B.}\ \bibnamefont {Ouada}}, \ and\
  \bibinfo {author} {\bibfnamefont {F.}~\bibnamefont {Gadea}},\ }\href
  {\doibase 10.1021/jp101588v} {\bibfield  {journal} {\bibinfo  {journal} {J.
  Phys. Chem.A}\ }\textbf {\bibinfo {volume} {114}},\ \bibinfo {pages} {6657}
  (\bibinfo {year} {2010})}\BibitemShut {NoStop}%
\bibitem [{\citenamefont {Jendoubi}\ \emph {et~al.}(2012)\citenamefont
  {Jendoubi}, \citenamefont {Berriche}, \citenamefont {Ben~Ouada},\ and\
  \citenamefont {Gadea}}]{Jendoubi2012}%
  \BibitemOpen
  \bibfield  {author} {\bibinfo {author} {\bibfnamefont {I.}~\bibnamefont
  {Jendoubi}}, \bibinfo {author} {\bibfnamefont {H.}~\bibnamefont {Berriche}},
  \bibinfo {author} {\bibfnamefont {H.}~\bibnamefont {Ben~Ouada}}, \ and\
  \bibinfo {author} {\bibfnamefont {F.}~\bibnamefont {Gadea}},\ }\href
  {\doibase 10.1021/jp209106w} {\bibfield  {journal} {\bibinfo  {journal} {J.
  Phys. Chem. A}\ }\textbf {\bibinfo {volume} {116}},\ \bibinfo {pages}
  {2945–2960} (\bibinfo {year} {2012})}\BibitemShut {NoStop}%
\bibitem [{\citenamefont {Aymar}\ and\ \citenamefont
  {Dulieu}(2012)}]{Aymar2012}%
  \BibitemOpen
  \bibfield  {author} {\bibinfo {author} {\bibfnamefont {M.}~\bibnamefont
  {Aymar}}\ and\ \bibinfo {author} {\bibfnamefont {O.}~\bibnamefont {Dulieu}},\
  }\href {\doibase 10.1088/0953-4075/45/21/215103} {\bibfield  {journal}
  {\bibinfo  {journal} {J. Phys. B:At. Mol. Opt. Phys.}\ }\textbf {\bibinfo
  {volume} {45}},\ \bibinfo {pages} {215103} (\bibinfo {year}
  {2012})}\BibitemShut {NoStop}%
\bibitem [{\citenamefont {Habli}(shed)}]{Habli2013}%
  \BibitemOpen
  \bibfield  {author} {\bibinfo {author} {\bibfnamefont {H.}~\bibnamefont
  {Habli}},\ }\emph {\bibinfo {title} {Etude spectroscopique théorique
  ab-initio au-delà de l'approximation de Born Oppenheimer de l'hydrure de
  calcium CaH et de son ion CaH+}},\ \href@noop {} {Ph.D. thesis},\ \bibinfo
  {school} {University of Monastir(Tunisia)} (\bibinfo {year} {2013
  (\textit{unpublished})})\BibitemShut {NoStop}%
\bibitem [{\citenamefont {Belayouni}\ \emph {et~al.}(2016)\citenamefont
  {Belayouni}, \citenamefont {Ghanmi},\ and\ \citenamefont
  {Berriche}}]{Belayouni2016}%
  \BibitemOpen
  \bibfield  {author} {\bibinfo {author} {\bibfnamefont {S.}~\bibnamefont
  {Belayouni}}, \bibinfo {author} {\bibfnamefont {C.}~\bibnamefont {Ghanmi}}, \
  and\ \bibinfo {author} {\bibfnamefont {H.}~\bibnamefont {Berriche}},\ }\href
  {\doibase 10.1139/cjp-2015-0801} {\bibfield  {journal} {\bibinfo  {journal}
  {Can. J. Phys.}\ }\textbf {\bibinfo {volume} {94}},\ \bibinfo {pages} {791}
  (\bibinfo {year} {2016})}\BibitemShut {NoStop}%
\bibitem [{\citenamefont {ElOualhazi}\ and\ \citenamefont
  {Berriche}(2016)}]{ElOualhazi2016}%
  \BibitemOpen
  \bibfield  {author} {\bibinfo {author} {\bibfnamefont {R.}~\bibnamefont
  {ElOualhazi}}\ and\ \bibinfo {author} {\bibfnamefont {H.}~\bibnamefont
  {Berriche}},\ }\href {\doibase 10.1021/acs.jpca.5b10209} {\bibfield
  {journal} {\bibinfo  {journal} {J. Phys. Chem. A}\ }\textbf {\bibinfo
  {volume} {120}},\ \bibinfo {pages} {452–465} (\bibinfo {year}
  {2016})}\BibitemShut {NoStop}%
\bibitem [{\citenamefont {Aymar}\ \emph {et~al.}(2011)\citenamefont {Aymar},
  \citenamefont {Gu{\'e}rout},\ and\ \citenamefont {Dulieu}}]{Aymar2011}%
  \BibitemOpen
  \bibfield  {author} {\bibinfo {author} {\bibfnamefont {M.}~\bibnamefont
  {Aymar}}, \bibinfo {author} {\bibfnamefont {R.}~\bibnamefont {Gu{\'e}rout}},
  \ and\ \bibinfo {author} {\bibfnamefont {O.}~\bibnamefont {Dulieu}},\ }\href
  {\doibase 10.1063/1.3611399} {\bibfield  {journal} {\bibinfo  {journal} {J.
  Chem. Phys.}\ }\textbf {\bibinfo {volume} {135}},\ \bibinfo {pages} {064305}
  (\bibinfo {year} {2011})}\BibitemShut {NoStop}%
\bibitem [{\citenamefont {Durand}\ and\ \citenamefont
  {Barthelat}(1975)}]{Durand1975}%
  \BibitemOpen
  \bibfield  {author} {\bibinfo {author} {\bibfnamefont {P.}~\bibnamefont
  {Durand}}\ and\ \bibinfo {author} {\bibfnamefont {J.-C.}\ \bibnamefont
  {Barthelat}},\ }\href {\doibase 10.1007/BF00963468} {\bibfield  {journal}
  {\bibinfo  {journal} {Chim. Acta}\ }\textbf {\bibinfo {volume} {38}},\
  \bibinfo {pages} {283–302} (\bibinfo {year} {1975})}\BibitemShut {NoStop}%
\bibitem [{\citenamefont {M{\"u}ller}\ \emph {et~al.}(1984)\citenamefont
  {M{\"u}ller}, \citenamefont {Flesch},\ and\ \citenamefont
  {Meyer}}]{Müller1984}%
  \BibitemOpen
  \bibfield  {author} {\bibinfo {author} {\bibfnamefont {W.}~\bibnamefont
  {M{\"u}ller}}, \bibinfo {author} {\bibfnamefont {J.}~\bibnamefont {Flesch}},
  \ and\ \bibinfo {author} {\bibfnamefont {W.}~\bibnamefont {Meyer}},\ }\href
  {\doibase 10.1063/1.447083} {\bibfield  {journal} {\bibinfo  {journal} {J.
  Chem. Phys}\ }\textbf {\bibinfo {volume} {80}},\ \bibinfo {pages} {3297}
  (\bibinfo {year} {1984})}\BibitemShut {NoStop}%
\bibitem [{\citenamefont {Mitroy}\ and\ \citenamefont
  {Zhang}(2008)}]{Mitroy2008}%
  \BibitemOpen
  \bibfield  {author} {\bibinfo {author} {\bibfnamefont {J.}~\bibnamefont
  {Mitroy}}\ and\ \bibinfo {author} {\bibfnamefont {J.}~\bibnamefont {Zhang}},\
  }\href {\doibase 10.1140/epjd/e2007-00320-5} {\bibfield  {journal} {\bibinfo
  {journal} {Eur. Phys. J. D}\ }\textbf {\bibinfo {volume} {46}},\ \bibinfo
  {pages} {415} (\bibinfo {year} {2008})}\BibitemShut {NoStop}%
\bibitem [{\citenamefont {Johnson}(1978)}]{JohnsonJCP78}%
  \BibitemOpen
  \bibfield  {author} {\bibinfo {author} {\bibfnamefont {B.~R.}\ \bibnamefont
  {Johnson}},\ }\href {\doibase 10.1063/1.436421} {\bibfield  {journal}
  {\bibinfo  {journal} {J. Chem. Phys.}\ }\textbf {\bibinfo {volume} {69}},\
  \bibinfo {pages} {4678} (\bibinfo {year} {1978})}\BibitemShut {NoStop}%
\bibitem [{\citenamefont {Tomza}(2015)}]{TomzaPRA15b}%
  \BibitemOpen
  \bibfield  {author} {\bibinfo {author} {\bibfnamefont {M.}~\bibnamefont
  {Tomza}},\ }\href {\doibase 10.1103/PhysRevA.92.062701} {\bibfield  {journal}
  {\bibinfo  {journal} {Phys. Rev. A}\ }\textbf {\bibinfo {volume} {92}},\
  \bibinfo {pages} {062701} (\bibinfo {year} {2015})}\BibitemShut {NoStop}%
\bibitem [{nis()}]{nist2018}%
  \BibitemOpen
  \href@noop {} {}\bibinfo {note} {{NIST Atomic Spectra Database
  http://physics.nist.gov/PhysRefData/ASD}}\BibitemShut {NoStop}%
\bibitem [{\citenamefont {Zemke}\ \emph {et~al.}(1978)\citenamefont {Zemke},
  \citenamefont {Crooks},\ and\ \citenamefont {Stwalley}}]{Zemke1978}%
  \BibitemOpen
  \bibfield  {author} {\bibinfo {author} {\bibfnamefont {W.~T.}\ \bibnamefont
  {Zemke}}, \bibinfo {author} {\bibfnamefont {J.~B.}\ \bibnamefont {Crooks}}, \
  and\ \bibinfo {author} {\bibfnamefont {W.~C.}\ \bibnamefont {Stwalley}},\
  }\href {\doibase 10.1063/1.435569} {\bibfield  {journal} {\bibinfo  {journal}
  {J. Chem. Phys.}\ }\textbf {\bibinfo {volume} {68}},\ \bibinfo {pages} {4628}
  (\bibinfo {year} {1978})}\BibitemShut {NoStop}%
\bibitem [{\citenamefont {Partridge}\ and\ \citenamefont
  {Langhoff}(1981)}]{Partridge1981}%
  \BibitemOpen
  \bibfield  {author} {\bibinfo {author} {\bibfnamefont {H.}~\bibnamefont
  {Partridge}}\ and\ \bibinfo {author} {\bibfnamefont {S.~R.}\ \bibnamefont
  {Langhoff}},\ }\href {\doibase 10.1063/1.441355} {\bibfield  {journal}
  {\bibinfo  {journal} {J. Chem. Phys.}\ }\textbf {\bibinfo {volume} {74}},\
  \bibinfo {pages} {2361} (\bibinfo {year} {1981})}\BibitemShut {NoStop}%
\bibitem [{\citenamefont {Pazyuk}\ \emph {et~al.}(1994)\citenamefont {Pazyuk},
  \citenamefont {Stolyarov},\ and\ \citenamefont {Pupyshev}}]{Pazyuk1994}%
  \BibitemOpen
  \bibfield  {author} {\bibinfo {author} {\bibfnamefont {E.}~\bibnamefont
  {Pazyuk}}, \bibinfo {author} {\bibfnamefont {A.}~\bibnamefont {Stolyarov}}, \
  and\ \bibinfo {author} {\bibfnamefont {V.}~\bibnamefont {Pupyshev}},\ }\href
  {\doibase 10.1016/0009-2614(94)00900-7} {\bibfield  {journal} {\bibinfo
  {journal} {Chem. Phys. Lett.}\ }\textbf {\bibinfo {volume} {228}},\ \bibinfo
  {pages} {219} (\bibinfo {year} {1994})}\BibitemShut {NoStop}%
\bibitem [{\citenamefont {Sayfutyarova}\ \emph {et~al.}(2013)\citenamefont
  {Sayfutyarova}, \citenamefont {Buchachenko}, \citenamefont {Yakovleva},\ and\
  \citenamefont {Belyaev}}]{SayfutyarovaPRA13}%
  \BibitemOpen
  \bibfield  {author} {\bibinfo {author} {\bibfnamefont {E.~R.}\ \bibnamefont
  {Sayfutyarova}}, \bibinfo {author} {\bibfnamefont {A.~A.}\ \bibnamefont
  {Buchachenko}}, \bibinfo {author} {\bibfnamefont {S.~A.}\ \bibnamefont
  {Yakovleva}}, \ and\ \bibinfo {author} {\bibfnamefont {A.~K.}\ \bibnamefont
  {Belyaev}},\ }\href {\doibase 10.1103/PhysRevA.87.052717} {\bibfield
  {journal} {\bibinfo  {journal} {Phys. Rev. A}\ }\textbf {\bibinfo {volume}
  {87}},\ \bibinfo {pages} {052717} (\bibinfo {year} {2013})}\BibitemShut
  {NoStop}%
\bibitem [{\citenamefont {Li}\ \emph {et~al.}(2019)\citenamefont {Li},
  \citenamefont {Mills}, \citenamefont {Puri}, \citenamefont {Petrov},
  \citenamefont {Hudson},\ and\ \citenamefont {Kotochigova}}]{LiPRA19}%
  \BibitemOpen
  \bibfield  {author} {\bibinfo {author} {\bibfnamefont {M.}~\bibnamefont
  {Li}}, \bibinfo {author} {\bibfnamefont {M.}~\bibnamefont {Mills}}, \bibinfo
  {author} {\bibfnamefont {P.}~\bibnamefont {Puri}}, \bibinfo {author}
  {\bibfnamefont {A.}~\bibnamefont {Petrov}}, \bibinfo {author} {\bibfnamefont
  {E.~R.}\ \bibnamefont {Hudson}}, \ and\ \bibinfo {author} {\bibfnamefont
  {S.}~\bibnamefont {Kotochigova}},\ }\href {\doibase
  10.1103/PhysRevA.99.062706} {\bibfield  {journal} {\bibinfo  {journal} {Phys.
  Rev. A}\ }\textbf {\bibinfo {volume} {99}},\ \bibinfo {pages} {062706}
  (\bibinfo {year} {2019})}\BibitemShut {NoStop}%
\bibitem [{\citenamefont {Krych}\ \emph {et~al.}(2011)\citenamefont {Krych},
  \citenamefont {Skomorowski}, \citenamefont {Paw{\l}owski}, \citenamefont
  {Moszynski},\ and\ \citenamefont {Idziaszek}}]{Krych2011}%
  \BibitemOpen
  \bibfield  {author} {\bibinfo {author} {\bibfnamefont {M.}~\bibnamefont
  {Krych}}, \bibinfo {author} {\bibfnamefont {W.}~\bibnamefont {Skomorowski}},
  \bibinfo {author} {\bibfnamefont {F.}~\bibnamefont {Paw{\l}owski}}, \bibinfo
  {author} {\bibfnamefont {R.}~\bibnamefont {Moszynski}}, \ and\ \bibinfo
  {author} {\bibfnamefont {Z.}~\bibnamefont {Idziaszek}},\ }\href {\doibase
  doi.org/10.1103/PhysRevA.83.032723} {\bibfield  {journal} {\bibinfo
  {journal} {Phys. Rev. A}\ }\textbf {\bibinfo {volume} {83}},\ \bibinfo
  {pages} {032723} (\bibinfo {year} {2011})}\BibitemShut {NoStop}%
\bibitem [{\citenamefont {Levine}(2009)}]{Levine09}%
  \BibitemOpen
  \bibfield  {author} {\bibinfo {author} {\bibfnamefont {R.~D.}\ \bibnamefont
  {Levine}},\ }\href@noop {} {\emph {\bibinfo {title} {Molecular reaction
  dynamics}}}\ (\bibinfo  {publisher} {Cambridge University Press},\ \bibinfo
  {year} {2009})\BibitemShut {NoStop}%
\bibitem [{\citenamefont {Jones}\ \emph {et~al.}(2006)\citenamefont {Jones},
  \citenamefont {Tiesinga}, \citenamefont {Lett},\ and\ \citenamefont
  {Julienne}}]{JonesRMP06}%
  \BibitemOpen
  \bibfield  {author} {\bibinfo {author} {\bibfnamefont {K.~M.}\ \bibnamefont
  {Jones}}, \bibinfo {author} {\bibfnamefont {E.}~\bibnamefont {Tiesinga}},
  \bibinfo {author} {\bibfnamefont {P.~D.}\ \bibnamefont {Lett}}, \ and\
  \bibinfo {author} {\bibfnamefont {P.~S.}\ \bibnamefont {Julienne}},\ }\href
  {\doibase 10.1103/RevModPhys.78.483} {\bibfield  {journal} {\bibinfo
  {journal} {Rev. Mod. Phys.}\ }\textbf {\bibinfo {volume} {78}},\ \bibinfo
  {pages} {483} (\bibinfo {year} {2006})}\BibitemShut {NoStop}%
\bibitem [{\citenamefont {Zemke}\ and\ \citenamefont
  {Stwalley}(2004)}]{Zemke2004}%
  \BibitemOpen
  \bibfield  {author} {\bibinfo {author} {\bibfnamefont {W.~T.}\ \bibnamefont
  {Zemke}}\ and\ \bibinfo {author} {\bibfnamefont {W.~C.}\ \bibnamefont
  {Stwalley}},\ }\href {\doibase 10.1063/1.1630299} {\bibfield  {journal}
  {\bibinfo  {journal} {J. Chem. Phys.}\ }\textbf {\bibinfo {volume} {120}},\
  \bibinfo {pages} {88} (\bibinfo {year} {2004})}\BibitemShut {NoStop}%
\bibitem [{\citenamefont {Fedorov}\ \emph {et~al.}(2017)\citenamefont
  {Fedorov}, \citenamefont {Barnes},\ and\ \citenamefont
  {Varganov}}]{Fedorov2017}%
  \BibitemOpen
  \bibfield  {author} {\bibinfo {author} {\bibfnamefont {D.~A.}\ \bibnamefont
  {Fedorov}}, \bibinfo {author} {\bibfnamefont {D.~K.}\ \bibnamefont {Barnes}},
  \ and\ \bibinfo {author} {\bibfnamefont {S.~A.}\ \bibnamefont {Varganov}},\
  }\href {\doibase 10.1063/1.4986818} {\bibfield  {journal} {\bibinfo
  {journal} {J. Chem. Phys.}\ }\textbf {\bibinfo {volume} {147}},\ \bibinfo
  {pages} {124304} (\bibinfo {year} {2017})}\BibitemShut {NoStop}%
\end{thebibliography}%

\end{document}